\documentclass[%
reprint,
superscriptaddress,
showpacs,preprintnumbers,
 amsmath,amssymb,
 aps,
]{revtex4-1}

\usepackage{graphicx}
\usepackage{dcolumn}
\usepackage{bm}

\usepackage{references}
\usepackage{hyperref} 

\usepackage{equationarray}
\usepackage[normalem]{ulem}
\usepackage[usenames]{xcolor}
\usepackage{booktabs}

\newcommand{\be}{\begin{equation}}
\newcommand{\ee}{\end{equation}}
\newcommand{\ba}{\begin{eqnarray}}
\newcommand{\ea}{\end{eqnarray}}
\newcommand{\bd}{\begin{displaymath}}
\newcommand{\ed}{\end{displaymath}}

\usepackage{ulem} 

\begin{document}

\preprint{APS/123-QED}

\title{Thermal properties of supernova matter: The bulk homogeneous phase }

\author{Constantinos Constantinou}
\email{cconstan@helios.phy.ohiou.edu }
\affiliation{Department of Physics and Astronomy, Ohio University, Athens, OH 45701}
\affiliation{Department of Physics and Astronomy, Stony Brook University, Stony Brook, NY 11794-3800}

\author{Brian Muccioli}
\email{bm956810@ohio.edu}
\affiliation{Department of Physics and Astronomy, Ohio University, Athens, OH 45701}

\author{Madappa Prakash}
\email{prakash@phy.ohiou.edu}
\affiliation{Department of Physics and Astronomy, Ohio University, Athens, OH 45701}

\author{James M. Lattimer}
\email{james.lattimer@stonybrook.edu}
\affiliation{Department of Physics and Astronomy, Stony Brook University, Stony Brook, NY 11794-3800}

\date{\today}

\begin{abstract}
We investigate the thermal properties of the potential model equation
of state of Akmal, Pandharipande and Ravenhall.  This equation of
state approximates the microscopic model calculations of Akmal and
Pandharipande, which feature a neutral pion condensate.  We
treat the bulk homogeneous phase for isospin asymmetries ranging from
symmetric nuclear matter to pure neutron matter and for temperatures
and densities relevant for simulations of core-collapse supernovae,
proto-neutron stars, and neutron star mergers.  Numerical results of
the state variables are compared with those of a typical Skyrme energy
density functional with similar properties at nuclear densities, but
which differs substantially at supra-nuclear densities. Analytical
formulas, which are applicable to non-relativistic potential models
such as the equations of state we are considering, are derived for all
state variables and their thermodynamic derivatives.  A highlight of
our work is its focus on thermal response functions in the degenerate
and non-degenerate situations, which allow checks of the numerical
calculations for arbitrary degeneracy.  These functions are sensitive
to the density dependent effective masses of neutrons and protons,
which determine the thermal properties in all regimes of degeneracy.
We develop the ``thermal asymmetry free energy'' and establish its
relation to the more commonly used nuclear symmetry energy. We also 
explore the role of the pion condensate at supra-nuclear densities and 
temperatures. Tables of matter properties as functions of baryon density, 
composition (i.e., proton fraction) and temperature are being produced which 
are suitable for use in astrophysical simulations of supernovae and 
neutron stars.  \\

\noindent Keywords: Supernova matter, potential models, thermal effects.

\end{abstract}

\pacs{21.65.Mn,26.50.+x,51.30.+i,97.60.Bw}
\maketitle


\section{INTRODUCTION}
\label{Sec:Intro}

The equation of state (EOS) of dense, hot matter is an essential
ingredient in modeling neutron stars and hydrodynamical simulations of
astrophysical phenomena such as core-collapse supernova explosions,
proto-neutron stars, and compact object mergers.  In broad terms, two
major regions for the EOS can be identified at relatively low
temperatures or entropies.  At sub-nuclear densities ($n$ of $10^{-7}$ to
$\sim 0.1~{\rm fm}^{-3}$), matter is in an inhomogeneous mixture of
nucleons (neutrons and protons), light nuclear clusters (alpha
particles, deuterons, tritons etc.), and heavy nuclei.  Leptons,
mainly electrons, are also present to balance the nuclear charges.
Uniform matter, and heavy nuclei become progressively more
neutron-rich as the density rises.  Above about 0.01 fm$^{-3}$, nuclei
deform in resonse to competition between surface and Coulomb energies,
which may also lead to pasta-like geometrical configurations.  By the
density 0.1 fm$^{-3}$, the inhomogeneous phase gives way to a uniform
phase of nucleons and electrons.  Above the nuclear saturation
density, $n_0\simeq0.16$ fm$^{-3}$, the uniform phase may become
populated with more exotic matter, including Bose (pion or kaon)
condensates, hyperons and deconfined quark matter.  The appearance of
Bose condensates and deconfined quark matter may be through first-order
or continuous phase transitions.

At large-enough temperatures below $n_0$, the inhomogeneous phase
disappears and is again replaced by a uniform phase of nucleons and
electrons.  At sufficiently high temperatures at every density,
thermal populations of hadrons and pions should appear.  

The composition and thermodynamic properties of matter at a given
density $n$, temperature $T$, and overall charge fraction
(parametrized by the electron concentration $Y_e=n_e/n$) is determined by
minimizing the free energy density.  In all realistic situations,
matter is charge-neutral, but the net baryonic charge is non-zero and
equalized by the net leptonic charge.  It can generally be assumed
that baryonic species are in strong interaction equilibrium, but
equilibrium does not always exist for leptonic species which are
subject to weak interactions.  In circumstances in which dynamical
timescales are long compared to weak interaction timescales, the free
energy minimization is also made with respect to $Y_e$.  Such matter
is said to be in beta equilibrium and its properties are a function of
only density and temperature, and, if neutrinos are trapped in 
matter, the total number of leptons per baryon.  Below $n_0$, where
generally the only baryons are neutrons and protons and the only
leptons are electrons and possibly neutrinos, charge neutrality
dictates that the number of electrons per baryon $Y_e$ equals the
proton fraction $x=n_p/n$, but at higher densities, the charge fractions of
muons, hyperons, Bose condensates and quarks, if present, have to be
included.  Beta equilibrium may not occur during gravitational
collapse or dynamical expansion, such as occurs in Type II
supernovae and neutron star mergers.

The free energy can be calculated using a variety of methods, but it
is generally a complicated functional of the main physical variables $n$,
$T$ and $Y_e$ and cannot be expressed analytically.  In order to
efficiently describe the EOS, it is customary to build
three-dimensional tables of its properties.  An essential criterion is
that full EOS tables be thermodynamically consistent so as not to
generate spurious and unphysical entropy during hydrodynamical
simulations.  Beginning with the work of Lattimer and Swesty
(hereafter referred to as LS)\cite{LS}, examples of such tables
include the works of Shen et al \cite{Shen1}, Shen et al
\cite{Shen2a,Shen2b}, and others \cite{Steiner,Hempel}.  We refer the
reader to Refs. \cite{SLM94,Oconnor,sys07,Steiner} in which
comparisons of outcomes in supernova simulations, for pre-bounce
evolution and black hole formation, respectively, have been made using
different EOSs.  A parallel study \cite{bbj13} of neutron star mergers
with different EOSs has also been undertaken.

The EOS, in addition to controlling the global hydrodynamical
evolution, also determines weak interaction rates including those of
electron capture and beta decay reactions and neutrino-matter
interactions. These reaction rates depend sensitively on the
properties of matter, including the magnitudes of the neutron and
proton chemical potentials and effective nucleon masses, among other
aspects. Also of considerable importance are the specific heats and
susceptabilities of the constituents, which determine, respectively,
the thermal and transport properties of matter.  Thermal properties,
especially, may be easier to diagnose from neutrino observations of
supernovae: the timescale for black hole formation, in cases where
that happens, appears to be an important example~\cite{bbj13}.

One of the most realistic descriptions of the properties of
interacting nucleons is the potential model Hamiltonian density of
Akmal, Pandharipande, and Ravenhall (APR hereafter)~\cite{apr}, which
reproduces the microscopic potential model calculations of Akmal and
Pandharipande (AP) \cite{ap}.  An interesting feature of the AP model
is the occurrence of a neutral pion condensate at supra-nuclear
densities for all proton fractions.  The AP model is especially
relevant because it satisfies several important global criteria that
have been gleaned from nuclear physics experiments and astrophysical
studies of neutron stars, especially those concerning the neutron star
maximum mass and their typical radii.

Both isospin-symmetric and isospin-asymmetric properties of cold
baryonic (neutron-proton) matter in the vicinity of $n_0$ are of
considerable importance, as they govern the masses of nuclei,
nucleon-pairing phenomena, collective motions of nucleons within
  nuclei, the transition density from inhomogeneous to homogeneous
bulk matter, the radii of neutron stars, and many observables in
medium-energy heavy-ion collisions~\cite{Steiner05}.  One of the most
important isospin-symmetric properties at $n_0$ is the density derivative of the pressure $P$, or,
the incompressibility $K_0$ of matter  which is now rather well-determined:
$K_0=9(dP/dn)_{n_0,x=1/2,T=0}\simeq230\pm30$ MeV
from Refs. \cite{Garg04,Colo04} and $240\pm20$ MeV from Ref. \cite{shlomo06}  

Another isospin-symmetric nuclear constraint stems from the 
thermal properties of nuclei and bulk matter.  Fermi liquid theory 
holds that the thermal properties of the equation of state are 
largely controlled by the nucleon effective masses.  In short, 
experiments indicate that nucleon effective masses are reduced from 
their bare values ($m$) at $n_0$ for symmetric matter to 
approximately $m^*_0/m\simeq0.8\pm0.1$ \cite{bohigas79,krivine80} and microscopic 
theory suggests they further decrease at higher densities. The extraction 
of $m^*_0$ from nuclear level densities is complicated by 
uncertain contributions from the surface energy as well as possible 
energy dependences in $m^*$.

Additionally, of great significance is the influence of isospin-asymmetry on the properties of 
nucleonic matter, not only on the effective masses, but also on its energy 
$E(n,x,T)$, particularly the symmetry energy
parameter $S_v = 1/8(\partial^2E/\partial x^2)_{n_0,x=1/2,T=0}$
and its stiffness parameter $L = 3/8(\partial^2E/\partial n\partial x^2)_{n_0,x=1/2,T=0}$.
Starting from the Bethe-Weizacker mass formula~\cite{W35,BB36} and its
modernization \cite{moller95,pearson01} for nuclei containing a 
fraction $x$ of protons, most mass formulas characterize the symmetry energy 
of nucleonic matter by these two parameters.
From a variety of experiments, including measurements of nuclear
binding energies, neutron skin thicknesses of heavy nuclei, dipole
polarizabilities, and giant dipole resonance energies \cite{L,tsang12} 
$S_v$ lies in the range 30-35 MeV and $L$ lies in the range 40-60 MeV.
Recent developments in the prediction of the properties of
pure neutron matter by Gandolfi, Carlson and Reddy \cite{gcr12} and
by Hebeler and Schwenk \cite{kths13} suggest very similar values for $S_v$ 
and $L$ compared to those derived from nuclear experiments.

It is worth noting that there exists a phenomenological
relation~\cite{lp01} between neutron star radii and zero temperature
neutron star matter pressures near $n_0$, which is nearly that of pure 
neutron matter and largely a function of the $L$ parameter \cite{L}. 
Astrophysical observations of photospheric radius
expansion in X-ray bursts \cite{ogp09} and quiescent low-mass X-ray
binaries \cite{gswr13} have been used \cite{slb10,slb13} to conclude
that the radii of neutron stars with masses in the range 1.2-1.8$~{\rm
  M}_\odot$ are between 11.5 km and 13 km, and therefore predict that
$L\simeq45\pm10$ MeV, although the astrophysical model dependence of
this result may significantly enlarge its uncertainty.
Nevertheless, this range overlaps that from nuclear experiments and 
also that from neutron matter theory, suggesting that systematic 
dependencies are not playing a major role in the astrophysical determinations.

A potentially more important astrophysical constraint originates from 
mass measurements of neutron stars.  A consequence of general relativity 
is the existence of a maximum neutron star mass for every equation of state. 
Causality arguments, together with current radius estimates, indicate this 
is in the range of 2-2.8~${\rm M}_\odot$ \cite{L}. The largest precisely measured 
neutron star masses are $1.97\pm0.04~{\rm M}_\odot$ \cite{Demorest} and 
$2.01\pm0.04~{\rm M}_\odot$ \cite{Antoniadis}.  It is likely that the true maximum mass is
at least a few tenths of a solar mass larger than these
measurements. 

An important issue concerns the quality and relevance of 
experimental information that could constrain the thermal properties of 
dense matter. Calibrating the thermal properties of bulk matter
from experimental results involves disentangling the effects of
several overlapping energy scales (associated with shell and pairing
effects, collective motion, etc.) that determine the properties of
finite sized nuclei.  The level densities of nuclei (inferred through
data on, for example, neutron evaporation spectra and the disposition
of single particle levels in the valence shells of nuclei \cite{egidy88,egidy05} 
depend on the Landau effective masses, $m_{n,p}^*$, of neutrons
and protons.  These masses are sensitive to both the
momentum and energy dependence of the nucleon self-energy leading to
the so-called the $k$-mass and the $\omega$-mass, emphasized, for
example, in Refs. \cite{NY81,FFP81,Prakash83,Mahaux85}.  For bulk
matter, in which the predominant effect is from the $k$-mass,
$m_{n,p}^*/m=0.7\pm 0.1$ has been generally preferred. The specific
heat and entropy of nuclei receive substantial contributions from
low-lying collective excitations, as shown in Refs.  \cite{VV83,VV85},
a subject that needs further exploration to pin down the role of
thermal effects in bulk matter.

Following the suggestions in Refs. ~\cite{S83,BS83}, the liquid-gas
phase transition has received much attention with the finding that the
transition temperature for nearly isospin symmetric matter lies in the
range 15-20 MeV~\cite{DasGupta01}. Although the critical temperature 
depends on the incompressibility parameter $K_0$, it is also sensitive 
to the specific heat of bulk matter in the vicinity of $n_0$ which depends 
on the effective masses. Further information about the effective masses 
can be ascertained from fits of the optical model potential to data \cite{myers66}, albeit it
at low momenta.   
The density and the saturating aspect of the high momentum dependence of the real part of the optical 
model potential has been crucial in explaining
the flow of momentum and energy observed in intermediate energy ($< 1$ GeV) collisions of
heavy-ions, preserving at the same time the now well-established value
of the incompressibility $K_0=230\pm 30$ MeV, as demonstrated in
Refs. \cite{Prakash88,Gale90,Danielewicz00}.  Notwithstanding these
activities, further efforts are needed to calibrate the finite
temperature properties of nucleonic matter to reach at least the level 
of accuracy to which the zero temperature properties have been assessed.

Relatively few EOSs have been constructed from underlying interactions
satisfying all these important constraints \cite{Steiner}. However, 
AP and APR satisfy nearly all of them.  For APR, $K_0=266$
MeV, $S_2\simeq 32.6$ MeV, $L\simeq58.5$ MeV, and $m^*_0/m=0.7$, within 
two standard deviations of the experimental ranges. The maximum
neutron star mass supported by the APR model is in excess of $2{\rm
  M}_\odot$, and the radius of a $1.4M_\odot$ star is about 12 km.
Despite its obvious positive characteristics, no three-dimensional
tabular EOS has been constructed with the APR equation of state.
Furthermore, its finite temperature properties for arbitrary
degeneracy and proton fractions, including the effects of its pion
condensate, have not been studied to date. 

The chief motivation for the present study is to perform a detailed analysis
of the EOS of AP through a study of the properties predicted by its APR
parametrization.  Particular attention is paid to the density
dependence of nucleon effective masses which govern both the
qualitative and quantitative behaviors of its thermal properties.
Another objective of the present work is to document the analytic
relations describing the thermodynamic properties of potential models.
These are essential ingredients in the generation of EOS tables based
upon modern energy density functionals that employ 
Skyrme-like energy density functionals.  Importantly, the
analytic expressions developed here can be utilized to update LS-type
liquid droplet EOS models that take the presence of nuclei at
subnuclear densities and subcritical temperatures into account.  This
would represent a significant improvement to existing EOS tables in
that they could be replaced with ones including realistic effective
masses.

Some aspects of the thermal properties of hot, dense matter have 
been explored in Ref.~ \cite{Prak87} for isospin symmetric matter, but the comparative thermal 
properties of different Skyrme-like interactions remain largely unexplored.  
In view of the lack of systematic studies contrasting
the predicted thermodynamic properties of the APR model with those of
other Skyrme energy density functionals, we are additionally motivated
to perform such studies for one particular case, that of the SKa force
due to Kohler \cite{ska}.  This is one of the EOS's tabulated in the suite of
  EOS's provided in Ref.~\cite{LattimerEOSs}
		which is reproduced here in detail.   
The methods developed here are general and can be advantageously used for other 
Skyrme-like energy density functions in current use.

For both the APR and Ska models, we compute the EOS for uniform matter
for temperatures ranging up to 50 MeV, baryon number densities in the
range $10^{-7}$ fm$^{-3}$ to 1 fm$^{-3}$, and proton fractions between
0 (pure neutron matter) and 0.5 (isospin-symmetric nuclear matter).
Ideal gas photonic and leptonic contributions (both electrons and
muons) are included for all models.  
The results presented here for densities below 0.1 fm$^{-3}$ in the homogeneous phase 
serve only to gauge differences from the more realistic situation in which supernova matter contains 
an inhomogeneous phase.  
Work toward extending calculations to realistically describe the low
density/temperature inhomogeneous phase containing finite nuclei is in
progress at various levels of sophistication 
(Droplet model, Hartree, Hartree-Fock, Hartree-Fock-Boguliobov, etc.) 
beginning with an LS-type liquid droplet model approach and will be
reported separately.
In addition, hyperons and a possible phase transition to deconfined quark matter
are not considered in this work.
. 
The organization of this paper is as follows. In
Sec. \ref{Sec:pmodels}, we briefly discuss some of the features of the
APR and Ska Hamiltonians, and the ingredients involved in their
construction.  We then present their single-particle energy spectra
and potentials using a variational procedure in Sec III.  In Sec. IV,
properties of cold, isospin-symmetric matter and consequences for small
deviations from zero isospin asymmetry are examined. Analyses of
results for the two models include those of energies, pressures,
neutron and proton chemical potentials, and inverse susceptibilities.
Section V contains our study of the behavior of all the relevant state
variables for the APR and SKa models at finite temperature. The
numerical results, valid for all regimes of degeneracy, are juxtaposed
with approximate ones in the degenerate and non-degenerate limits for
which analytical expressions have been derived. Contributions from
leptons and photons are also summarized in this section.  In Sec. VI,
we address the transition from a low-density to a high-density phase
in which a neutral pion condensate is present using a Maxwell
construction.  The numerical results of this section constitute the
equation of state of supernova matter for the APR model in the bulk
homogeneous phase.  Our summary and conclusions are given in
Sec. VII. The appendices contain ancillary material employed in this
work. In Appendix A, we provide a detailed derivation of the
single-particle energy spectra for the potential models used.  General
expressions for all the state variables of the APR model valid for all
neutron-proton asymmetries are collected in Appendix B.  
The formalism to include contributions from leptons (electrons and positrons) and photons is presented 
in Appendix C, wherein both the exact and analytical representations are summarized. 
Numerical
methods used in our calculations of the Fermi-Dirac integrals for
arbitrary degeneracy are summarized in Appendix D.
Appendix E contains thermodynamically consistent prescriptions to render 
EOS's causal when they become acausal at some high density for both zero and finite temperature cases.

\section{POTENTIAL MODELS}
\label{Sec:pmodels}

In this work, we study the thermal properties of uniform matter predicted by potential
models. We focus on an interaction derived from the work of Akmal and Pandharipande
(hereafter AP) \cite{ap}, using an approximation developed by Akmal, Pandharipande 
and Ravenhall (hereafter APR)~\cite{apr}, and a Skyrme \cite{Skyrme} force developed by 
K\"{o}hler (Ska henceforth) \cite{ska}. We pay special attention to the finite 
temperature properties of these two models for the physical conditions expected in supernovae 
and neutron star mergers, which has heretofore not received much attention.

The Hamiltonian density of Ska \cite{ska} is a typical example of the approach 
based on effective zero-range forces pioneered by Skyrme \cite{Skyrme}, which are 
typically called Skyrme forces. These were further developed to describe properties of
bulk matter and nuclei in Ref. \cite{vb}. Skyrme forces are easier to use in this 
context than finite-range forces (see, e.g., Ref \cite{Gogny}). To date, a vast number 
of variants of this approach exist in the literature \cite{Dutra} which have varying 
success in accounting for properties of nuclei and neutron stars. The strength parameters 
of the Skyrme-like energy density functionals are calibrated at nuclear and sub-nuclear 
densities to reproduce the properties of many nuclei, their behavior at high densities 
being constrained largely by neutron-star data.

The Hamiltonian density of APR is a parametric fit to the AP microscopic model 
calculations in which the nucleon-nucleon interaction is modeled by the Argonne
$v18$ 2-body potential ~\cite{v18}, the Urbana UIX 3-body potential ~\cite{uix}, 
and a relativistic boost potential $\delta v$ ~\cite{dv} which is a kinematic correction 
when the interaction is observed in a frame other than the rest-frame of the nucleons. 
These microscopic potentials accurately fit scattering data in vacuum and thus 
incorporate the long scattering lengths of nucleons at low energy.  Additionally, they 
have also been successful in accounting for the binding energies and spectra of light nuclei. 
An interesting feature of AP, incorporated in the Skyrme-like parametrization of the APR 
model, is that at supra-nuclear densities a neutral pion condensate appears. Despite the 
softness induced by the pion condensate in the high density equation of state, the APR model 
is capable of supporting a neutron star of 2.19 M$_\odot$, in excess of the recent 
accurate measurements of the masses of PSR J1614-2230 ($1.97\pm0.04~{\rm M}_\odot$) 
\cite{Demorest} and PSR J0348+0432 ($2.01\pm0.04~{\rm M}_\odot$) \cite{Antoniadis}.  

Being non-relativistic potential models, both the APR and Ska models have the potential to 
become acausal (that is, the speed of sound exceeds the speed of light) at high density. 
A practical fix to keep their behaviors causal which is thermodynamically consistent is 
possible and is adopted in this work (see Appendix  E).  

Our choice of these two models was motivated by several considerations, including the facts that 
(i) both models yield similar results for the equilibrium density, binding energy, symmetry energy, 
and compression modulus of symmetric matter, as well as for the maximum mass of neutron stars 
and (ii) the two models differ significantly in other properties such as their Landau 
effective masses (important for thermal properties), derivatives of their symmetry energy at 
nuclear density (important for the high density behavior of isospin asymmetry energies), 
skewness (i.e., the derivative of the compression modulus) at nuclear density, and their 
predicted radii corresponding to the maximum mass configuration. The impact 
of the different features of these two models for their thermal properties is one of the main 
foci of our work here. The methods used to explore their thermal effects are applicable 
and easily adapted to other Skyrme-like energy density functionals.

\subsection{Hamiltonian density of APR}

Explicitly, the APR Hamiltonian density is given by \cite{apr}
\begin{eqnarray}
\mathcal{H}_{APR} & = & \left[\frac{\hbar^2}{2m}+(p_3+(1-x)p_5)ne^{-p_4 n}\right]\tau_n  \nonumber \\
                  &   & +\left[\frac{\hbar^2}{2m}+(p_3+x p_5)ne^{-p_4 n}\right]\tau_p  \nonumber \\
                  &   & +g_1(n)[1-(1-2x)^2)]+g_2(n)(1-2x)^2,  
\label{HAPR}
\end{eqnarray}
where $n=n_n+n_p$ is the baryon density, $x=n_p/n$ is the proton fraction, and 
\begin{eqnarray}
n_i & = & \frac{1}{\pi^2}\int dk_i~\frac{k_i^2}{1+e^{(\epsilon_{k_i}-\mu_i)/T}}  \\
\tau_i & = & \frac{1}{\pi^2}\int dk_i~\frac{k_i^4}{1+e^{(\epsilon_{k_i}-\mu_i)/T}}
\end{eqnarray}
are the number densities and kinetic energy densities of nucleon species $i=n,p$, respectively. 
The quantities $\epsilon_{k_i}$, $\mu_i$ and $T$ are the single-particle spectra, chemical potentials and temperature (with Boltzmann's constant $k_B$ set to unity), respectively.
The first two terms on the right-hand side of this expression are due to kinetic energy and 
momentum-dependent interactions while the last terms are due to density-dependent interactions. 
Compared to a classical Skyrme interaction, such as Ska (described below), this model has a more 
complex density dependence in the single-particle potentials and effective masses. Due to the 
occurrence of a neutral pion condensation at supra-nuclear densities, the  
potential energy density functions  $g_1$ and $g_2$ take different forms on either side 
of the transition density. In the low density phase (LDP)
\begin{eqnarray}
g_{1L} & = & -n^2\left[ p_1+p_2n+p_6n^2+(p_{10}+p_{11}n)e^{-p_9^2n^2}\right]  \\
g_{2L} & = & -n^2\left( \frac{p_{12}}{n}+p_7+p_8n+p_{13}e^{-p_9^2n^2} \right),
\end{eqnarray}
whereas, in the high density phase (HDP)
\begin{eqnarray}
g_{1H} & = g_{1L}-n^2\left[p_{17}(n-p_{19})+p_{21}(n-p_{19})^2 \right] e^{p_{18}(n-p_{19})} \nonumber \\ \\
g_{2H} & = g_{2L}-n^2\left[p_{15}(n-p_{20})+p_{14}(n-p_{20})^2 \right] e^{p_{16}(n-p_{20})}\,.  \nonumber \\
\end{eqnarray}

The values of the parameters $p_1$ through $p_{21}$, as well as their dimensions which ensure 
that $\mathcal{H}_{APR}$ has units of $\mbox{MeV~fm}^{-3}$, are presented in Table I.  
Alternate choices for the underlying microscopic physics lead to different fits to the 
above generic form, so even though $p_{13}, p_{14}$ and $p_{21}$ are all 0 in our case, we 
carry the terms containing these coefficients in the algebra of Appendix B.

\begin{table}[!h]
\begin{center}
\begin{tabular}{|l|c||l|c|}
\hline
$p_1 $ & $337.2~\mbox{MeV~fm}^3$       &   $p_{14}$  &     0                        \\
$p_2 $ & $ -382.0~\mbox{MeV~fm}^6$     &   $p_{15}$  &    $287.0~\mbox{MeV~fm}^6$   \\        
$p_3 $ &  $ 89.8~\mbox{MeV~fm}^5$      &   $p_{16}$  &    $-1.54~\mbox{fm}^3$       \\
$p_4 $ & $  0.457~\mbox{fm}^3$         &   $p_{17}$  &   $175.0~\mbox{MeV~fm}^6 $   \\            
$p_5 $ &$-59.0~\mbox{MeV~fm}^5$        &   $p_{18}$  &   $-1.45~\mbox{fm}^3$    \\   
$p_6 $ &  $-19.1~\mbox{MeV~fm}^9$      &   $p_{19}$  &   $0.32~\mbox{fm}^{-3}$       \\
$p_7 $ & $214.6~\mbox{MeV~fm}^3$       &   $p_{20}$  &  $0.195~\mbox{fm}^{-3}$       \\            
$p_8 $ & $  -384.0~\mbox{MeV~fm}^6$    &   $p_{21}$  &   0                          \\
$p_9 $ & $  6.4~\mbox{fm}^6$           &   &  \\
$p_{10} $ & $  69.0~\mbox{MeV~fm}^3$    &   &  \\
$p_{11}$ & $ -33.0~\mbox{MeV~fm}^6$     &   &  \\
$p_{12} $ & $  0.35~\mbox{MeV}$         &   &  \\
$p_{13} $ & 0                           &   &  \\
\hline
\end{tabular}
\caption[Parameter values for $\mathcal{H}_{APR}$]{Parameter values for the Hamiltonian density 
of Akmal, Pandharipande, and Ravenhall \cite{apr}. Values in the last column are specific to the 
high density phase (HDP). The 
dimensions are such that the Hamiltonian density is in MeV fm$^{-3}$.}
\end{center}
\end{table}

The trajectory in the $n-x$ plane, for any temperature, along which the transition from the LDP to the HDP 
occurs is obtained by solving
\begin{eqnarray}
&&  g_{1L}[1-(1-2x)^2]+g_{2L}(1-2x)^2  \nonumber \\
    && \hspace{40pt} = g_{1H}[1-(1-2x)^2]+g_{2H}(1-2x)^2\,. 
    \label{transition}
\end{eqnarray}
The solution gives a transition density $n_t=0.32 ~\mbox{fm}^{-3}$ for symmetric nuclear matter $(x=1/2)$ and 
$n_t=0.195 ~\mbox{fm}^{-3}$ for pure neutron matter $(x=0)$. 
For intermediate values of $x$, the transition density is approximated to high accuracy by the polynomial fit
\be
n_t(x) =0.1956+ 0.3389~x + 0.2918~x^2 - 1.2614~x^3 + 0.6307~x^4\, .
\label{polfit}
\ee
In calculations of subsequent sections, the transition from the LDP to the HDP at zero and finite temperatures will be made through the use of the above polynomial fit. The mixed phase region is determined via a Maxwell construction for the numerical purposes of which $n_t$ is used as an input. We show in Sec. VI that while the transition is independent of $T$ for any $x$, the two densities which define the boundary of the phase-coexistence region do exhibit a weak dependence on temperature.

\subsection{Hamiltonian density of Ska}

The Hamiltonian density of Ska \cite{ska} based on the Skyrme energy density functional approach is expressed as
\begin{eqnarray}
\mathcal{H}_{Ska} & = & \frac{\hbar^2}{2m_n}\tau_n+\frac{\hbar^2}{2m_p}\tau_p  \nonumber  \\
                  &   & +n(\tau_n+\tau_p)\left[\frac{t_1}{4}\left(1+\frac{x_1}{2}\right)  
                                               +\frac{t_2}{4}\left(1+\frac{x_2}{2}\right)\right] \nonumber \\
                  &   & +(\tau_n n_n+\tau_p n_p)\left[\frac{t_2}{4}\left(\frac{1}{2}+x_2\right)
                                               -\frac{t_1}{4}\left(\frac{1}{2}+x_1\right)\right] \nonumber \\
                  &   & +\frac{t_o}{2}\left(1+\frac{x_o}{2}\right)n^2
                        -\frac{t_o}{2}\left(\frac{1}{2}+x_o\right)(n_n^2+n_p^2)  \nonumber  \\
                  &   & \left[\frac{t_3}{12}\left(1+\frac{x_3}{2}\right)n^2
                        -\frac{t_3}{12}\left(\frac{1}{2}+x_3\right)(n_n^2+n_p^2)\right]n^{\epsilon} \nonumber \\
\label{HSKA}
\end{eqnarray}
Terms involving $\tau_i$ with $i=n,p$ are purely kinetic in origin whereas terms involving $n\tau_i$ and $n_i\tau_i$ arise from the exchange part of the nucleon-nucleon interaction. The latter determine the density dependence of the effective masses (see below). The remaining terms, dependent on powers of the individual and total densities give the potential part of the energy density. The various strength parameters are calibrated to desired properties of bulk matter and of nuclei chiefly close to the empirical nuclear equilibrium density. Many other parametrizations of the Skyrme-like energy density functional also exist \cite{Dutra} and are characterized by different values of observable physical quantities (see below).  
 The parameters $t_o$ through $t_3$, $x_o$ through $x_3$, and $\epsilon$ for the Ska model~\cite{ska} are listed in Table II.

\begin{table}[!h]
\begin{center}
\begin{tabular}{|c|c|c|c|}
\hline
i & $t_i$ & $x_i$ & $\epsilon$ \\
\hline
0 & $-1602.78 ~\mbox{MeVfm}^6$ & 0.02 & 1/3 \\
1 & $570.88 ~\mbox{fm}^3$ & 0 &              \\ 
2 & $-67.7 ~\mbox{fm}^3$  & 0 &              \\
3 & $8000.0 ~\mbox{MeVfm}^7$ & -0.286 &     \\
\hline
\end{tabular}
\caption[Parameter values for $\mathcal{H}_{Ska}$.]{Parameter values for the Ska Hamiltonian density~\cite{ska}.The 
dimensions are such that the Hamiltonian density is in MeV fm$^{-3}$.}
\end{center} 
\end{table}

\section{SINGLE-PARTICLE ENERGY SPECTRA}

The single-particle energy spectra $\epsilon_{k_i},~(i=n,p)$ that appear in the Fermi-Dirac (FD) 
distribution functions \\ $n_{k_i} = \left[{1+e^{(\epsilon_{k_i}-\mu_i)/T}}\right]^{-1}$ are obtained from functional 
derivatives of the Hamiltonian density (see appendix A for derivation): 
\ba
\epsilon_{k_i} & = & k_i^2\frac{\partial \mathcal{H}}{\partial \tau_i} + \frac{\partial \mathcal{H}}{\partial n_i}.
\label{spectra}
\ea
The ensuing results can be expressed as 
\ba
\epsilon_{k_n} &=& \frac{\hbar^2k^2}{2m} + U_n(n,k) \nonumber \\
\epsilon_{k_p} &=& \frac{\hbar^2k^2}{2m} + U_p(n,k)\,,
\label{spectranp}
\ea
where $m$ is the nucleon mass in vacuum, and $U_n$ and $U_p$ are the neutron and proton single-particle momentum-dependent potentials, respectively.
Utilizing these spectra, the Landau effective masses $m_i^*$ are 
\ba
m_i^* \equiv k_{F_i} \left[ \left| \frac {\partial \epsilon_{k_i}} {\partial k}  \right|_{k_{F_i}}  \right]^{-1} \,,
\label{effmi}
\ea
where $k_{F_i}$ are the Fermi-momenta of species $i$. Physical quantities such as the thermal energy, thermal pressure, susceptibilities, specific heats at constant volume and pressure, and entropy all depend sensitively on these effective masses as  
highlighted in later sections.

\subsection*{APR single-particle potentials}

From Eq. (\ref{HAPR}) and Eq. (\ref{spectranp}), the explicit forms of the single-particle potentials for the LDP Hamiltonian density of APR are 
\ba 
U_{nL}(n,k) &=& (p_3+Y_np_5)ne^{-p_4n}k^2        \nonumber \\
                   &+& \left\{\left[p_3+p_5-p_4n(p_3+Y_np_5)\right]\tau_n \right. \nonumber \\
                   &+&     \left.\left[p_3-p_4n(p_3+Y_pp_5)\right]\tau_p\right\} e^{p_4n}  \nonumber  \\
                   &+&     4Y_p\frac{g_{1L}}{n} + 2(Y_n-Y_p)\frac{g_{2L}}{n}   \nonumber \\
                   &+&     4Y_nY_pf_{1L} + (Y_n-Y_p)^2f_{2L}  \nonumber \\
U_{pL}(n,k) &=& (p_3+Y_pp_5)ne^{-p_4n}k^2        \nonumber \\
                   &+&      \left\{\left[p_3+p_5-p_4n(p_3+Y_pp_5)\right]\tau_p \right. \nonumber \\
                   &+&     \left.\left[p_3-p_4n(p_3+Y_np_5)\right]\tau_n\right\} e^{p_4n}  \nonumber  \\
                   &+&     4Y_n\frac{g_{1L}}{n} + 2(Y_p-Y_n)\frac{g_{2L}}{n}   \nonumber \\
                   &+&     4Y_nY_pf_{1L} + (Y_n-Y_p)^2f_{2L} \,, 
\label{Unps} 
\ea
with $Y_p=x$ and $Y_n=1-x$, and where
\ba
f_{1L} =  \frac{dg_{1L}}{dn} - \frac{2g_{1L}}{n}~~ {\rm and}~~
f_{2L}  =  \frac{dg_{2L}}{dn} - \frac{2g_{2L}}{n}.
\ea
In the HDP,
\ba
U_{nH}(n,k) & = & U_{nL}(n,k)-\frac{4Y_p(Y_n-Y_p)}{n}(\delta g_1-\delta g_2) \nonumber \\
                             &+&4Y_nY_p\delta f_1 + (Y_n-Y_p)^2\delta f_2    \\
U_{pH}(n,k) & = & U_{pL}(n,k)+\frac{4Y_n(Y_n-Y_p)}{n}(\delta g_1-\delta g_2) \nonumber \\
                             &+&4Y_nY_p\delta f_1 + (Y_n-Y_p)^2\delta f_2.   
\ea
The functions $\delta g_1,~\delta g_2,~\delta f_1$, and $\delta f_2$ are defined in Appendix B.
The corresponding effective masses from Eq. (\ref{effmi}) are 
\ba
\frac {m_i^*}{m} = \left[ 1 + \frac {2m}{\hbar^2} (p_3+Y_ip_5)ne^{-p_4 n}   \right]^{-1} \,,
\label{effmAPR}
\ea
where $Y_i = (1-x)$ for neutrons ($i=n$) and $Y_i=x$ for protons ($i=p$).  Subsuming the $k^2$-dependent parts of $U_i(n,k)$ in 
Eq. (\ref{Unps}) into the kinetic energy terms in Eqs. (\ref{spectranp}), the single-particle energies may be expressed as 
\ba
\epsilon_{k_i} &=& \frac{\hbar^2k^2}{2m_i^*} + V_i(n) \,,
\label{newspectranp}
\ea
where the functional forms of $V_i(n)$ are readily ascertained from the relations in Eq. (\ref{Unps}).  The quadratic momentum-dependence 
of the single particle spectra, albeit density and concentration dependent through the effective masses, is akin to that of free Fermi gases. Consequently, the thermal state variables
can be calculated as for free Fermi gases, but with attendant modifications arising from the density-dependent effective masses as will be discussed later.

\subsection*{Skyrme single-particle potentials}

Explicit forms of the single-particle potentials for the Ska Hamiltonian are given by
\ba
U_n(n,k) &=& (X_1+Y_nX_2)nk^2      
                   + (X_1+X_2)\tau_n+X_1\tau_p \nonumber \\
                   &+& 2n(X_3+Y_pX_4)   
                   + n^{1+\epsilon}[(2+\epsilon)X_5    \nonumber  \\
                   &+& 2Y_n+\epsilon({Y_n}^2 +{Y_p}^2)]  \nonumber \\
U_p(n,k) &=& (X_1+Y_pX_2)nk^2     
                   + (X_1+X_2)\tau_p+X_1\tau_n \nonumber \\
                   &+& 2n(X_3+Y_nX_5)   
                   + n^{1+\epsilon}[(2+\epsilon)X_6    \nonumber  \\
                   &+& 2Y_p+\epsilon({Y_n}^2 +{Y_p}^2)]  \,,
\label{UnpsSka}
\ea
where
\ba
X_1   &=& \frac{1}{4} \left[ t_1 \left(1+\frac{x_1}{2}\right) + t_2 \left(1+\frac{x_2}{2} \right) \right] \nonumber\\
X_2  &=& \frac{1}{4} \left[t_2 \left(\frac{1}{2}+x_2 \right)-t_1\left( \frac{1}{2}+x_1 \right) \right] \nonumber\\
X_3 &=& \frac{t_0}{2} \left(1+\frac{x_0}{2}\right)\,; \quad X_4 = -\frac{t_0}{2} \left(\frac{1}{2}+x_0\right) \nonumber\\
X_5 &=& \frac{t_3}{12} \left(1+\frac{x_3}{2} \right)\,; \quad X_6 = -\frac{t_3}{12} \left(\frac{1}{2}+x_3 \right)\,. 
\ea
From Eq. (\ref{effmi}), the density-dependent Landau effective masses are
\ba
\frac {m_i^*}{m} = \left[ 1 + \frac {2m}{\hbar^2} (X_1+Y_iX_2)n   \right]^{-1} \,.
\label{effmSka}
\ea
The single-particle spectra have therefore the same structure as in Eq. (\ref{newspectranp}), but with
the potential terms $V_i(n)$ inferred from Eq. (\ref{UnpsSka}).

\section{ZERO TEMPERATURE PROPERTIES}

At temperature T=0, nucleons are restricted to their lowest available quantum states. Therefore, the 
Fermi-Dirac distribution functions that appear in the integrals of the number density and the kinetic 
energy density become step-functions:
\begin{eqnarray}
n_{ki} & = & \theta(\epsilon_{ki}-\epsilon_{Fi}),
\end{eqnarray}
where $\epsilon_{Fi}$ is the energy at the Fermi surface for species $i$. Consequently,
\begin{eqnarray}
n_i & = & \frac{1}{\pi^2}\int_0^{k_{Fi}}k_i^2dk_i = \frac{k_{Fi}^3}{3\pi^2}  \\
\tau_i & = & \frac{1}{\pi^2}\int_0^{k_{Fi}}k_i^4dk_i = \frac{k_{Fi}^5}{5\pi^2} = \frac 35 n_ik_{Fi}^2 \,.
\end{eqnarray}
Thus, the kinetic energy densities can be written as simple functions of the number density $n$ and 
the proton fraction $x$ :
\begin{eqnarray}
\tau_p & = & \frac{1}{5\pi^2}(3\pi^2n_p)^{5/3} = \frac{1}{5\pi^2}(3\pi^2nx)^{5/3} \\
\tau_n & = & \frac{1}{5\pi^2}(3\pi^2n_n)^{5/3} = \frac{1}{5\pi^2}(3\pi^2n(1-x))^{5/3}.
\end{eqnarray}
We can therefore write 
\[\mathcal{H}(n_p,n_n,\tau_p,\tau_n;T=0)=\mathcal{H}(n,x)\,,\] 
and use standard thermodynamic relations to get the various quantities of interest, 
some examples of which are listed below beginning with $x=1/2$ for isospin symmetric nuclear matter. General expressions for arbitrary $x$ are provided in Appendix B.
  
\subsection{Isospin symmetric nuclear matter}

\subsection*{The APR Hamiltonian}

It is convenient to write $\mathcal{H}_{APR}$ as the sum of a kinetic part $\mathcal{H}_k$, a part consisting of momentum-dependent interactions $\mathcal{H}_m$, and a density-dependent interactions part $\mathcal{H}_d$. The energy per particle of symmetric nuclear matter $E$ can then be similarly decomposed as    
\begin{equation}
E \equiv \frac{\mathcal{H_{APR}}}{n} =  {E_k} +   {E_m} +   {E_d} \,,
\end{equation}
where
\ba
 {E_k} &=& \frac 35 \frac {\hbar^2k_F^2}{2m}\,; \quad k_F = (3\pi^2n/2)^{1/3} \nonumber \\
 {E_m} &=& \frac 35 nk_F^2e^{-p_4 n} (p_3 + p_5/2 ) \nonumber \\
 {E_{dL}} &=& \frac {g_{1L}}{n} \,, \quad E_{dH} = \frac{g_{1H}}{n} = E_{dL}+\frac{\delta g_1}{n} \,. 
\ea
The corresponding pressure is
\ba 
P &=& n^2\frac{\partial E}{\partial n} = P_k + P_m + P_d \nonumber \\
P_k &=& \frac 23 n {E_k}\,, P_m = \left( \frac 53 - p_4n \right) n  {E_m} \nonumber \\
P_{dL} &=& n \left( {E_d} + f_{1L} \right) \nonumber \\
P_{dH} &=& P_{dL}-\delta g_1+n\delta f_1 \,.
\ea

The nucleon chemical potential takes the form
\ba
\mu &=& \frac {\partial \mathcal {H}}{\partial n} = \mu_k + \mu_m + \mu_d \nonumber \\
\mu_k &=&   \frac 53 {E_k} = \frac {\hbar^2k_F^2}{2m} \nonumber\\ 
\mu_m &=& nk_F^2e^{-p_4 n} \left\{ p_5 \left( \frac 45 - \frac {p_4n}{2}  \right)
+ p_3 \left ( \frac {8}{3} - p_4 n\right) \right\}   \nonumber \\
\mu_{dL} &=& \frac {dg_{1L}}{dn}\,,\quad \mu_{dH} = \mu_{dL}+\delta f_1 \,.
\ea

The inverse susceptibility is given by 
\ba
\chi^{-1 } &=& \frac {\partial\mu}{\partial n} =  \chi_k^{-1 } + \chi_m^{-1 } + \chi_d^{-1 } \nonumber \\
 \chi_k^{-1 } &=& \frac 23 \frac {\mu_k}{n}  \nonumber \\   
 \chi_m^{-1 } &=& -p_4\mu_m + \frac 35 k_F^2 e^{-p_4n} \nonumber \\
 &*& \left\{ \frac 43 p_5 \left( \frac {10}{3} - p_4n \right) + \frac 23 p_3 \left( \frac {25}{3} -4p_4 n \right)  \right\}  \nonumber \\ 
\chi_{dL}^{-1 } &=& 8 \frac {f_{1L}}{n} + 4 h_{1L}  \nonumber \\
\chi_{dH}^{-1} &=& \chi_{dL}^{-1}-\frac{2}{n^2}(\delta g_1-\delta g_2)+\delta h_1 \,,
\ea 
where
\ba
h_{1L} =  \frac{df_{1L}}{dn} - \frac{2f_{1L}}{n}
\ea
and $\delta h_1$ can be found in App.B.
The nuclear matter incompressibility is given by 
\ba 
K &=& 9\frac{dP}{dn} = K_k + K_m + K_d \nonumber \\
K_k &=& 10  {E_k} = 6 \frac  {\hbar^2k_F^2}{2m} \nonumber \\
K_m &=& \left( 40-48p_4n + 9 p_4^2n^2 \right)  {E_m} \nonumber \\
K_{dL} &=& 18  {E_d} + 9 \left[ 4f_{1L} + n h_{1L} \right]  \nonumber \\
K_{dH} &=& K_{dL} + 9 n \delta h_1 \,.     
\ea 
The speed of sound can be written in terms of $\mu$ and $K$ or $\chi^{-1}$ as 
\be
\left(\frac{c_s}{c}\right)^2 = \frac{K}{9(\mu+m)} = \frac{n\chi^{-1}}{\mu+m}
\label{cs}
\ee
From this relation, it can be shown that the APR model becomes acausal ($c_s/c = 1$) at $n = 0.841$ fm$^{-3}$ in the case of symmetric matter.

The speed of sound $c_s$, and the response functions $K$ and $\chi$ are generated by density fluctuations. 
Evidently, they are not independent of each other (relationships between them in 
the case of general asymmetry are given in Appendix  B). Each quantity, however,  is useful in its own 
right for a number of applications. For example, $c_s$ is necessary in implementing causality (see Appendix E), $K$ is 
essential to the calculation of the liquid-gas phase transition (Sec.V), and $\chi$ is required in the 
numerical scheme by which the mixed-phase region, at the onset of pion condensation, is constructed (Sec.VI). 
At finite temperature, this group also includes the specific heats at constant volume and pressure, $C_V$ 
and $C_P$. The latter can be used to identify phase transitions, address causality at finite $T$ and, 
furthermore, are related to hydrodynamic time-scales as in the collapse to black holes.

\subsection*{The Skyrme Hamiltonian}

Similarly to the APR Hamiltonian we write $\mathcal{H}_{Ska}$ as the sum of a kinetic part $\mathcal{H}_k$, momentum-dependent interactions $\mathcal{H}_m$, and a density-dependent interactions $\mathcal{H}_d$. The energy per particle is then given by
\begin{equation}
E \equiv \frac{\mathcal{H}_{Ska}}{n} =  {E_k} +   {E_m} +   {E_d} \,,
\end{equation}
where
\ba
 {E_k} &=& \frac 35 \frac {\hbar^2k_F^2}{2m}\,, \quad  {E_m} = \frac 35 nk_F^2 \left(X_1+\frac{1}{2}X_2 \right)\, \nonumber \\
  {E_d} &=& n \left[X_3+\frac{1}{2}X_4+n^\epsilon \left( X_5 + \frac{1}{2}X_6 \right) \right] \,.
\ea
Contributions to the pressure arise from  
\ba 
P_k &=& \frac 23 n {E_k}\,, P_m = \frac 53 n  {E_m} \nonumber \\
P_d &=& n \left( {E_d} + \epsilon n^{\epsilon+1}\left(X_5+\frac 12 X_6 \right) \right) \,.
\ea
The nucleon chemical potential receives contributions from 
\ba
\mu_k &=&   \frac 53  {E_k} \,, \quad \mu_m = \frac 83 {E_m} \nonumber \\
\mu_d &=& 2 {E_d}+\epsilon\left(X_5+\frac 12 X_6\right)n^{\epsilon+1} \,.
\ea
The inverse susceptibility is composed of terms involving 
\ba
 \chi_k^{-1 } &=& \frac 23 \frac {\mu_k}{n}\,, \quad \chi_m^{-1 } = \frac{25}{12} \frac {\mu_m}{n} + \frac{4m}{\hbar^2} X_2 \mu_k
 \nonumber \\   
 \chi_d^{-1 } &=& \frac{\mu_d}{n}+n^\epsilon\left[\left(X_5+\frac 12 X_6\right)\epsilon+X_6\right] + X_4  \,.
\ea 
The nuclear matter incompressibility is determined by the terms 
\ba 
K_k &=& 10  {E_k}\,, \quad K_m = 40 {E_m} \nonumber \\
K_d &=& 18  {E_d} + 9 \epsilon\left(\epsilon+3\right)n^{1+\epsilon}\left[X_5+\frac 12 X_6\right] \,.     
\ea 
Combining the above results with Eq.(\ref{cs}) we find that Ska violates causality for baryon densities above 
$n = 1.028$ fm$^{-3}$.

\subsection{Isospin asymmetric matter}

Here, we focus on the energetics of matter with neutron
excess beginning with some general considerations that are model
independent. The neutron-proton asymmetry is commonly 
characterized by the parameter $\alpha = (n_n-n_p)/n$ which is 
connected to the proton fraction $x$ through the simple relation 
$\alpha=1-2x$.

The expansion of the energy per particle $E(n,\alpha) = \mathcal{H}/n$ 
of isospin asymmetric matter in powers $\alpha$, is given by:
\ba
E(n,\alpha) &=& E(n,0)+\sum_{l=2,4,\ldots}S_l(n)\alpha^l 
\label{Ealpha}
\ea
where 
\be
S_l = \left.\frac{1}{l!}\frac{\partial^l E(n,\alpha)}{\partial \alpha^l}\right|_{\alpha=0} ~~;~~l=2,4,\ldots
\ee
Similarly, the pressure of isospin-asymmetric matter can be written as
\ba
P(n,\alpha) &=&n^2\frac{\partial E(n,\alpha)}{\partial n}     \\
            &=& P(n,0)+\frac{n}{3}\sum_{l=2,4,\ldots}L_l(n)\alpha^l 
\label{Palpha}
\ea
where 
\be
L_l = 3n\frac{dS_l(n)}{dn}
\ee
Evaluating Eqs. (\ref{Ealpha})-(\ref{Palpha}) for pure neutron matter at the
saturation density $n_0$ of symmetric matter to $O(\alpha^2)$ gives
\ba
E(n_0,1) &\simeq& E_0 +S_v   \\
P(n_0,1) &\simeq& \frac{Ln_0}{3} 
\ea
where $E_0 = E(n_0,0)$ is the saturation energy of nuclear matter, $S_v=S_2(n_0)$ 
is its symmetry energy parameter that characterizes the energy cost involved in restoring isospin 
symmetry from small deviations, and $L=L_2(n_0)$ is its stiffness parameter. By the definition 
of $n_0$, $P(n_0,0)=0$.

Only even powers of $\alpha$ survive in the two series in Eqs. \ref{Ealpha} and \ref {Palpha} above 
because the two nucleon species are treated symmetrically in
the Hamiltonian. Furthermore, due to the near complete isospin invariance 
of the nucleon-nucleon interaction, the density dependent potential terms are 
generally carried only up to $O(\alpha^2)$; that is, $S_l(n)$ and $L_l(n)$ for $l>2$ 
receive contributions just from the kinetic energy and the momentum-dependent interactions. 
Finally, as demonstrated in Refs. \cite{lagaris81, wff88, bombaci91, gerry98},   
$S_2(n)\gg S_4(n),S_6(n),\ldots$ and hence coefficients with $l=2$ suffice in describing 
bulk matter even when $\alpha \sim 1$. 

While the full calculations are rather involved, the dominance of $S_2(n)$ can be 
illustrated in a simple manner by turning to the isospin-asymmetric free gas whose 
kinetic energy can be expressed as
\ba
E^{kin} = \frac{1}{3} E_F~\left[ \frac{1}{2} \left\{ (1+\alpha)^{5/3} + (1-\alpha)^{5/3} \right\} - 1 \right]\,,  
\ea
where
\ba
E_F = \frac{\hbar^2k_F^2}{2m} = \frac{\hbar^2}{2m}\left(\frac{3\pi^2n}{2}\right)^{2/3}
\ea
is the Fermi energy of non-interacting nucleons in symmetric nuclear matter. 
Through a Taylor expansion of terms involving $\alpha$ (terms in odd powers of $\alpha$ canceling), 
the various contributions from kinetic energy are 
\ba
S_2^{kin}(n) = \frac{1}{3} E_F,~
S_4^{kin}(n) = \frac{1}{81} E_F,~ 
S_6^{kin}(n) = \frac{7}{2187}E_F\ldots \nonumber \\
\label{s_kin}
\label{Skins}
\ea 
the series converging rapidly to the exact result of $(E_F/3)~(2^{2/3} -1)$. 
At the empirical nuclear equilibrium density of  $n_0=0.16~{\rm fm}^{-3}$, $S_2^{kin}(n_0) \simeq 12.28$ MeV, whereas
its associated stiffness parameter is $L^{kin} = (2/3)E_{F_0} \simeq 24.56$ MeV.

As mentioned earlier, in the presence of interactions, $S_4(n),S_6(n),\ldots$ 
are modified solely by the momentum-dependent terms which, predominantly, give rise 
to the effective mass while preserving the relative sizes of the $S_l$'s and their derivatives (For 
APR, at $n_0$, $S_2/S_4 \simeq 35$ and $L_2/L_4 \simeq 18$ whereas for Ska, $S_2/S_4 \simeq 29$ 
and $L_2/L_4 \simeq 17$.). Thus, we can write
\be
P(n,\alpha) \simeq n^2\left[E'(n,0)+\alpha^2S_2'(n)\right]  \label{pnalpha}
\ee
where the primes denote derivatives with respect to the density $n$.

By expanding $E'(n,0)$ and $S_2'(n)$ about the saturation density $n_0$ of symmetric matter 
(noting that $E'(n_0,0)=0$), we obtain
\ba
E'(n,0) &\simeq& \frac{K_0}{9n_0}\delta + \frac{Q_0}{54n_0}\delta^2  \label{en0}\\
S_2'(n) &\simeq& \frac{L}{3n_0} + \frac{K_{S_2}}{9n_0}\delta + \frac{Q_{S_2}}{54n_0}\delta^2 \label{s2pr}
\ea
where $\delta = (n/n_0)-1$, and 
\ba
K_0 &=& 9n_0^2\left.\frac{d^2E(n,0)}{dn^2}\right|_{n_0}\,,    \quad
Q_0 = 27n_0^3\left.\frac{d^3E(n,0)}{dn^3}\right|_{n_0}   \\
L &=& 3n_0\left.\frac{dS_2(n)}{dn}\right|_{n_0}\,,  \qquad
K_{S_2} = 9n_0^2\left.\frac{d^2S_2(n)}{dn^2}\right|_{n_0}   \\
Q_{S_2} &=& 27n_0^3\left.\frac{d^3S_2(n)}{dn^3}\right|_{n_0}    
\ea
The skewness $\mathcal{S}$ is related to $K_0$ and $Q_0$ via  
\ba
\mathcal {S} &=& k_F^3\left.\frac{d^3 E}{dk_F^3}\right|_{\alpha=0,n_0} 
              = 6 K_0 + Q_0
\ea
and the symmetry term $K_{\tau}$ of the liquid drop formula for the isospin asymmetric 
incompressibility ~\cite{kt} is related to $S_v$, $L$, $K_0$, and $K_{S_2}$ via
\begin{eqnarray}
K_{\tau} &=& K_{S_2} -\frac{LS_v}{K_0}.
\end{eqnarray}
At the equilibrium density $n_{0\alpha}$ of isospin asymmetric matter,
\begin{equation}
P(n_{0\alpha},\alpha) = 0 = E'(n_{o\alpha},0)+S_2'(n_{0\alpha})\alpha^2 .
\label{Pn0alpha}
\end{equation}
The insertion of Eqs. (\ref{en0})-(\ref{s2pr}) into Eq. (\ref{Pn0alpha}), while retaining terms 
up to $O(\delta)$, leads to \cite{blaizot81,prak85}
\be
\delta_{\alpha} \equiv \frac{n_{0\alpha}}{n_0}-1 = - \frac {3L}{K_0}\alpha^2 \equiv -C\alpha^2 
\label{delal}
\ee
to lowest  order in $\alpha^2$. This relation allows us to trace the loci of the minima of 
the energy per particle for changing asymmetries. Further improvement to cover higher values 
of $\alpha$ requires keeping terms to $O(\delta^2)$ in Eqs.(\ref{en0})-(\ref{s2pr}):
\ba
\delta_{\alpha} &=&  \frac{3K_0}{Q_0}\frac{\left(1+\frac{K_{S_2}}{K_0}\alpha^2\right)}
                                                     {\left(1+\frac{Q_{S_2}}{Q_0}\alpha^2\right)} \nonumber \\
                &*& \left\{-1+\left[1-\frac{2LQ_0\alpha^2\left(1+\frac{Q_{S_2}}{Q_0}\alpha^2\right)}
                                           {K_0^2\left(1+\frac{K_{S_2}}{K_0}\alpha^2\right)^2}\right]^{1/2}
                      \right\}.  \label{delal2}
\ea
In this expression, we have discarded terms involving $L_4$ because, as we mentioned earlier, these 
are very small and make no significant contributions. Additionally, for APR, $K_{S_2}/K_0 
\sim 0.4$ and $Q_{S_2}/Q_0 \sim -1.2$. The large $(>1)$ magnitude of $|Q_{S_2}/Q_0|$ means that for 
$\alpha \ge 0.7$ (which was the reason for going beyond $\alpha^2$ in the first place), we incur 
significant error upon expanding Eq. (\ref{delal2}) in a Taylor series in $\alpha$. This problem does 
not arise for Ska where $K_{S_2}/K_0 \sim 0.3$ and $Q_{S_2}/Q_0 \sim -0.6$. In the latter 
case, Eq. (\ref{delal2}) can be reduced to the simple form
\be
\delta = -\frac{3L}{K_0}\alpha^2\left[1+\left(\frac{Q_0L}{2K_0^2}
             -\frac{K_{S_2}}{K_0}\right)\alpha^2\right]. \label{delez}
\ee
We stress that Eq. (\ref{delez}) is applicable only in situations where $|K_{S_2}/K_0|$ and $|Q_{S_2}/Q_0|$ 
are much smaller than 1. If this condition does not hold (such as in APR), the more general 
expression (\ref{delal2}) must be used.

Finally, we calculate the incompressibility at the saturation density $n_{0\alpha}$ of 
asymmetric matter in terms of symmetric matter equilibrium properties, to $O(\alpha^2)$ (see also, Refs. \cite{blaizot81,prak85}). 
Using Eq. (\ref{pnalpha}) we get, for general $n$, 
\ba
K(n,\alpha) &=& 9 \frac{\partial P(n,\alpha)}{\partial n}  \\
            &=& K(n,0)\left(1+A(n)\alpha^2\right)
\ea
where
\ba
K(n,0) &=& 9\left[2nE'(n,0)+n^2E''(n,0)\right]  \\
A(n) &=& \frac{9}{K(n,0)}\left[2nS_2'(n)+n^2S_2''(n)\right]
\ea
At $n=n_{0\alpha}$, 
\ba
K(n_{0\alpha}) &\simeq& K(n_0,0)+\left.\frac{dK(n,0)}{dn}\right|_{n_0}(n_{0\alpha}-n_0)   \\
  &=& K_0 + \left(4K_0+\frac{Q_0}{3}\right)\delta_{\alpha}  \\
  &\simeq& K_0\left[1-\frac{12L}{K_0}\left(1+\frac{Q_0}{12K_0}\right)\alpha^2\right]  \\
  &\equiv& K_0(1+B\alpha^2)
\ea
and
\ba
A(n_{0\alpha}) &\simeq& \frac{9}{K_0}\left(2n_0\left.\frac{dS_2(n)}{dn}\right|_{n_0}
                           +n_0^2\left.\frac{d^2S_2(n)}{dn^2}\right|_{n_0}\right)  \\
 &=& \frac{9}{K_0}\left(2n_0\frac{L}{3n_0}+n_0^2\frac{K_{S_2}}{9n_0^2}\right)  \\
 &=& \frac{6L}{K_0}\left(1+\frac{K_{S_2}}{6L}\right) \equiv A.
\ea
Hence, to ${\cal{O}}(\alpha^2)$,  
\ba
K(n_{0\alpha},\alpha) 
    &\simeq& K_0[1+(A+B)\alpha^2]   \\
    &\equiv& K_0(1+\tilde A \alpha^2)\,,
\ea
where the coefficient $A$ represents modifications to the compressibility evaluated at $n_0$ 
due to changing asymmetry, whereas the coefficient $B$ encodes alterations due to the shift of the 
saturation point of matter as the asymmetry varies.

\subsection{Results and  analysis} 

In this section, the zero temperature results obtained from the APR and Ska  Hamiltonians are presented.  
Columns 2 and 3 in
Table \ref{satprops} contain the key symmetric nuclear matter properties for both models at their respective equilibrium densities
(nearly the same). Note that while the  
energy per particle $E(n_0) \equiv E_0$ and  the compression modulus $K_0$ for both models are similar, the effective masses
$m_0^*/m$ are somewhat different near nuclear densities. Significant differences are seen in the skewness parameters 
$\mathcal {S}$, the Ska model being more asymmetric than the APR model at its equilibrium density.    \\

\begin{table}[!h]
\begin{center}
\begin{tabular}{|c|c|c|c|c|}
\hline
     & APR & Ska & Experiment & Reference\\
\hline
 $n_0$(fm$^{-3}$) & 0.160 & 0.155 & $0.17\pm0.02$ & \cite{day78,jackson74,myers66,myers96}  \\
 $E_0$ (MeV) & -16.00 & -15.99 & $-16\pm1$        & \cite{myers66,myers96}                  \\
 $K_0$ (MeV) & 266.0 & 263.2 & $230\pm30$         & \cite{Garg04,Colo04}                    \\
 &&&$240\pm20$  &\cite{shlomo06}\\
 $Q_0$ (MeV) & -1054.2 & -300.2 & $-700\pm500$    & \cite{Farine97}                         \\
 $S_v$ (MeV) & 32.59 & 32.91 & 30-35              & \cite{L,tsang12}                        \\
 $L$   (MeV) & 58.46 & 74.62 & 40-70              & \cite{L,tsang12}                        \\
 $K_{S_2}$ (MeV) & -102.6 & -78.46 & $-100\pm200$ & This work                               \\
 $Q_{S_2}$ (MeV) & 1217.0 & 174.5  &    ?         &                                         \\
 $\mathcal{S}$ (MeV)& 541.8 & 1278.9 & $680\pm530$& This work                               \\ 
 $m_0^*/m$   &  0.70  &  0.61  & $0.8\pm0.1$      & \cite{bohigas79,krivine80}              \\   
\hline
\end{tabular}
\caption[Saturation properties of symmetric nuclear matter.]{
Entries in this table are at the equilibrium density $n_0$ of symmetric nuclear matter for the APR and Ska models.
$E_0$ is the energy per particle, $K_0$ is the compression modulus, $Q_0$ is related to the third 
derivative of $E$, $\mathcal{S}$ is the skewness, $m_0^*/m$ is 
the ratio of the Landau effective mass to mass in vacuum, $S_v$ is the nuclear symmetry energy parameter, 
and $L$, $K_{S_2}$, and $Q_{S_2}$ are related to the first, second, and third derivative of the symmetry 
energy, respectively.}
\label{satprops}
\end{center}
\end{table}

\begin{figure*}[htb]
\centering
\begin{minipage}[b]{0.49\linewidth}
\centering
\includegraphics[width=10cm]{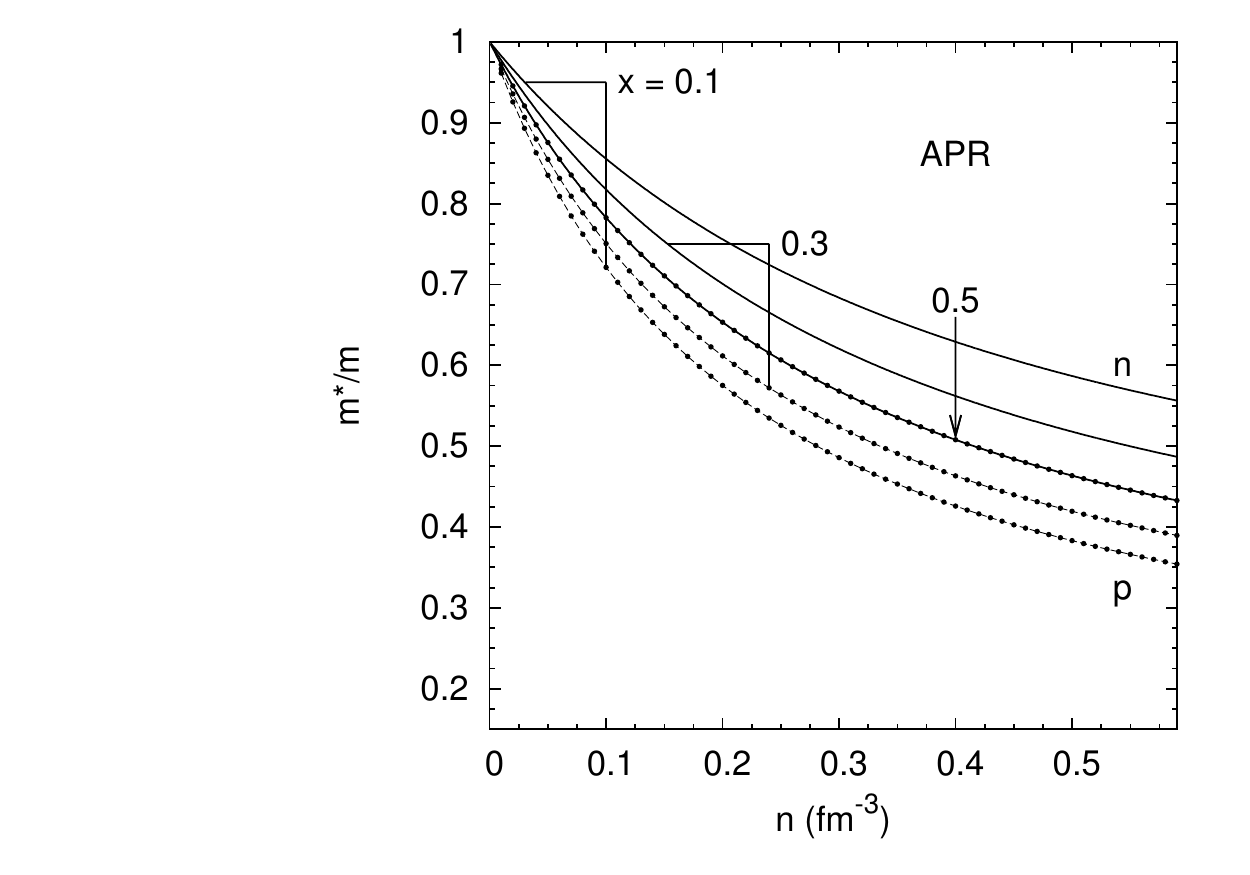}
\end{minipage}
\begin{minipage}[b]{0.49\linewidth}
\centering
\includegraphics[width=10cm]{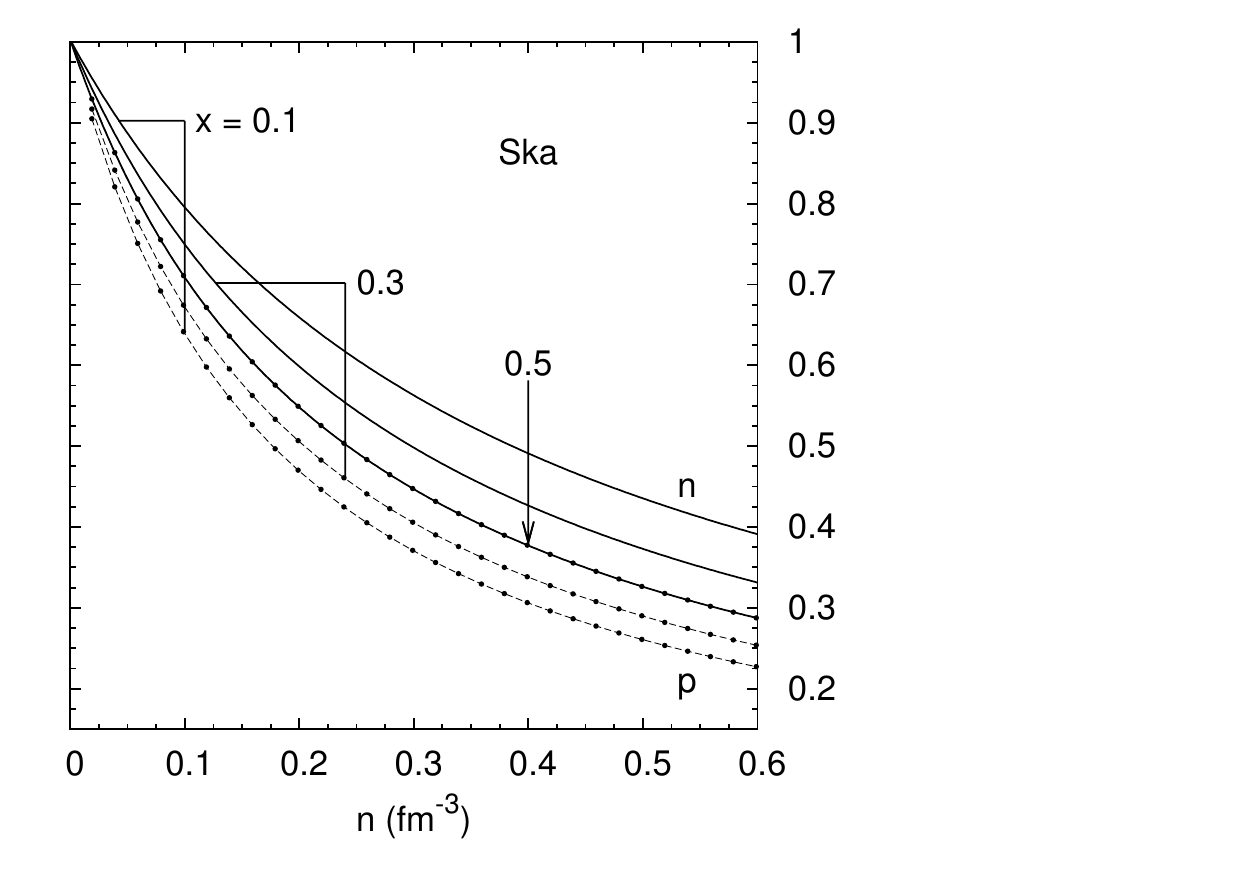}
\end{minipage}
\vskip -0.5cm
\caption{Left panel: Ratios of the neutron (solid) and proton (dotted) Landau effective masses 
to the vacuum mass versus baryon density $n$ for the APR model from Eq. (\ref{effmAPR}). 
Right panel:   Same as the left panel but for the Ska  model from Eq. (\ref{effmSka}). 
Values of the proton fraction $x$ are as indicated in the figure.} 
\label{APRSKA_Ms}
\end{figure*}

Among the most important quantities to be discussed are the 
nucleon Landau effective masses as they are critical to the thermal 
properties of the equation of state. We show ratios of the neutron
and proton Landau effective masses to the vacuum mass versus baryon
density $n$ for values of $x=0.5,~0.3$ and 0.1, respectively, in Fig. \ref{APRSKA_Ms}. 
The left panel is for the APR model from Eq. (\ref{effmAPR}) and the right panel 
contains similar results for the Ska model from Eq. (\ref{effmSka}). At the equilibrium density $n_0$ of symmetric nuclear matter, $m^*_0$ for Ska is 
smaller than for APR, and since $|X_2|<2X_1$ and $|p_5|<2p_3$, this 
means that $m^*$ is also smaller for Ska at every $x$ at $n_0$. Therefore, 
defining $a_{Ska}=X_1+Y_iX_2$ and $a_{APR}=p_3+Y_ip_5$, we must have $a_{Ska}>a_{APR}e^{-bn_0}$ 
for any $Y_i\in[0,1]$ from Eqs. (\ref{effmAPR}) and (\ref{effmSka}). It then follows 
from $p_4>0$ that $m^*_i$ is smaller for Ska at all densities for every value of 
$x\in[0,1]$ and for both neutrons and protons. Furthermore, since $p_5<0$ and $X_2<0$, 
we have that $m^*_n(n,x)>m^*_0>m^*_p(n,x)$ for $n>0$ and $x<1/2$.  

\begin{table}[!h]
\begin{center}
\begin{tabular}{|c|c|c||c|c|}
\hline
 n (fm$^{-3}$)   &AP(SNM)& APR(SNM) &AP(PNM)& APR(PNM) \\
\hline
 0.04   & -6.48  &  -5.63  &   6.45 &   6.42  \\
 0.08   & -12.13 &  -11.56 &   9.65 &   9.58  \\
 0.12   & -15.04 &  -14.98 &  13.29 &  13.28  \\
 0.16   & -16.00 &  -16.00 &  17.94 &  17.99  \\
 0.20   & -15.09 &  -15.16 &  22.92 &  23.57  \\
 0.24   & -12.88 &  -12.96 &  27.49 &  28.04  \\
 0.32   & -5.03  &  -5.14  &  38.82 &  39.41  \\
 0.40   &  2.13  &   2.62  &  54.95 &  54.72  \\
 0.48   &  15.46 &  15.14  &  75.13 &  74.59  \\
 0.56   &  34.39 &  32.92  &  99.74 &  99.45  \\
 0.64   &  58.35 &  56.22  & 127.58 & 129.57  \\
 0.80   & 121.25 & 119.97  & 205.34 & 206.22  \\
 0.96   & 204.02 & 207.14  & 305.87 & 305.06  \\
\hline
\end{tabular}
\caption{AP vs APR energies in MeV for symmetric nuclear matter (SNM) and pure neutron matter (PNM) extracted from Ref. \cite{apr}.} 
\label{APvsAPR} 
\end{center}
\end{table}

\begin{figure}[!h]
\begin{center}
\includegraphics[width=10cm]{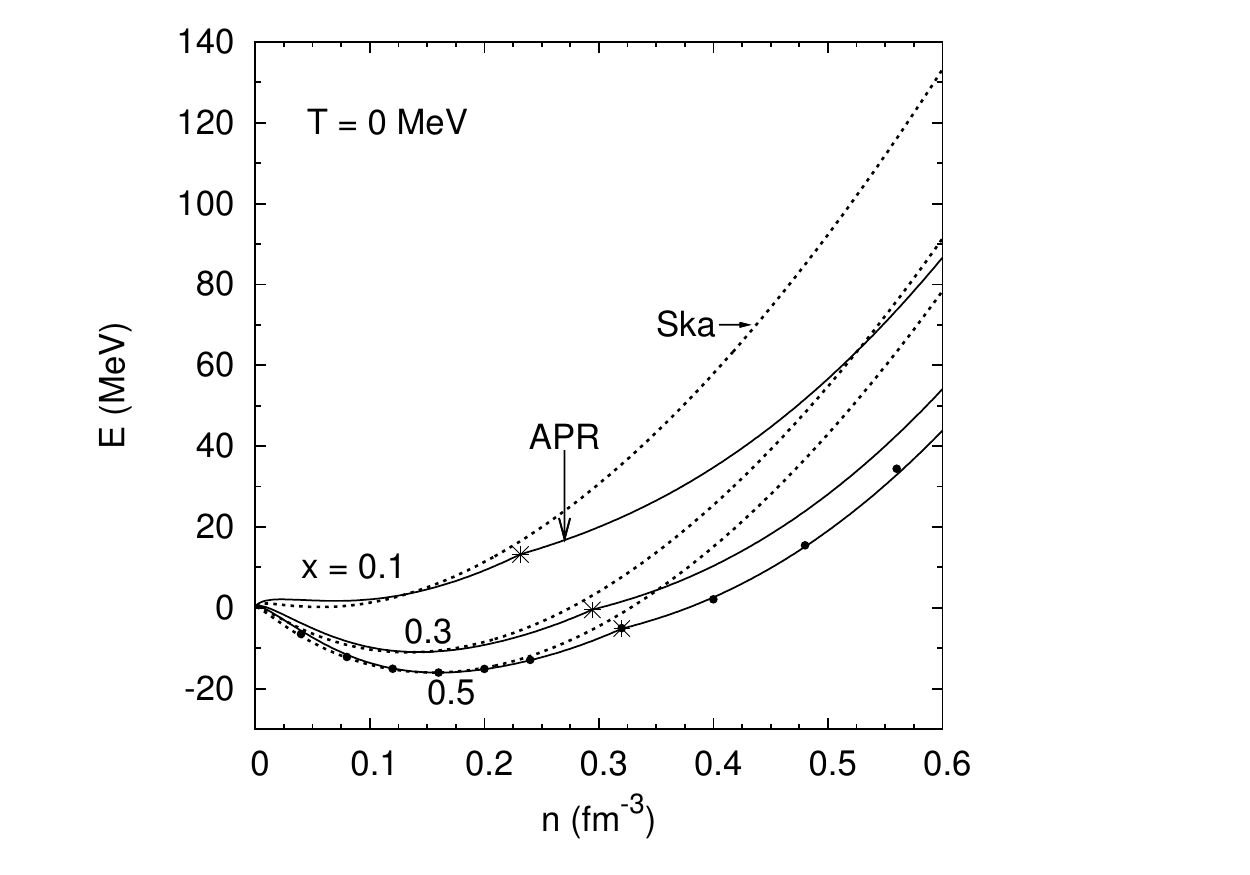}
\end{center}
\vskip -1cm
\caption{Zero temperature energy per particle $E$ versus baryon number density for the APR (solid curves) 
using Eqs. (\ref{EA1})-(\ref{EA2}) and Ska (dashed curves) 
models at the indicated values of the proton fraction $x$. The crosses on the APR curve for $x=1/2$ show values from column 6 of Table VI in Ref. ~\cite{apr}.  
Although not shown here, we have verified that similar agreement is obtained with the APR results in column 5 of Table VII in Ref.~\cite{apr} for pure neutron matter (x=0). 
The cusps in the APR curves are due to the onset of neutral pion condensation.} 
\label{APRSKA_0T_EoA}
\end{figure}

Figure \ref{APRSKA_0T_EoA} shows the energy particle $E$ as a function of baryon density $n$ 
for values of $x=0.5,~0.3$ and $0.1$ for the two models. Our calculated results of APR (solid curves)
agree well with those tabulated in Table VI and VII of Ref.~\cite{apr}  (shown by crosses for $x=0.5$ in this figure). We also contrast the microscopic AP results for pure neutron matter and symmetric nuclear matter with those obtained from the APR fit in Table \ref{APvsAPR}. (As noted in the introduction, results below $n\simeq 0.1~{\rm {fm}^{-3}}$ can be used to establish differences from the inhomogeneous phase of supernova matter containing nuclei, light nuclear clusters, etc.) The asterisks in Fig.  \ref{APRSKA_0T_EoA} show the densities at which the transition from the low density phase (LDP) to the high density phase (HDP) occurs due to pion condensation.   
While there is good agreement between the results of the two models
up to and slightly beyond the equilibrium density, the Ska model is seen to have both higher energies and pressures (slopes of the energy) than the APR model at high densities for all
values of $x$. This feature essentially stems from the emergence of the pion condensate in the HDP of APR
which softens the corresponding EOS. Both equations of state
become acausal at high densities; a scheme to retain causality will be outlined later.

Rows 5 and 6 in Table \ref{satprops} list the symmetry energy $S_v$ and its slope parameter $L$ for the two models. Although $S_v$ for both the models are similar, values of $L$ differ significantly. 
The higher value of $L$ for the Ska model leads to a greater energy and pressure of isospin asymmetric 
matter than for the APR model near nuclear saturation densities, a feature that persists to 
higher densities. \\

The density dependent symmetry energy $S_2(n)$ can in general be written as
$S_2 = S_{2k} + S_{2m} + S_{2d}$ with $S_{2k}$ as in Eq. (\ref{Skins}). 
Contributions from the momentum-dependent  and density-dependent parts, $S_{2m}$ and $S_{2d}$, 
depend on the model used. For the APR model,  
\ba
S_{2m} &=& \frac 13 k_F^2 n e^{-p_4n} \left( p_3 + 2p_5 \right)\,, \nonumber \\
S_{2d} &=& \frac 1n (-g_1 + g_2) \,,
\label{S2mAPR}
\ea
whereas for the Ska model
\ba
S_{2m} = \frac 13 k_F^2 n (X_1+2X_2)~{\rm and}~
S_{2d}  =  \frac n2 (X_4 + X_6 n^\epsilon) \,.
\label{S2mSKA}
\ea
Note that the terms $S_4(n)$ and $S_6(n)$ receive contributions from the momentum-dependent 
interaction part as well because of terms involving $n_i \tau_i$ in the $\mathcal {H}$'s of Eqs. (\ref{HAPR}) and (\ref{HSKA}).
Explicitly, 
\ba
S_{4m}  &= &  \frac{1}{3^4} k_F^2 n e^{-p_4n} \left(p_3-p_5\right)\,, \nonumber \\
S_{6m}  & =& \frac{7}{3^7} k_F^2 n e^{-p_4n} \left(p_3-\frac 15 p_5\right)\,
\label{S46APR}
\ea
for the APR model, and for the Ska model
\ba
S_{4m}  &= & \frac{1}{3^4} k_F^2 n \left(X_1-X_2\right)\,, \nonumber \\
S_{6m}  & =& \frac{7}{3^7} k_F^2 n \left(X_1-\frac{2}{5} X_2\right)\,. 
\label{S46SKA}
\ea

\begin{figure*}[htb]
\centering
\begin{minipage}[b]{0.49\linewidth}
\centering
\includegraphics[width=10cm]{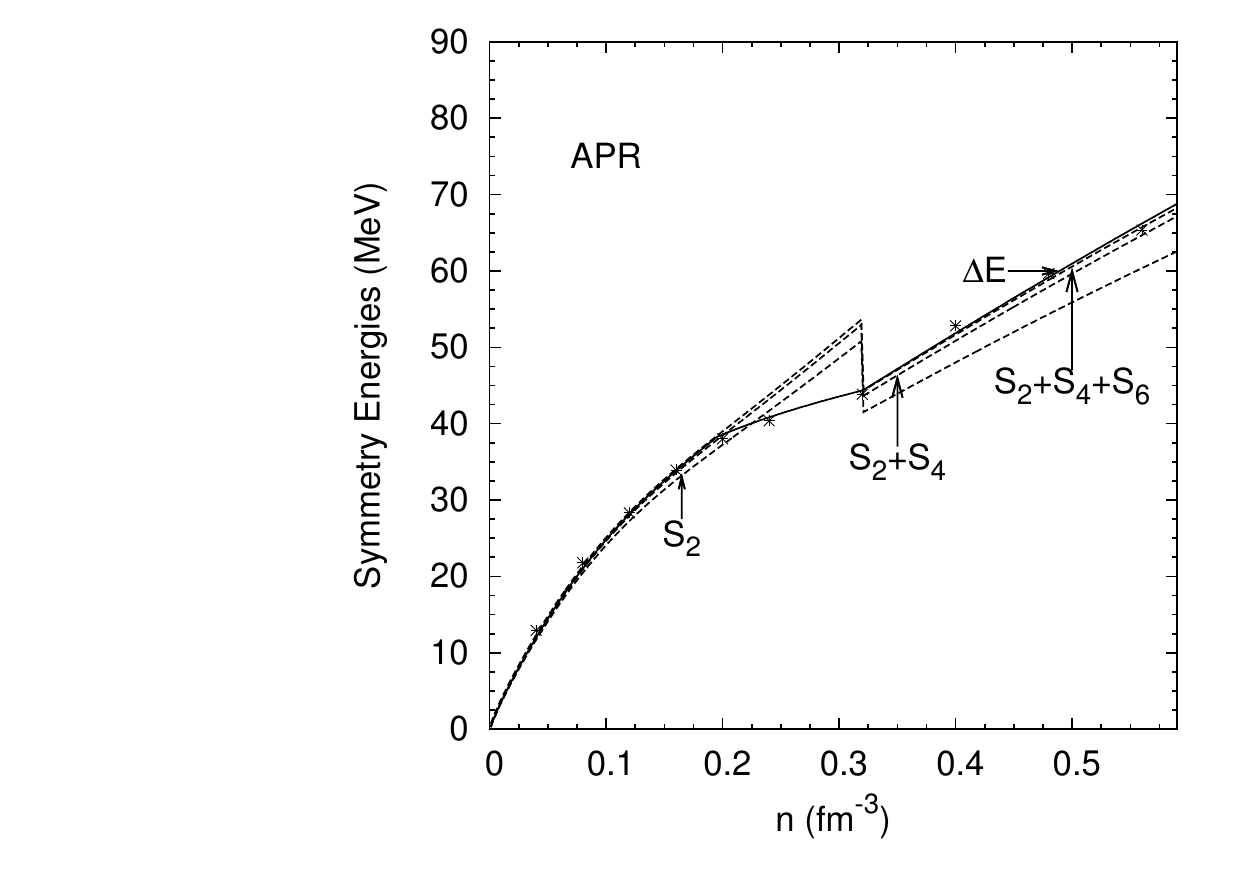}
\end{minipage}
\begin{minipage}[b]{0.49\linewidth}
\centering
\includegraphics[width=10cm]{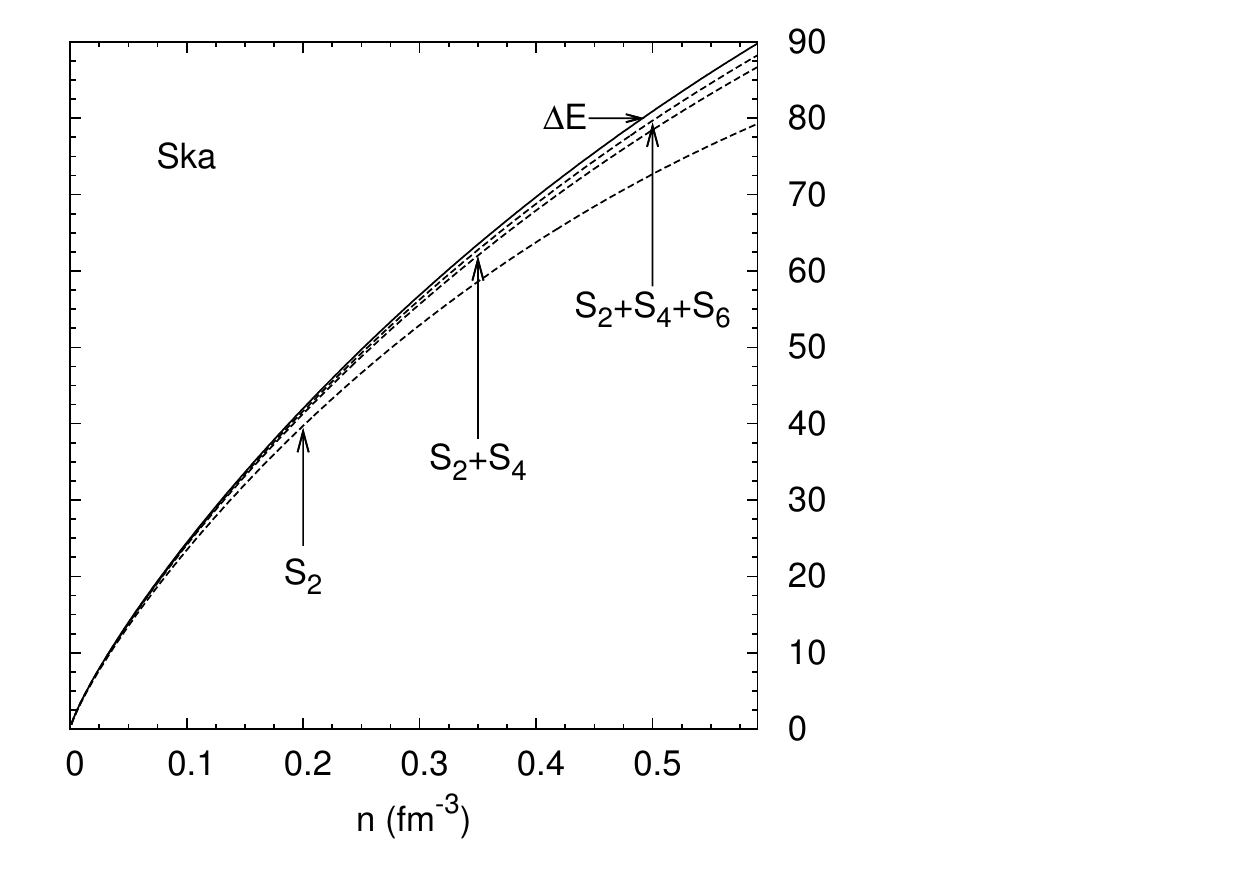}
\end{minipage}
\vskip -0.5cm
\caption{Left panel: Symmetry energies for APR (from Eqs. (\ref{s_kin}),(\ref{S2mAPR}) and (\ref{S46APR})) vs 
baryon density $n$. 
Right panel: Same as in the left panel but for Ska (Eqs. (\ref{s_kin}),(\ref{S2mSKA}) and (\ref{S46SKA})).}
\label{APR_SymE}
\end{figure*}

In Fig. \ref{APR_SymE}, the extent to which the functions  $S_2(n)$ (which we call the symmetry 
energy), $S_4(n)$ and $S_6(n)$ from Eqs. (\ref{Skins}), (\ref{S46APR}), and
(\ref{S46SKA}) contribute to the difference between pure neutron matter and nuclear matter 
energy, $\Delta E(n)=E(n,\alpha=1) - E(n,\alpha=0)$ (for which we reserve the term 
"asymmetry energy") is examined. The left (right) panel
shows results for the APR (Ska) model. The symmetry energy $S_2(n)$ adequately accounts for 
the total $\Delta E(n)$ up to twice $n_0$. However, for densities well in excess of $n_0$, 
contributions from $S_4(n)~S_6(n)~\cdots$ become important although $S_2(n)$ remains dominant. The jumps in 
the symmetry energies for APR at $n=p_{19}=0.32$ fm$^{-3}$ (at which transition from the LDP to 
HDP occurs for $x=0.5$) are due to the definitions of $S_2(n),~S_4(n),~S_6(n)~\cdots$ which
involve derivatives taken at $x=0.5$. As the transition to the HDP occurs at lower values 
of $n$ as $x$ decreases toward $x=0$, the conventional definitions of $S_2(n),~S_4(n),S_6(n)~\cdots$ 
fail to capture the true behavior of $\Delta E(n)$ in the presence of a phase transition. 
That is to say, 
\be
S(n) \equiv \sum_{l=2,4,\ldots}S_l(n) \ne \Delta E(n)
\ee
in the vicinity of a phase transition driven by density and composition, regardless of the order to which the sum is carried out.

Results for the coefficients $A,~B,~C,$ and $\tilde A$ 
that describe the isospin asymmetry dependence to $\mathcal {O}(\delta_{\alpha})$ of the equilibrium 
density and compression moduli for the APR and Ska models are displayed in Table \ref{Asymcoeffs}. 
Since asymmetry lowers the equilibrium density, transitions occurring at supra-nuclear densities do 
not affect these results. One observes that even though $\mathcal{H}_{APR}$ and $\mathcal{H}_{Ska}$ are calibrated to very similar values of the symmetry energy and the compression modulus, these asymmetry coefficients vary significantly.  

\begin{table}[h]
\begin{center}
\renewcommand{\arraystretch}{1.5}
\begin{tabular}{|c|c|c|c|c|}
\hline
Model & $A$ & $B$ & $C$ & $\tilde A=A+B$ \\
\hline
APR & 0.933 & -1.766 & 0.659 & -0.833  \\
Ska & 1.403 & -3.079 & 0.851 & -1.676  \\
\hline
\end{tabular}
\end{center}
\caption[Asymmetry Coefficients.]{Results for the coefficients that describe the isospin asymmetry 
dependence to $\mathcal {O}(\delta_{\alpha})$ of the equilibrium density and compression moduli.}
\label{Asymcoeffs}
\end{table}

\begin{figure*}[htb]
\centering
\begin{minipage}[b]{0.49\linewidth}
\centering
\includegraphics[width=10cm]{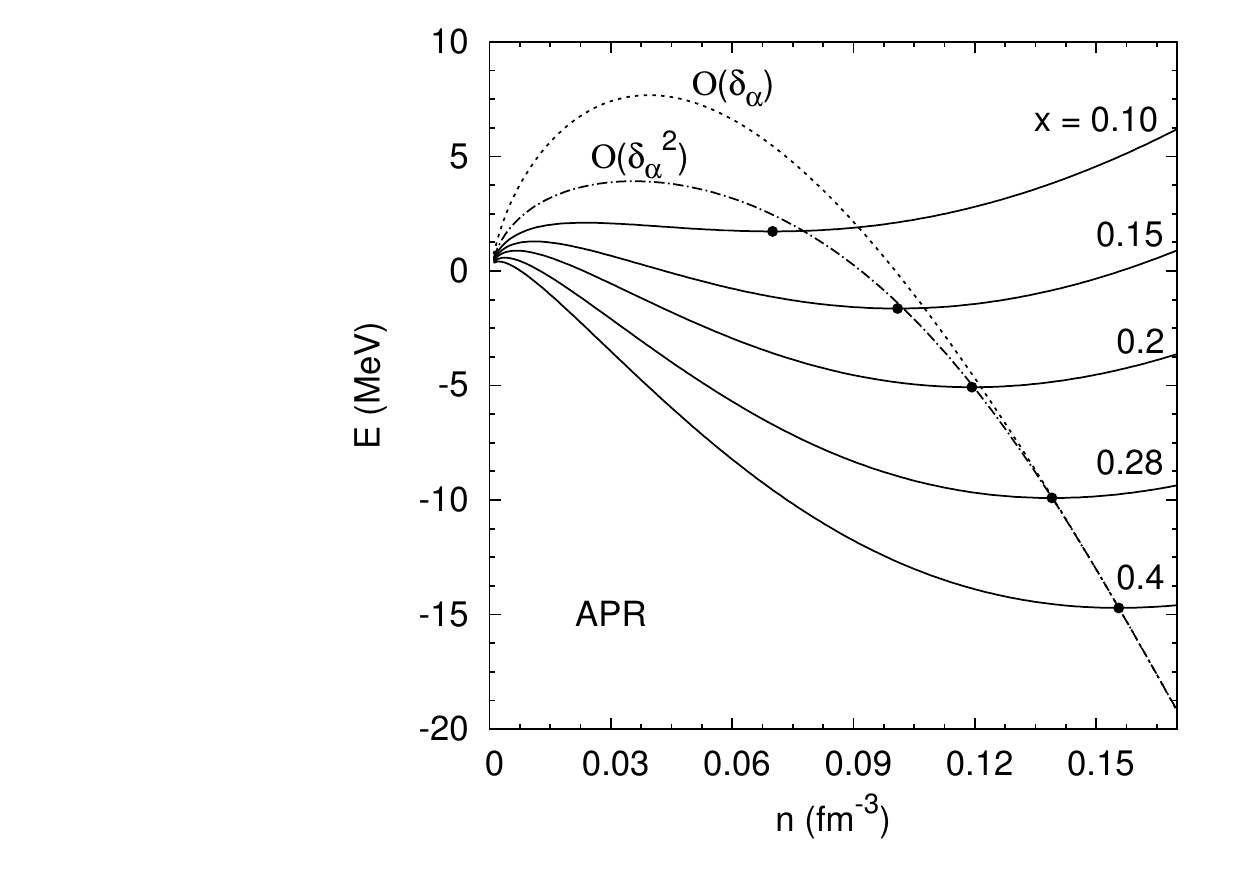}
\end{minipage}
\begin{minipage}[b]{0.49\linewidth}
\centering
\includegraphics[width=10cm]{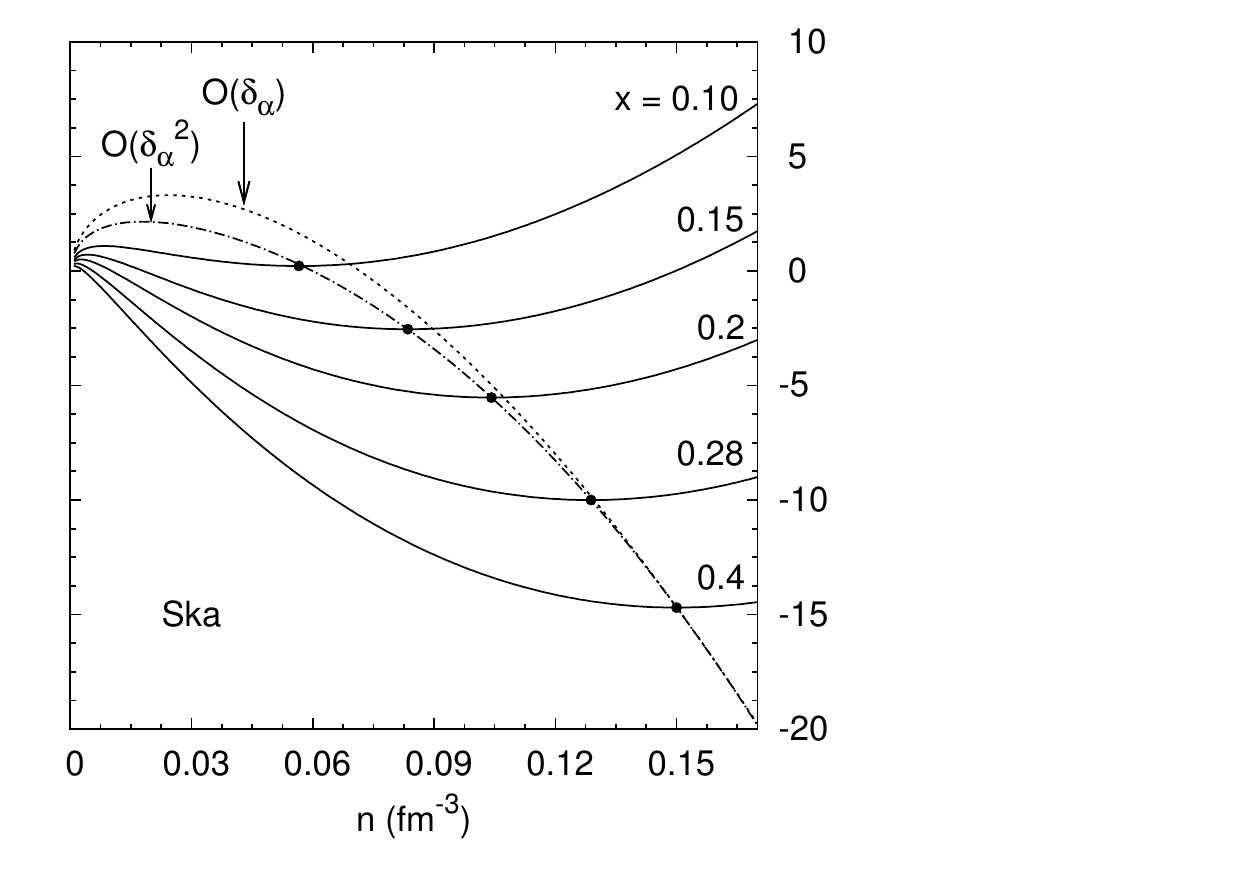}
\end{minipage}
\vskip -0.5cm
\caption{Left panel: Loci of minima in the energy per particle versus baryon density for the APR (left panel) and Ska (right panel) models for 
different proton fractions. The dark circles are exact results from  Eqs. (\ref{HAPR}) and (\ref{HSKA}).  The dotted curves show $\mathcal {O}(\delta_{\alpha})$ results from Eq. (\ref{delal}), whereas the 
$\mathcal {O}(\delta_{\alpha}^2)$ (Eq. (\ref{delal2})) contributions are shown as dashed lines. } 
\label{APRSKA_loci}
\end{figure*}

The extent to which Eq. (\ref{delal}), inserted into Eq. (\ref{Ealpha}) expanded to $\mathcal {O}(\alpha^2)$, 
adequately describes the loci of energy minima in the energy per particle of subnuclear matter 
for arbitrary $\alpha$ is demonstrated in Fig. \ref{APRSKA_loci} for the two models. 
The dark circles show locations of the minima resulting from the exact calculations using 
Eqs. (\ref{HAPR}) and (\ref{HSKA}) as the proton fraction $x$ is varied toward that of pure neutron matter. 
The leading order results shown by the dotted curves accurately trace the loci of minima down to
$x = 0.2$.  Considering the $\mathcal {O}(\delta_{\alpha}^2)$ contribution in Eq. (\ref{delal2}) improves agreement with the exact results even down to $x=0.1$. 

\begin{figure}[htb]
\includegraphics[width=9cm]{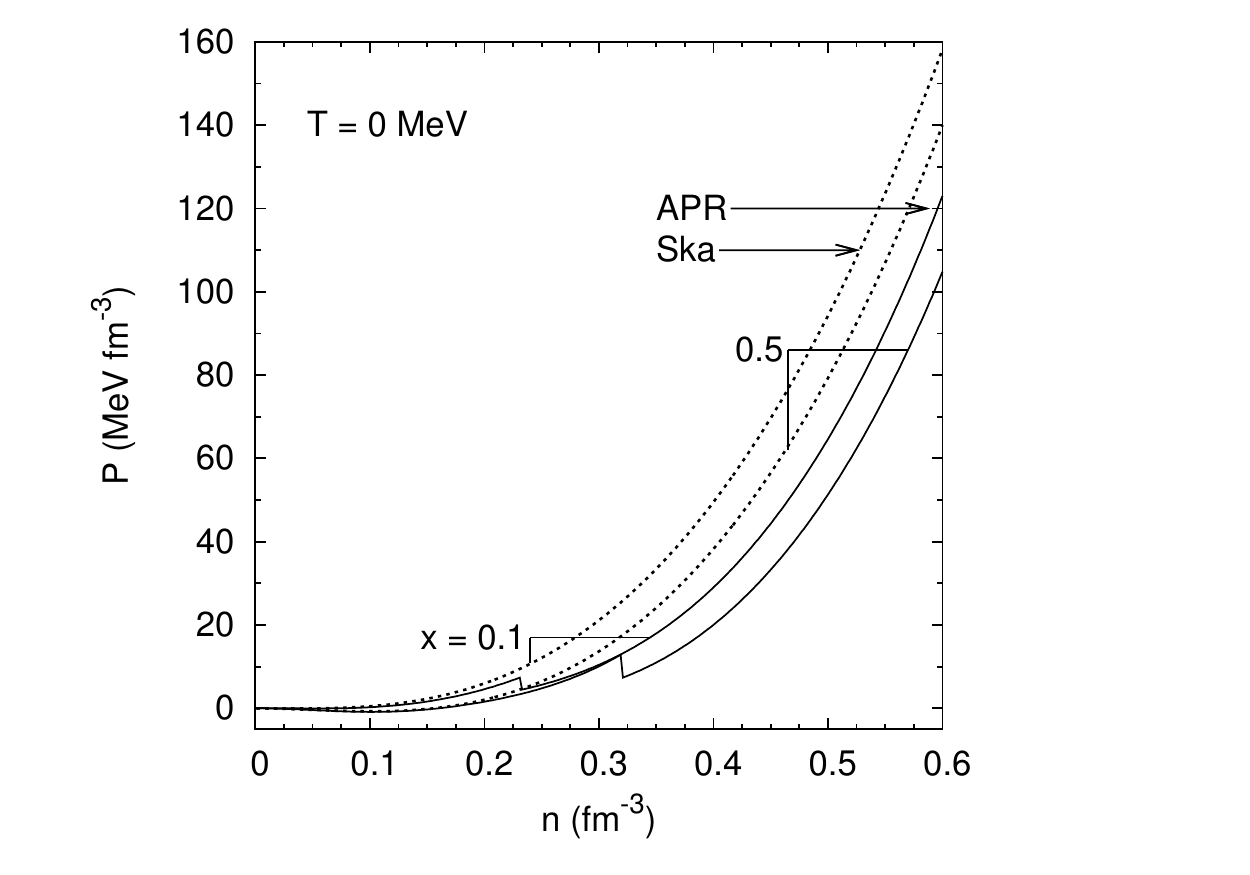}
\caption{Pressure versus baryon density for the APR (Eqs. (\ref{P1})-(\ref{P2})) and Ska models at different proton fractions. 
The jumps in the APR results are due to phase transition to a pion condensate at the values of $x$ indicated.}
\label{APRSKA_0T_P}
\end{figure}

In Fig. \ref{APRSKA_0T_P}, we show the pressure as a function of $n$ for representative values of $x$.
For all $x$, including for neutron matter (not shown), the Ska model has higher pressure than that 
for the APR model. As with the energy per particle shown in Fig. \ref{APRSKA_0T_EoA}, the larger 
stiffness of the Ska model relative to the APR model is caused by appearance of a pion condensate in 
the HDP of the latter. The distinctive jumps in pressure for the APR model are due to the phase transition to a pion condensate, i.e., from the LDP to the HDP which occurs at lower densities for increasingly 
asymmetric matter. 

\begin{figure*}[htb]
\centering
\begin{minipage}[b]{0.32\linewidth}
\centering
\includegraphics[width=7.5cm]{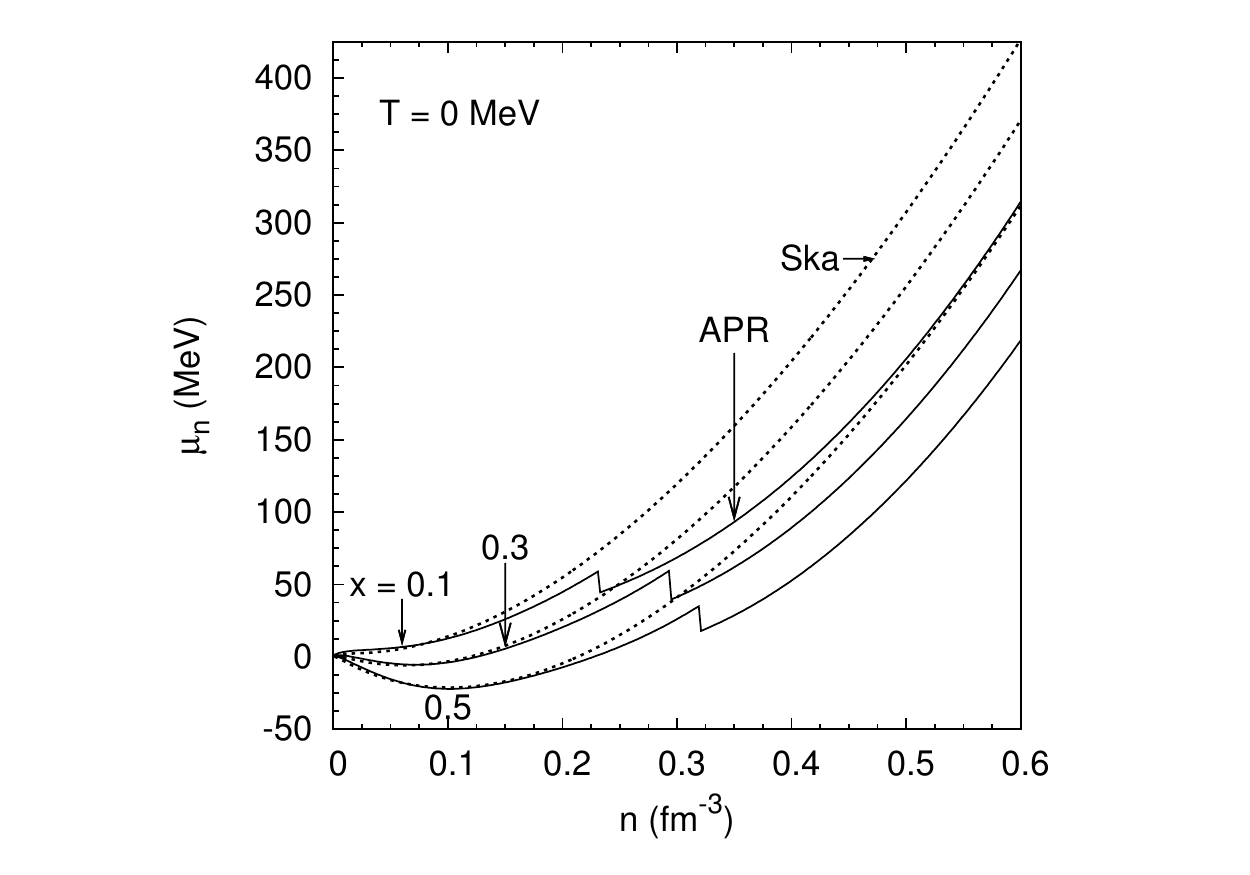}
\end{minipage}
\begin{minipage}[b]{0.32\linewidth}
\centering
\includegraphics[width=7.5cm]{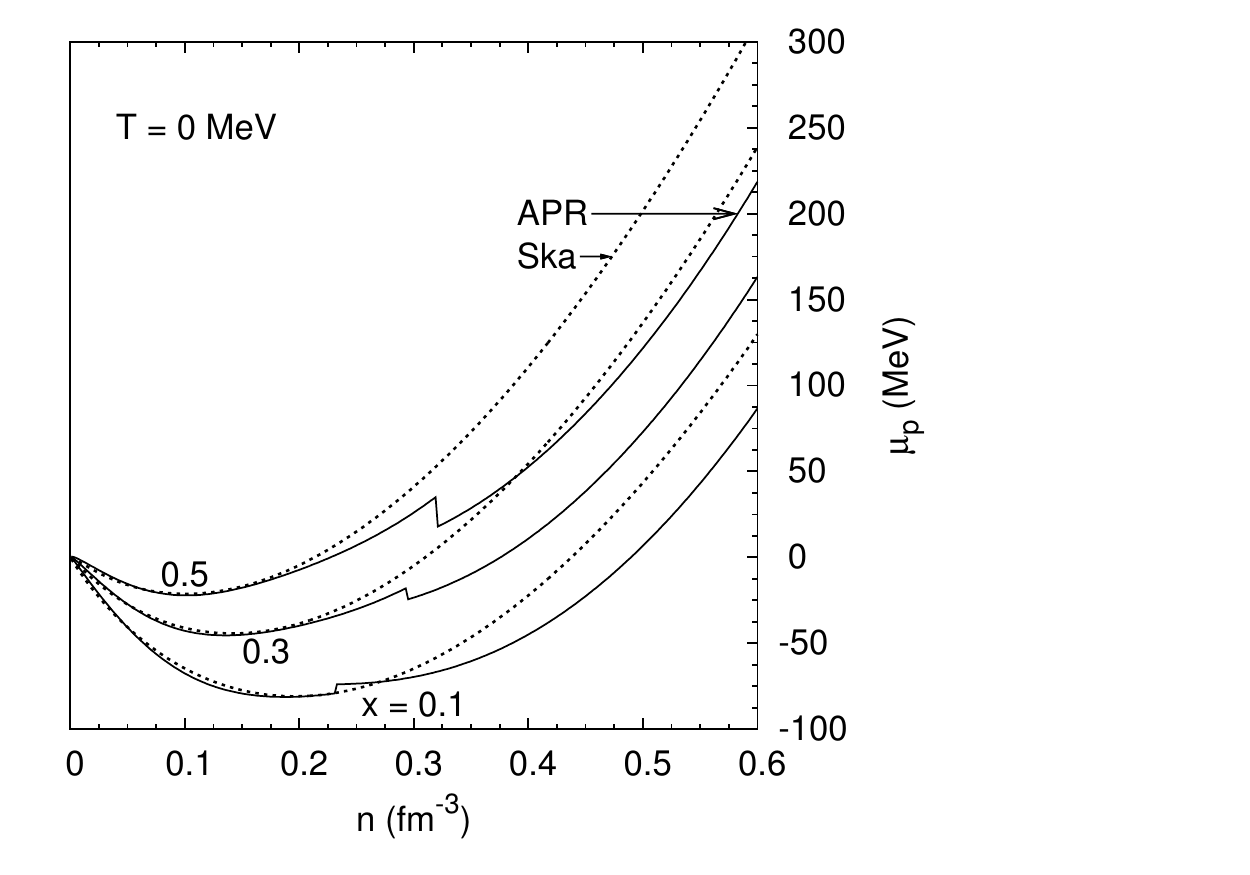}
\end{minipage}
\begin{minipage}[b]{0.32\linewidth}
\centering
\includegraphics[width=8cm]{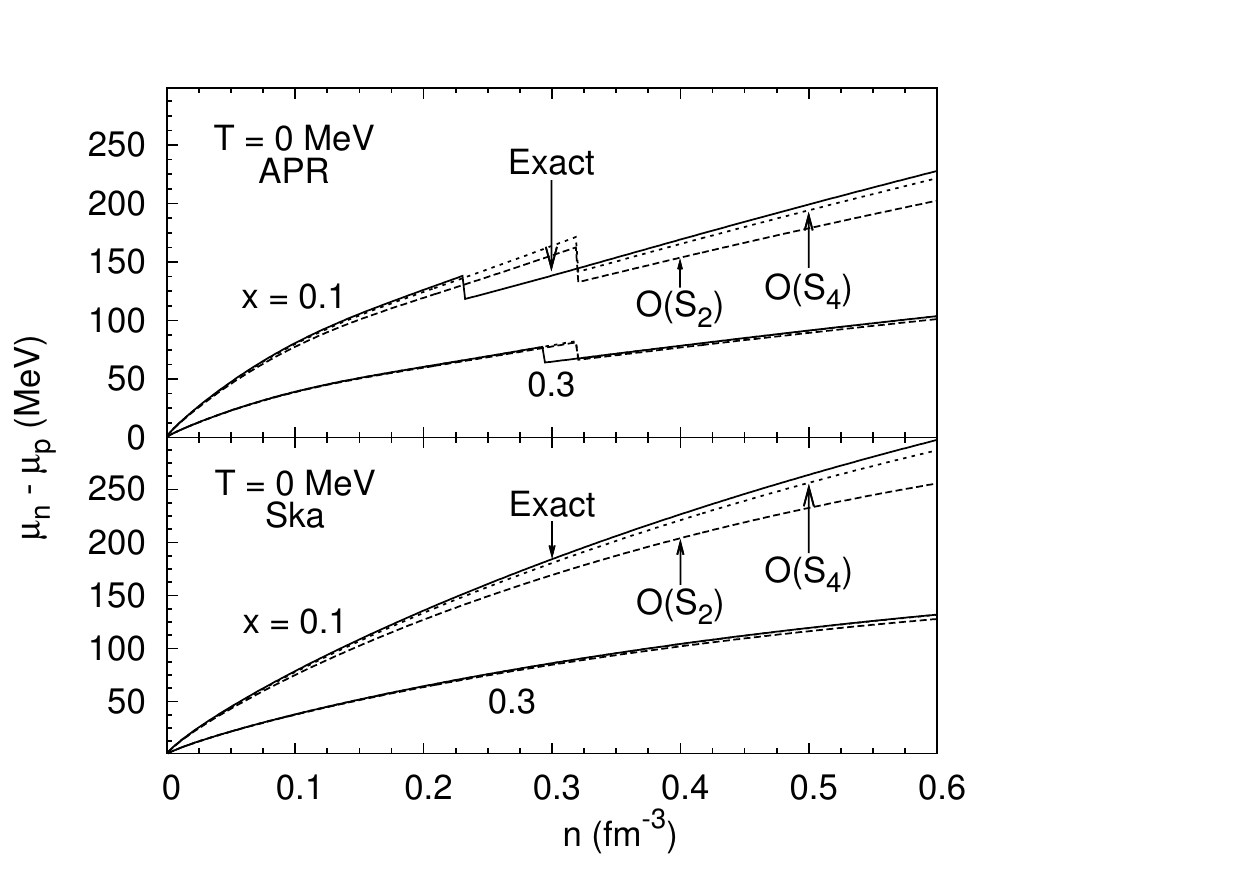}
\end{minipage}
\vskip -0.5cm
\caption{The first (second) panel shows the neutron (proton) chemical potential versus baryon density $n$ for the APR (Eqs. (\ref{MU1})-(\ref{MU2})) and Ska models for different values of $x$. The rightmost panel shows 
$\hat \mu =\mu_n - \mu_p$. The jumps in the APR results are due to phase transitions to a pion condensate.}
\label{APRSKA_0T_muNPs}
\end{figure*}

The neutron and proton chemical potentials, $\mu_n$ and $\mu_p$,  versus baryon density for the two models are shown in the first two panels of Fig. \ref{APRSKA_0T_muNPs}. 
Due to its relative stiffness, results for the Ska model are systematically larger than those for the APR model for all values of the proton fraction $x$.  It is worthwhile to mention here that $\hat \mu = \mu_n-\mu_p$  (with modifications from effects of temperature to be discussed in subsequent sections), shown in the rightmost panel of Fig.  \ref{APRSKA_0T_muNPs}, controls the reaction rates associated with electron captures and neutrino interactions in supernova matter.

\begin{figure*}[htb]
\centering
\begin{minipage}[b]{0.32\linewidth}
\centering
\includegraphics[width=7.5cm]{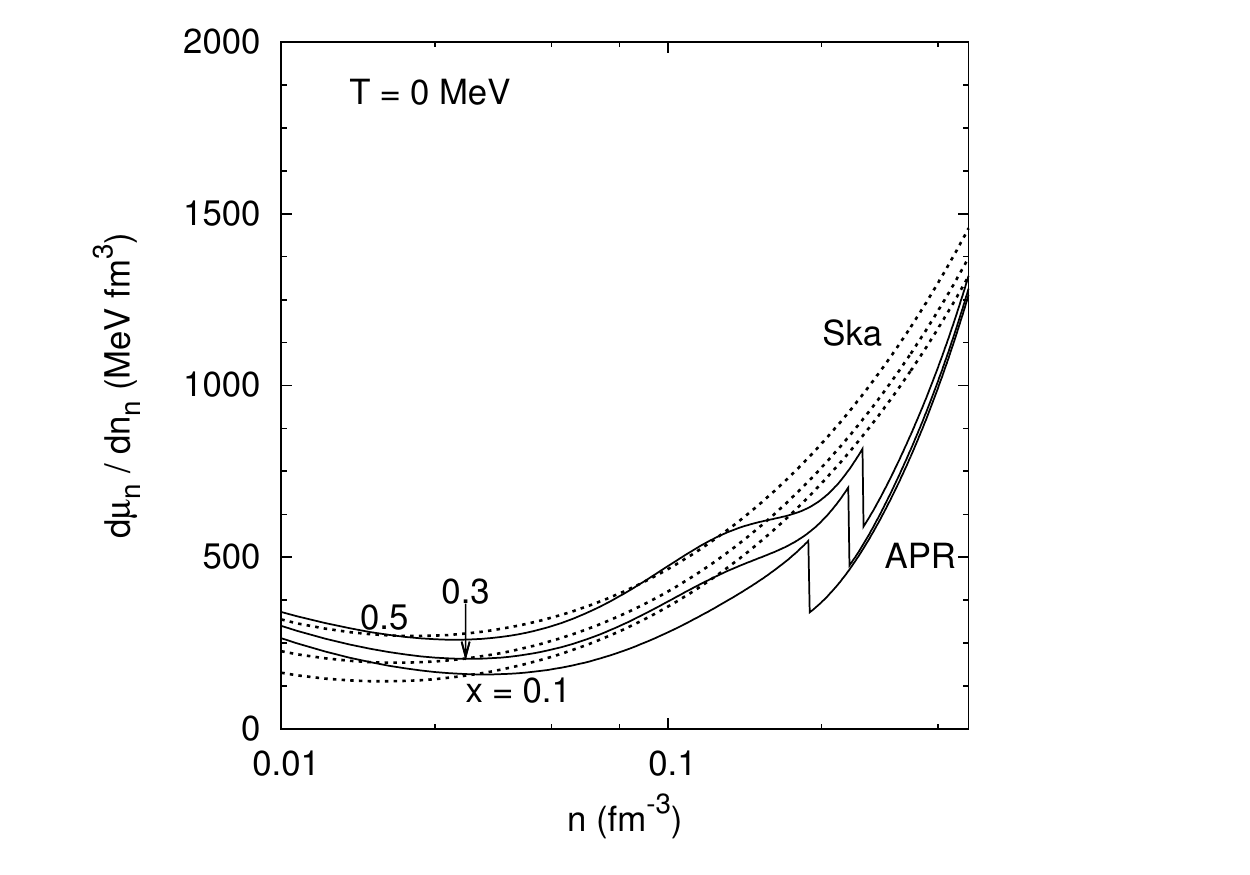}
\end{minipage}
\begin{minipage}[b]{0.32\linewidth}
\centering
\includegraphics[width=7.5cm]{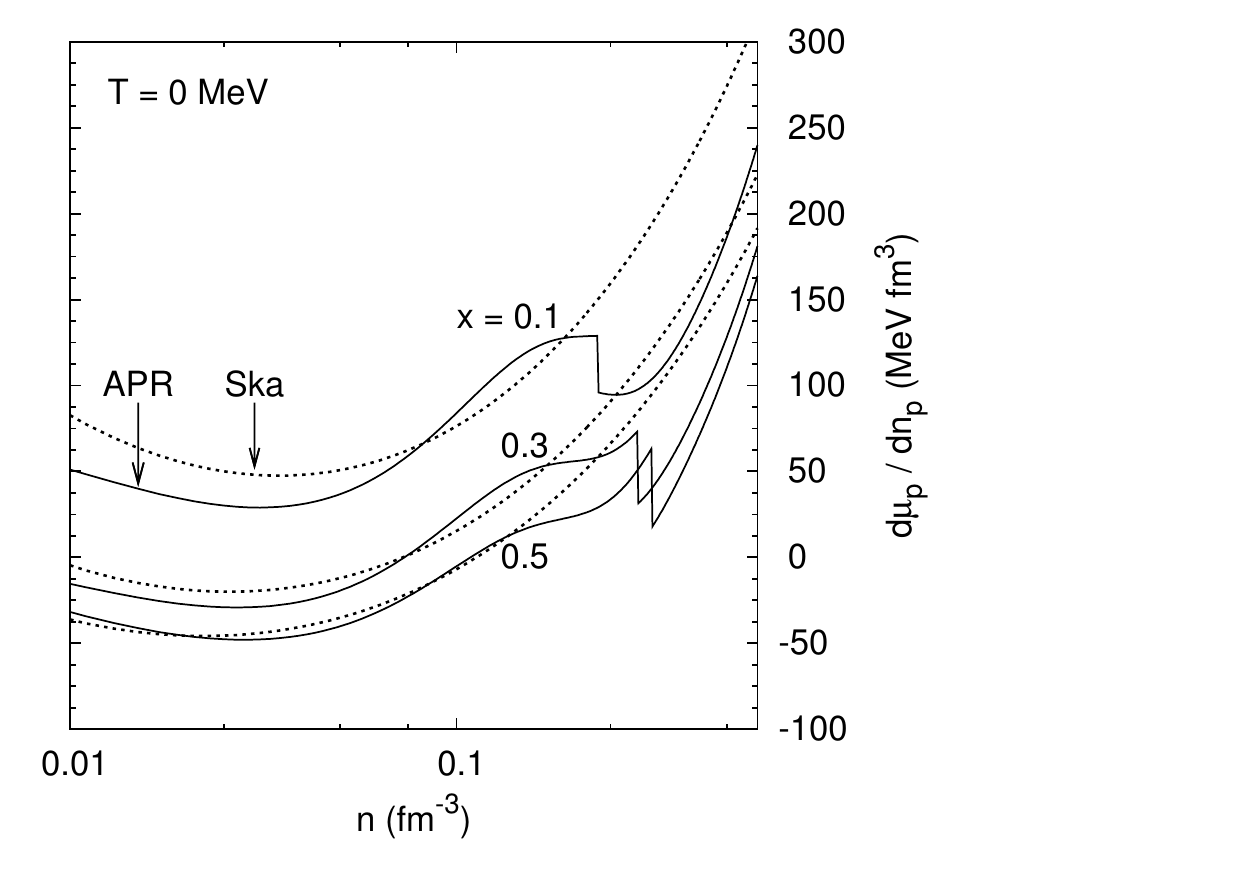}
\end{minipage}
\begin{minipage}[b]{0.32\linewidth}
\centering
\includegraphics[width=7.5cm]{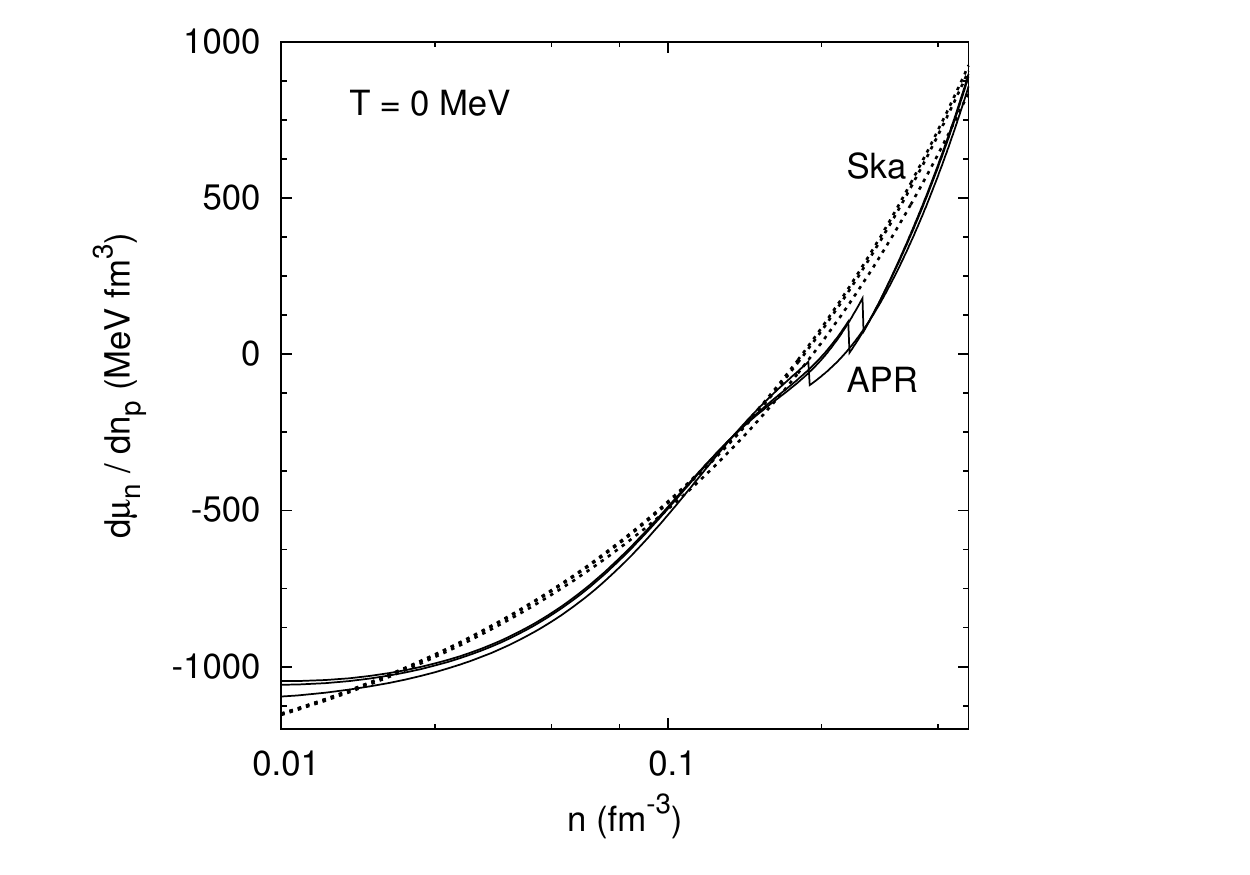}
\end{minipage}
\vskip -0.5cm
\caption{Neutron and proton inverse susceptibilities versus baryon density for the APR 
(Eqs. (\ref{CHI1})-(\ref{CHI2})) and Ska models at the indicated proton fractions $x$. 
Recall that $d\mu_n/dn_p = d\mu_p/dn_n$. 
Note that the cross derivatives have a very weak $x$-dependence. The jumps in the APR results are due 
to phase transitions to a pion condensate. }
\label{APRSKA_0T_dMudn}
\end{figure*}

The inverse susceptibilities are shown in Fig. \ref{APRSKA_0T_dMudn} for the APR and Ska models at representative proton fractions. The largest qualitative and quantitative differences between the 
two models occur at supra-nuclear densities for $d\mu_n/dn_n$ and $d\mu_p/dn_p$. The cross derivatives     $d\mu_n/dn_p = d\mu_p/dn_n$ are qualitatively similar for the two EOSs, but relatively small quantitative differences between the two models exist. In the case of the APR model, in which a pion condensate appears, these derivatives are required ingredients in the Maxwell construction which determines the phase boundary densities at which the pressure and an average chemical potential are equal (this ensures mechanical and chemical equilibria). These derivatives are also utilized in constructing the full dense matter 
tabular EOS as will be discussed later.

\section{FINITE TEMPERATURE PROPERTIES}

In this section, properties of the APR and Ska models at finite temperature $T$ are calculated.
At finite $T$, the Hamiltonian density is a function of four independent variables; namely, the number 
densities $n_i$ and the kinetic energy densities $\tau_i$ of the two nucleon species. These are, in turn, 
proportional to the $F_{1/2}$ and $F_{3/2}$ Fermi-Dirac (FD) integrals~\cite{pathria}, respectively:
\begin{eqnarray}
n_i & = & \frac{1}{2\pi^2}\left(\frac{2m_i^*T}{\hbar^2}\right)^{3/2}F_{1/2i}  
\label{nT}   \\
\tau_i & = & \frac{1}{2\pi^2}\left(\frac{2m_i^*T}{\hbar^2}\right)^{5/2}F_{3/2i} 
\label{tauT}   \\
\mbox{where} ~~~~~F_{\alpha i} & = & \int_0^{\infty}\frac{x_i^{\alpha}}{e^{-\psi_i}e^{x_i}+1}dx_i  \\
x_i & = & \frac{1}{T}\left(k_i^2\frac{\partial \mathcal{H}}{\partial \tau_i}\right)
      = \frac{1}{T}\frac{\hbar^2k_i^2}{2m_i^*} \equiv \frac{\varepsilon_{k_i}}{T}  \\
\psi_i & = & \frac{1}{T}\left(\mu_i-\frac{\partial \mathcal{H}}{\partial n_i}\right)
         = \frac{\mu_i-V_i}{T} \equiv \frac{\nu_i}{T}.
\end{eqnarray}
The quantity $\psi_i$, generally termed as the degeneracy parameter, is related
to the fugacity defined by $z_i = e^{\psi_i}$.
In the above equations, one must keep in mind that $m_i^*$ is a function of the number densities of both nucleon species $i=n,p$. Consequently, derivatives of the FD integrals with respect to the densities take the forms  
\ba
\frac{\partial F_{1/2i}}{\partial n_i} 
            & = & \frac{F_{1/2i}}{n_i}\left(1-
\frac{3}{2}\frac{n_i}{m_i^*}
            \frac{\partial m_i^*}{\partial n_i}
                          \right)   \\
\mbox{and} ~~~~~~\frac{\partial F_{1/2i}}{\partial n_j} & = & 
                        -\frac{3}{2}\frac{n_i}{m_j^*}
                        \frac{\partial m_i^*}{\partial n_j}F_{1/2i}.
\ea
FD integrals of different order are connected through their derivatives with respect to $\psi_i$:
\begin{eqnarray}
\frac{\partial F_{\alpha i}}{\partial \psi_i} & = & \alpha F_{(\alpha-1)i}.
\label{dfadpsi}
\end{eqnarray}
Therefore, 
\begin{eqnarray}
\frac{\partial F_{\alpha i}}{\partial n_i} & = & \frac{\partial F_{\alpha i}}{\partial F_{1/2i}}
                                                 \frac{\partial F_{1/2 i}}{\partial n_i}  \nonumber \\
     &=&  \frac{\partial F_{\alpha i}}{\partial \psi_i}
          \left(\frac{\partial F_{1/2 i}}{\partial \psi_i}\right)^{-1}
          \frac{\partial F_{1/2 i}}{\partial n_i}   \nonumber  \\
    & = & 2\alpha \frac{F_{(\alpha-1) i}}{F_{-1/2i}}\frac{\partial F_{1/2 i}}{\partial n_i}.  
\label{dfadni}
\end{eqnarray}
Similarly, cross derivatives with respect to density of Fermi integrals are given by 
\begin{eqnarray}
\frac{\partial F_{\alpha i}}{\partial n_j} & = &
      2\alpha \frac{F_{(\alpha-1) i}}{F_{-1/2i}}\frac{\partial F_{1/2 i}}{\partial n_j}.
\label{dfadnj}
\end{eqnarray}

Utilizing the relations 
\begin{eqnarray}
\frac{\partial}{\partial n} & = & \frac{\partial}{\partial n_n}\left.\frac{\partial n_n}{\partial n}\right|_x
             + \frac{\partial}{\partial n_p}\left.\frac{\partial n_p}{\partial n}\right|_x
      =(1-x)\frac{\partial}{\partial n_n}+x\frac{\partial}{\partial n_p}   \nonumber \\
\frac{\partial}{\partial x} & = & \frac{\partial}{\partial n_n}\left.\frac{\partial n_n}{\partial x}\right|_n
             + \frac{\partial}{\partial n_p}\left.\frac{\partial n_p}{\partial x}\right|_n
      =-n\frac{\partial}{\partial n_n}+n\frac{\partial}{\partial n_p} ,   \nonumber
\end{eqnarray}
the derivatives of $F_{\alpha i}$ with respect to $n$ and $x$ are obtained as 
\begin{eqnarray}
\frac{\partial F_{\alpha i}}{\partial n} & = & 2\alpha \frac{F_{(\alpha-1) i}}{F_{-1/2i}}
           \left[(1-x)\frac{\partial F_{\alpha i}}{\partial n_n}
                 +x\frac{\partial F_{\alpha i}}{\partial n_p}\right]  
\label{dfadn}  \\
\frac{\partial F_{\alpha i}}{\partial x} & = & 2\alpha \frac{F_{(\alpha-1) i}}{F_{-1/2i}}n
           \left[\frac{\partial F_{\alpha i}}{\partial n_p}
                 -\frac{\partial F_{\alpha i}}{\partial n_n}\right].
\label{dfadx}
\end{eqnarray}
Using Eqs.~ (\ref{dfadni})-(\ref{dfadx}),  we arrive at the following expressions for the density derivatives of the degeneracy parameter and the kinetic energy density:
\begin{eqnarray}
\frac{\partial \psi_i}{\partial n_i} & = & \frac{2}{F_{-1/2i}}\frac{\partial F_{1/2 i}}{\partial n_i}\,, \quad
\frac{\partial \psi_i}{\partial n_j}  =  \frac{2}{F_{-1/2i}}\frac{\partial F_{1/2 i}}{\partial n_j} \\
\frac{\partial \psi_i}{\partial n} & = & \frac{2}{F_{-1/2i}}\frac{\partial F_{1/2 i}}{\partial n}\,, \quad
\frac{\partial \psi_i}{\partial x} =  \frac{2}{F_{-1/2i}}\frac{\partial F_{1/2 i}}{\partial x} 
\ea
\ba
\frac{\partial \tau_i}{\partial n_i} & = & \frac{\tau_i}{n_i}\left[
          \frac{3F_{1/2i}^2}{F_{3/2i}F_{-1/2i}} \right.   \nonumber \\
         && \hspace{10pt} \left. +\frac{5}{2}\frac{n_i}{m_i^*}
                                                          \frac{\partial m_i^*}{\partial n_i}\left(
               1-\frac{9}{5}\frac{F_{1/2i}^2}{F_{3/2i}F_{-1/2i}}\right)\right]   
\label{dtaudni}  \\
\frac{\partial \tau_i}{\partial n_j} & = & \frac{5}{2}\frac{\tau_i}{m_i^*}
                                                 \frac{\partial m_i^*}{\partial n_j}\left(
               1-\frac{9}{5}\frac{F_{1/2i}^2}{F_{3/2i}F_{-1/2i}}\right)  
\label{dtaudnj}  \\
\frac{\partial \tau_i}{\partial n} & = & \tau_i \left[\frac{5}{2}\frac{1}{m_i^*}
                                                 \frac{\partial m_i^*}{\partial n}\right. \nonumber \\
&& \hspace{10pt} \left. +\frac{3F_{1/2i}}{F_{3/2i}F_{-1/2i}}\left((1-x)\frac{\partial F_{1/2 i}}{\partial n_n}
                                                    +x\frac{\partial F_{1/2 i}}{\partial n_p}\right)\right] 
\nonumber \\ 
\label{dtaudn}  \\
\frac{\partial \tau_i}{\partial x} & = & \tau_i \left[\frac{5}{2}\frac{1}{m_i^*}
                                          \frac{\partial m_i^*}{\partial x} \right. \nonumber \\
&& \hspace{10pt} \left. +\frac{3F_{1/2i}}{F_{3/2i}F_{-1/2i}}n\left(\frac{\partial F_{1/2 i}}{\partial n_p}
                                                    -\frac{\partial F_{1/2 i}}{\partial n_n}\right)\right] .
\label{dtaudx}
\end{eqnarray}
These relations will be used in subsequent discussions of the finite-temperature properties.
For a rapid evaluation of the FD integrals, 
two numerical techniques that give accurate results for varying degrees of degeneracy are described in 
Appendix D. 

\subsection{Thermal effects}

To infer the effects of finite temperature we focus on the thermal part of the various state variables; 
that is, the difference between the $T=0$ and the finite-$T$ expressions for a given thermodynamic 
function $X$:
\begin{equation}
X_{th} = X(n,x,T) - X(n,x,0)
\end{equation}
This subtraction scheme discards terms that do not depend on the kinetic energy density. 
The thermal energy is given by
\begin{eqnarray}
E_{th} & = & E(T) - E(0)   \nonumber  \\
               & = & \frac{1}{n}\sum_i\left[\frac{\hbar^2}{2m_i^*}\tau_i
                              -\frac{3}{5}\mathcal{T}_{Fi}n_i\right]
\label{eath}
\end{eqnarray}
where
\be
\mathcal{T}_{Fi} = \frac{\hbar^2k_{Fi}^2}{2m_i^*}.
\ee
The thermal pressure takes the form
\begin{eqnarray}
P_{th} & = & P(T)-P(0)  \nonumber  \\
       & = & \frac{2}{3}\sum_i Q_i\left[\frac{\hbar^2}{2m_i^*}\tau_i
                        -\frac{3}{5}\mathcal{T}_{Fi}n_i\right] , \label{pth} \\
\mbox{where} ~~~ Q_i & = & 1-\frac{3}{2}\frac{n}{m_i^*}  
\frac{\partial m_i^*}{\partial n} . \label{qi}
\end{eqnarray}
The quantities $Q_i$ are the consequence of the momentum-dependent interactions in the Hamiltonian which 
lead to the Landau effective mass. For a free gas, $Q_i = 1$ and $P_{th}=2E_{th}/3$ 
as usual. The entropy per particle can be written as
\begin{eqnarray}
S & = & \frac{1}{nT}\sum_i\left[\frac{5}{3}\frac{\hbar^2}{2m_i^*}\tau_i
                        +n_i(V_i-\mu_i)\right]  \nonumber \\
            & = & \frac{1}{n}\sum_i n_i\left[\frac{5}{3}\frac{F_{3/2i}}{F_{1/2i}}-\mbox{ln}z_i\right].
\label{entr}
\end{eqnarray}
The thermal free energy density can be expressed as
\begin{eqnarray}
\mathcal{F}_{th} & = & \mathcal{F}(T)-\mathcal{H}(0)  
                  =  \mathcal{H}(T)-nTS  -\mathcal{H}(0)  \nonumber \\
                 & = & \sum_i\left[\frac{\hbar^2}{2m_i^*}\tau_i-\frac{3}{5}\mathcal{T}_{Fi}n_i
                       -Tn_i\left(\frac{5}{3}\frac{F_{3/2i}}{F_{1/2i}}-\mbox{ln}z_i\right)\right]
\nonumber \\
\label{fden}
\end{eqnarray}
in terms of which the thermal contribution to the chemical potentials are
\begin{eqnarray}
\mu_{ith} & = & \mu_i(T)-\mu_i(0) = \left.\frac{\partial\mathcal{F}_{th}}{\partial n_i}\right|_{n_j} . 
\label{muth}  \\
\mbox{where} ~~~ \mu_i(T) &= &T\psi_i+V_i  
\end{eqnarray}
The total free energy
\be
F = \sum_i\left[\frac{\hbar^2}{2m_i^*}\frac{\tau_i}{n}
         -TY_i\left(\frac{5F_{3/2i}}{3F_{1/2i}}-\psi_i\right)\right]+F_d
\ee
can be expressed, with the aid of
\be
\tau_i = \frac{2m_i^*T}{\hbar^2}\frac{F_{3/2i}}{F_{1/2i}}n_i,
\ee
as 
\be 
F = \sum_i\left[TY_i\left(-\frac{2F_{3/2i}}{3F_{1/2i}}+\psi_i\right)\right]+F_d \label{totEfree}
\ee
The second derivative of the above with respect to the proton fraction $x$ evaluated at $x=1/2$ yields the 
symmetry energy at finite temperature:
\ba
S_2(T) &=& \frac{1}{8}\left.\frac{d^2F}{dx^2}\right|_{x=1/2}  \\
       &=&  -\frac{T}{3}\frac{F_{3/2}}{F_{1/2}^2}\left[\frac{dF_{1/2}}{dx} \right.\nonumber \\
&+&\left.\left(\frac{1}{2F_{1/2}}-\frac{3F_{1/2}}{4F_{3/2}F_{-1/2}}\right)\left(\frac{dF_{1/2}}{dx}\right)^2
                                             -\frac{1}{4}\frac{d^2F_{1/2}}{dx^2}\right]  \nonumber \\
       &+& S_{2d}
\label{s2t} 
\ea
where
\ba
F_{\alpha} &\equiv& F_{\alpha i}(x=0.5) \nonumber \\
\frac{dF_{1/2}}{dx} &\equiv& \left.\frac{dF_{1/2n}}{dx}\right|_{x=1/2} 
                          =\left. -\frac{dF_{1/2p}}{dx}\right|_{x=1/2}   \nonumber \\
               &=& -2F_{1/2}\left(1+\frac{3}{4m^*}\frac{dm^*}{dx}\right)  \\                     
\frac{d^2F_{1/2}}{dx^2} &\equiv& \left.\frac{d^2F_{1/2n}}{dx^2}\right|_{x=1/2} 
                          =\left. \frac{d^2F_{1/2p}}{dx^2}\right|_{x=1/2}   \nonumber \\
                &=&   \frac{6F_{1/2}}{m^*}\frac{dm^*}{dx}
               \left(1+\frac{1}{8m^*}\frac{dm^*}{dx}\right) \\ 
m^* &\equiv& m_n^*(x=1/2) = m_p^*(x=1/2)   \label{msym}\\
\frac{dm^*}{dx} &\equiv& \left.\frac{dm_n^*}{dx}\right|_{x=1/2} 
                          =\left. -\frac{dm_p^*}{dx}\right|_{x=1/2} \label{dmsym} 
\ea
Note that 
\be
\frac{d^2m^*}{dx^2} = \frac{2}{m^*}\frac{dm^*}{dx}.
\ee
Thus the thermal contributions to the symmetry energy are 
\be
S_{2,th} = S_2(T) - S_2(0) 
\label{s2th}
\ee
For the calculation of the 
specific heat at constant volume, we begin by writing the energy per particle as
\[
{E} = \frac{1}{n}\sum_i \frac{\hbar^2}{2m_i^*}\tau_i + n\mbox{-dependent terms}
\]
Then
\begin{eqnarray}
C_V  =  \left.\frac{\partial E}{\partial T}\right|_n   
     =  \frac{1}{n}\sum_i \frac{\hbar^2}{2m_i^*}\left.\frac{\partial\tau_i}{\partial T}\right|_{n_i}
                 \nonumber
\end{eqnarray}
The condition that $n_i$ are constant implies
\begin{eqnarray}
\frac{dn_i}{dT}  = 0  & = & \left.\frac{\partial n_i}{\partial T}\right|_{F_{1/2i}}
                          +\left.\frac{\partial n_i}{\partial F_{1/2i}}\right|_T
                                    \left.\frac{\partial F_{1/2i}}{\partial T}\right|_{n_i} \nonumber \\
\Rightarrow \left.\frac{\partial n_i}{\partial T}\right|_{F_{1/2i}} & = & 
      -\left.\frac{\partial n_i}{\partial F_{1/2i}}\right|_T
                                    \left.\frac{\partial F_{1/2i}}{\partial T}\right|_{n_i}
\end{eqnarray}
But 
\begin{eqnarray}
\left.\frac{\partial F_{1/2i}}{\partial T}\right|_{n_i}  = 
     \left.\frac{\partial \psi_i}{\partial T}\right|_{n_i}\frac{\partial F_{1/2i}}{\partial \psi_i}
  =  \frac{1}{2}F_{-1/2i}\left.\frac{\partial \psi_i}{\partial T}\right|_{n_i}
\end{eqnarray}
where Eq. (\ref{dfadpsi}) was used in obtaining the second equality. 
Solving for $\left.\frac{\partial \psi_i}{\partial T}\right|_{n_i}$ gives
\[\left.\frac{\partial \psi_i}{\partial T}\right|_{n_i} = 
   -\left.\frac{\partial n_i}{\partial T}\right|_{F_{1/2i}}\left(
            \left.\frac{\partial n_i}{\partial F_{1/2i}}\right|_T\frac{1}{2}F_{-1/2i}\right)^{-1}
\]
Using Eq. (\ref{nT}) for the derivatives of $n_i$ with respect to $T$ and $F_{1/2i}$ we get
\begin{equation}
\left.\frac{\partial \psi_i}{\partial T}\right|_{n_i} = 
   -\frac{3}{T}\frac{F_{1/2i}}{F_{-1/2i}}
\label{dpsidT}
\end{equation}
The $T$-derivative of Eq. (\ref{tauT}) is
\begin{eqnarray}
\left.\frac{\partial \tau_i}{\partial T}\right|_{n_i} & = &
        \tau_i\left(\frac{5}{2T} +\frac{1}{F_{3/2i}}
                                   \left.\frac{\partial F_{3/2i}}{\partial T}\right|_{n_i}\right) \nonumber \\
 & = &  \tau_i\left(\frac{5}{2T} +\frac{1}{F_{3/2i}}
                                   \left.\frac{\partial \psi_i}{\partial T}\right|_{n_i}
                                   \frac{\partial F_{3/2i}}{\partial \psi_i}\right)  \nonumber \\
 & = &  \tau_i\left(\frac{5}{2T} -\frac{9}{2T}
                                   \frac{F_{1/2i}^2}{F_{3/2i}F_{-1/2i}}\right)
\label{dtaudT}
\end{eqnarray}
where equations (\ref{dfadpsi}) and (\ref{dpsidT}) have been exploited for the last line. Thus
\begin{equation}
C_V = \frac{5}{2nT}\sum_i \frac{\hbar^2 \tau_i}{2m_i^*}
      \left(1-\frac{9}{5}\frac{F_{1/2i}^2}{F_{3/2i}F_{-1/2i}}\right)
\label{cv}
\end{equation}
The starting point of the calculation of the specific heat at constant pressure is
\be
C_P = C_V +\frac{T}{n^2}\frac{\left(\left.\frac{\partial P}{\partial T}\right|_n\right)^2}
                                {\left.\frac{\partial P}{\partial n}\right|_T}
\label{cp}
\ee
The temperature derivative of the pressure at fixed density is given by
\ba
\left.\frac{\partial P}{\partial T}\right|_n &=& 
    \frac{2}{3}\sum_i \frac{\hbar^2}{2m_i^*}Q_i\left.\frac{\partial \tau_i}{\partial T}\right|_n \nonumber \\
&=& \frac{5}{3T}\sum_i \frac{\hbar^2}{2m_i^*}Q_i\tau_i\left(1-\frac{9}{5}\frac{F_{1/2i}^2}{F_{3/2i}F_{-1/2i}}
                         \right)  
\ea
where Eq.(\ref{dtaudT}) was used in going from the first line to the second. 
The density derivative of the pressure at fixed temperature is
\ba
\left.\frac{\partial P}{\partial n}\right|_T &=& \frac{\hbar^2}{3}\frac{d}{dn}\left(
                 \sum_i\frac{Q_i\tau_i}{m_i^*}\right) + \frac{dP_d}{dn} \nonumber \\
&=& \frac{\hbar^2}{3}\sum_i\left[\frac{Q_i}{m_i^*}\frac{d\tau_i}{dn}+\frac{\tau_i}{m_i^*}\frac{dQ_i}{dn}
  -\frac{\tau_iQ_i}{m_i^{*2}}\frac{dm_i^*}{dn}\right] \nonumber \\
&+& \frac{dP_d}{dn}  
\ea
The density derivatives of the kinetic energy density are given in Eqs. (\ref{dtaudni})-(\ref{dtaudx}) and those of $m^*$, $Q$, and
$P_d$ in Appendix B.

Finally, the inverse susceptibilities are given by
\be
\chi_{ij,th}  =  \chi_{ij}(T)-\chi_{ij}(0)=\left(\frac{\partial \mu_{ith}}{\partial n_j}\right)^{-1} 
\label{xi} 
\ee
where
\ba
\chi_{ii}(T) & = & \left(\frac{\partial \mu_i}{\partial n_i}\right)^{-1}
                                = \left(T\frac{\partial \psi_i}{\partial n_i}
                                        +\frac{\partial V_i}{\partial n_i}\right)^{-1}  \nonumber \\
 & = & \left[T\left(\frac{\partial F_{1/2i}}{\partial \psi_i}\right)^{-1}
                              \frac{\partial F_{1/2i}}{\partial n_i}
                                        +\frac{\partial V_i}{\partial n_i}\right]^{-1}  \nonumber \\
 & = & \left[\frac{2T}{n_i}\frac{F_{1/2i}}{F_{-1/2i}}\left(1-\frac{3}{2}\frac{n_i}{m_i^*}
                \frac{\partial m_i^*}{\partial n_i}\right)
             +\frac{\partial V_i}{\partial n_i}\right]^{-1} ,  \nonumber \\
\label{xii} 
\ea
\ba
\chi_{ij}(T) = \left[-3T\frac{F_{1/2i}}{F_{-1/2i}}\frac{1}{m_i^*}
                \frac{\partial m_i^*}{\partial n_j}
             +\frac{\partial V_i}{\partial n_i}\right]^{-1} ;~ i\ne j.
\nonumber \\
\label{xij}
\ea

\subsection*{Results}

We now present numerical results.
Comparisons of these results with analytical results in degenerate and non-degenerate situations will be presented in the next sub-section. 

\begin{figure*}[!ht]
\centering
\begin{minipage}[b]{0.49\linewidth}
\centering
\includegraphics[width=10cm]{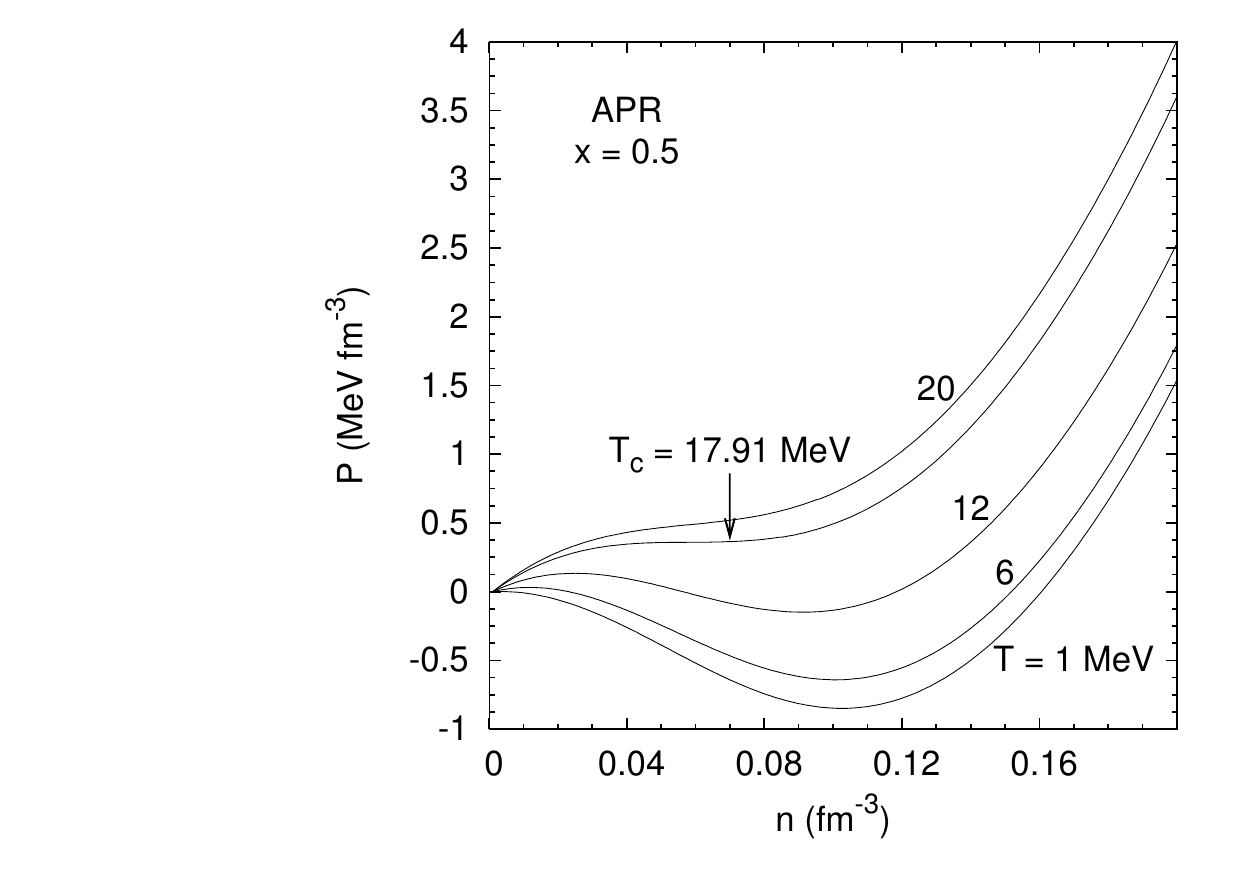}
\end{minipage}
\begin{minipage}[b]{0.49\linewidth}
\centering
\includegraphics[width=10cm]{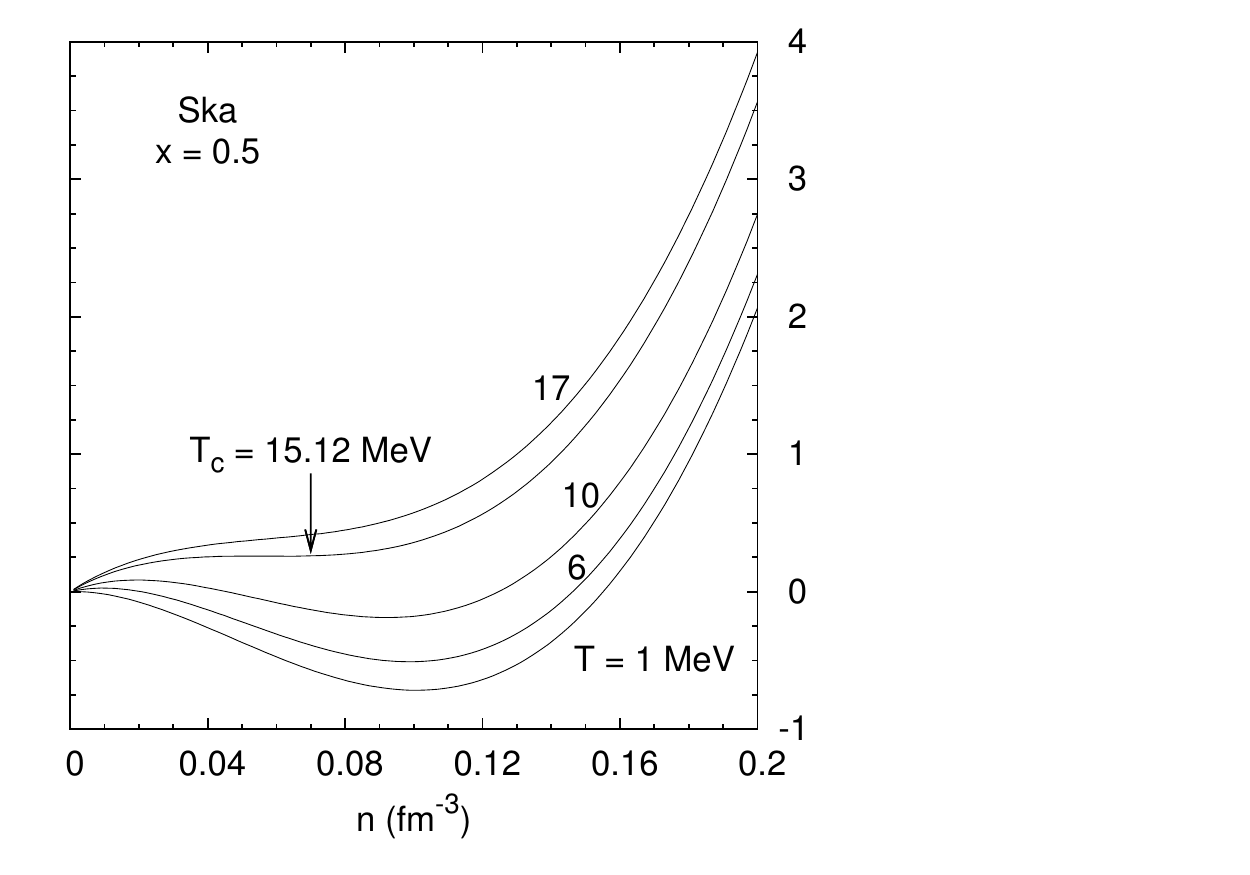}
\end{minipage}
\vskip -0.5cm
\caption{Pressure of isospin symmetric matter vs baryon density (from Eq. (\ref{pth})) for the APR (left) and Ska (right) models at the indicated temperatures. The point $(P,n)$ on the critical temperature curve of each 
model at which $dP/dn=d^2P/dn^2=0$ is indicated by the downward arrow.} 
\label{APRSKA_IsoTherm}
\end{figure*}

We begin by examining results of the total pressure (from Eq. (\ref{pth}))  as it varies with temperature and density in the sub-nuclear regime for isospin symmetric matter ($x=0.5$). Our results for the APR and Ska models are shown in Fig. \ref{APRSKA_IsoTherm}. The prominent feature in this figure is the onset of a liquid-gas phase transition, the critical temperature and density for which are obtained by the condition
\be
\left.\frac {dP}{dn}\right|_{n_c,T_c} = \left.\frac {d^2P}{dn^2}\right|_{n_c,T_c} = 0 \,. 
\ee
The critical temperatures (densities) for the APR and Ska models were found to be 17.91 MeV
(0.057 fm$^{-3}$) and 15.12 MeV (0.056 fm$^{-3}$), respectively, so that
\be
\frac {P_c}{n_cT_c} = 
\left\{
\begin{array}{ll} 
0.347 \,, & \qquad \mbox{for APR} \\
0.303 \,, & \qquad \mbox{for Ska}\,. 
\end{array} 
\right. 
\ee
These results provide an interesting contrast  with the value 0.375 for 
a Van der Waals-like equation of state  and the experimental values that lie in the range 0.27-0.31 for noble gases (see, e.g. Ref.~\cite{stanley}, p.69).

\begin{figure*}[!ht]
\centering
\begin{minipage}[b]{0.49\linewidth}
\centering
\includegraphics[width=10.75cm]{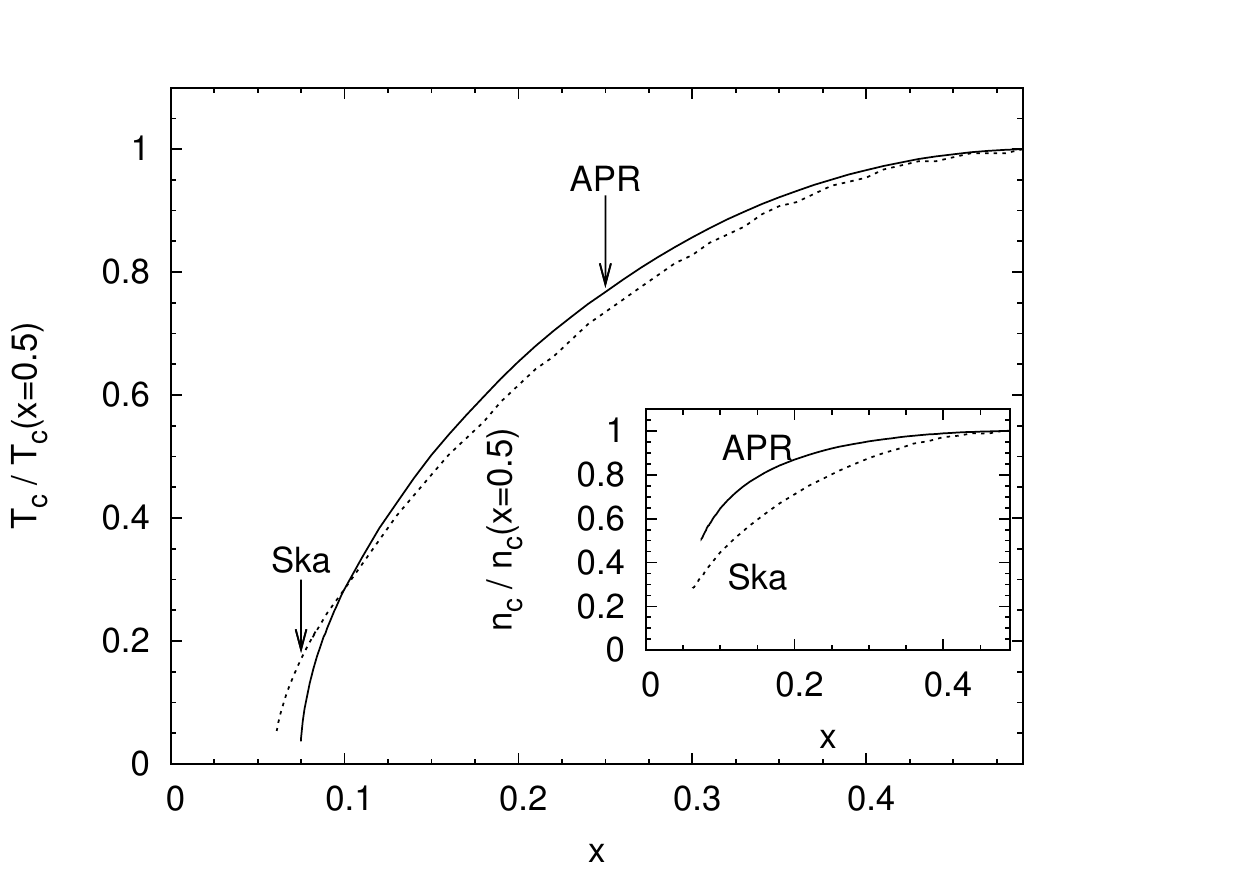}
\end{minipage}
\begin{minipage}[b]{0.49\linewidth}
\centering
\includegraphics[width=10cm]{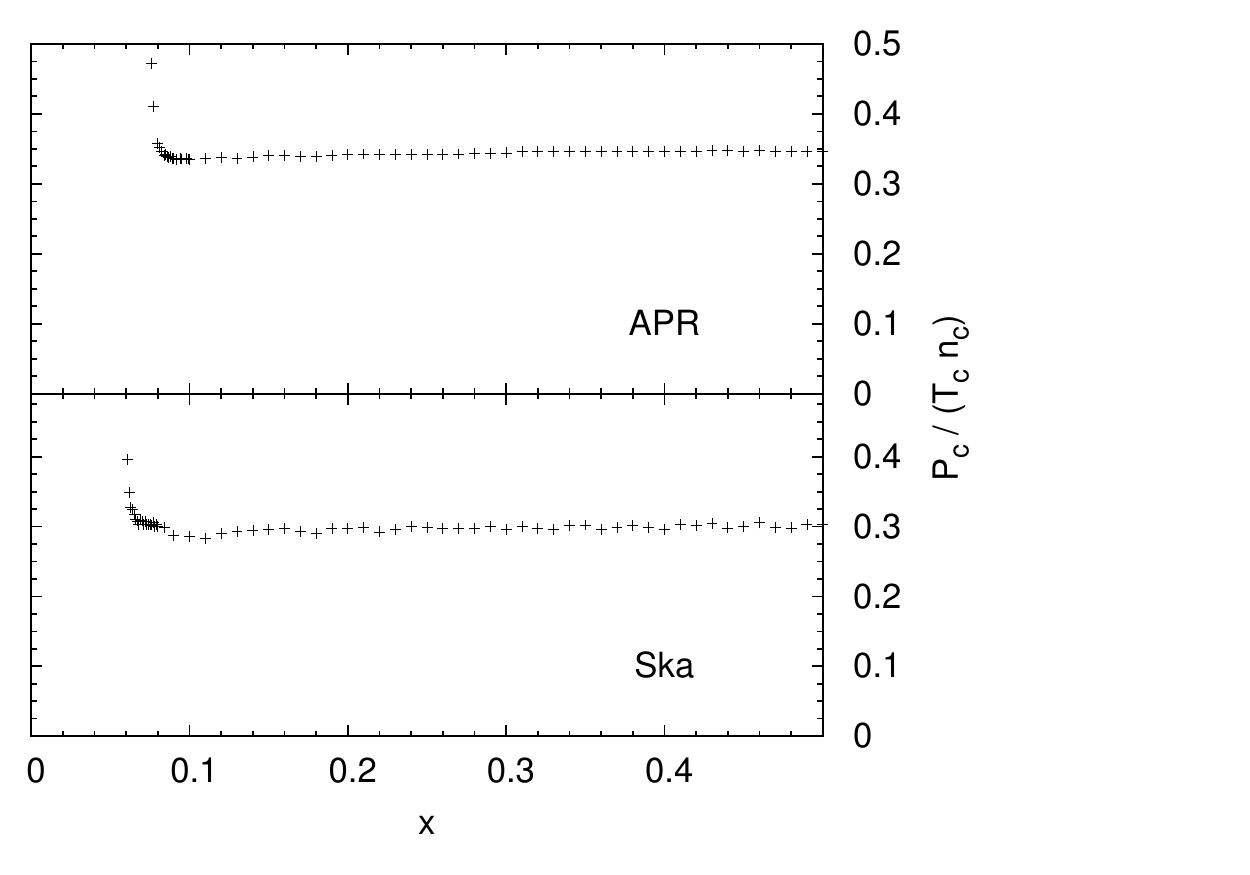}
\end{minipage}
\vskip -0.5cm
\caption{Left panel: Critical temperatures (scaled with their respective values for proton fraction $x=0.5$) vs $x$. The inset shows critical densities (scaled with their respective values for $x=0.5$)  vs $x$. Right panel: Critical parameter $P_c/n_cT_c$ vs $x$.} 
\label{APRSKA_critical}
\end{figure*}

In Fig. \ref{APRSKA_critical}, we show how the critical temperatures and densities vary as a function of proton fraction $Y_p$ in the left panel.  Both quantities are scaled to their respective values for symmetric nuclear matter ($Y_p=0.5$). The fall-off of the critical temperature with $Y_p$ is similar for the APR and Ska models, whereas the fall-off of the critical density with $Y_p$ for the Ska model is  steeper than for the APR model. The critical proton fractions beyond which the phase transition disappears are similar for both models, that for the APR model being slightly larger than for the SkA model. 
As is evident from the  right panel in this figure, $P_c/n_cT_c$ exhibits very little variation with  $Y_p$.   

The thermal properties are dominated by the behavior of the effective masses.  For all densities, at a given value of $x$, the APR effective masses are larger than for Ska.  As a result, thermal contributions to entropy, energy, pressure, free energy, etc. are larger in the case of APR at the same density.  This explains the relative behaviors in Figures 10, 11, 12 and 14. The reverse behavior is seen in the thermal part of chemical potentials in Figure 13. This behavior can be understood through the limiting cases 
(\ref{mudeg}) and (\ref{mund}) where the effective masses enter with an overall negative sign. 

\begin{figure}[!h]
\includegraphics[width=9cm]{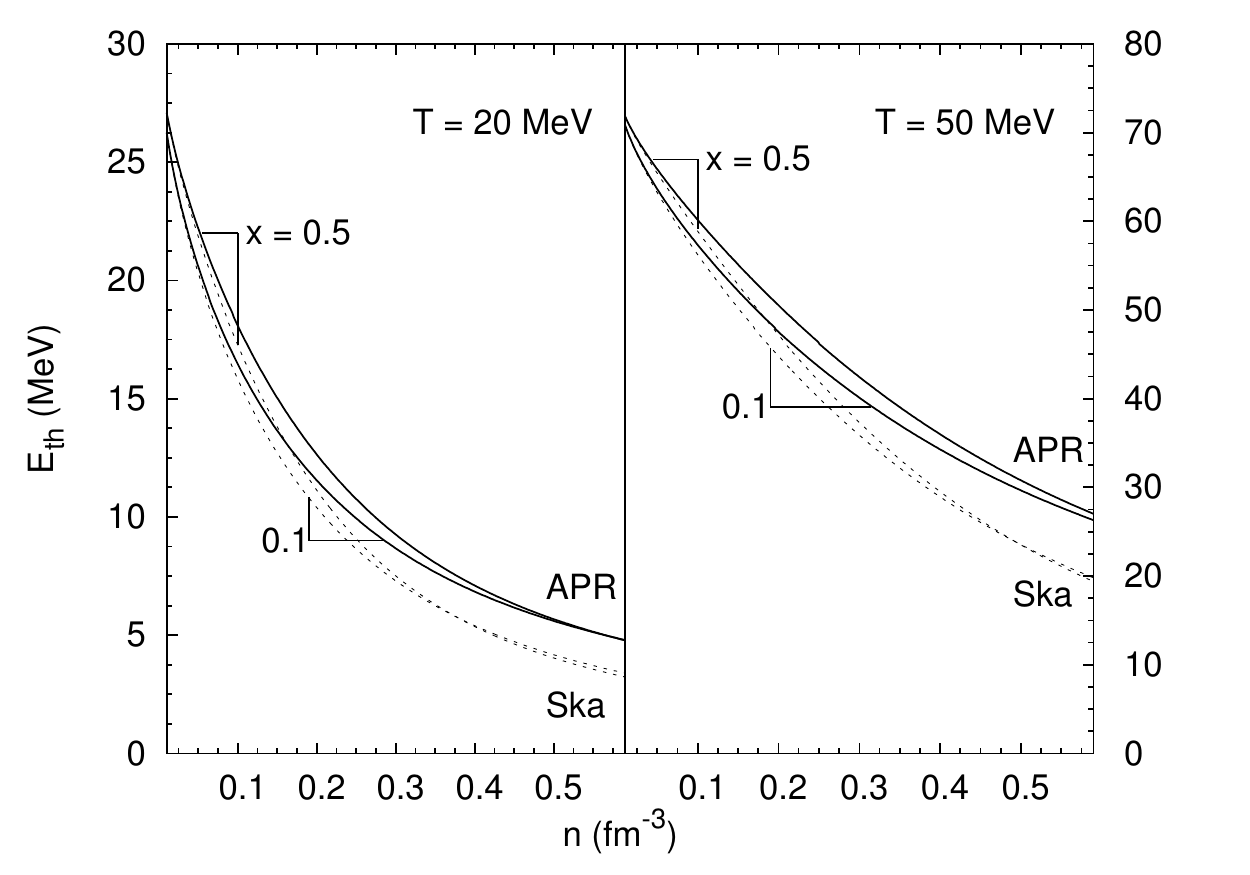}
\caption{Thermal energy per particle (Eq. (\ref{eath})) at the indicated proton fractions and temperatures for the APR and Ska models.}
\label{APRSKA_Eth}
\end{figure}

The thermal energy (from Eq. (\ref{eath})) is shown in Fig. \ref{APRSKA_Eth} for the two models at proton fractions $x$ of 0.5 and 0.1, and at temperatures $T$ of 20 and 50 MeV, respectively, for the two models.  Common to both models are the features that the thermal energy (i) decreases, and (ii)  is nearly independent of the proton fraction with increasing density. Maximal differences (with respect to $x$) are seen to be in the vicinity of $n_0=0.16~{\rm fm}^{-3}$ for both models.  Differences between the two models increase with increasing density, particularly for densities in excess of  $n_0$.  These common and different features arise due to  a combination of effects involving the dependence of the thermal energy on the effective masses as the degree of degeneracy changes with density as will be discussed in the next sub-section with analytical results in hand. 

\begin{figure}[!h]
\includegraphics[width=9cm]{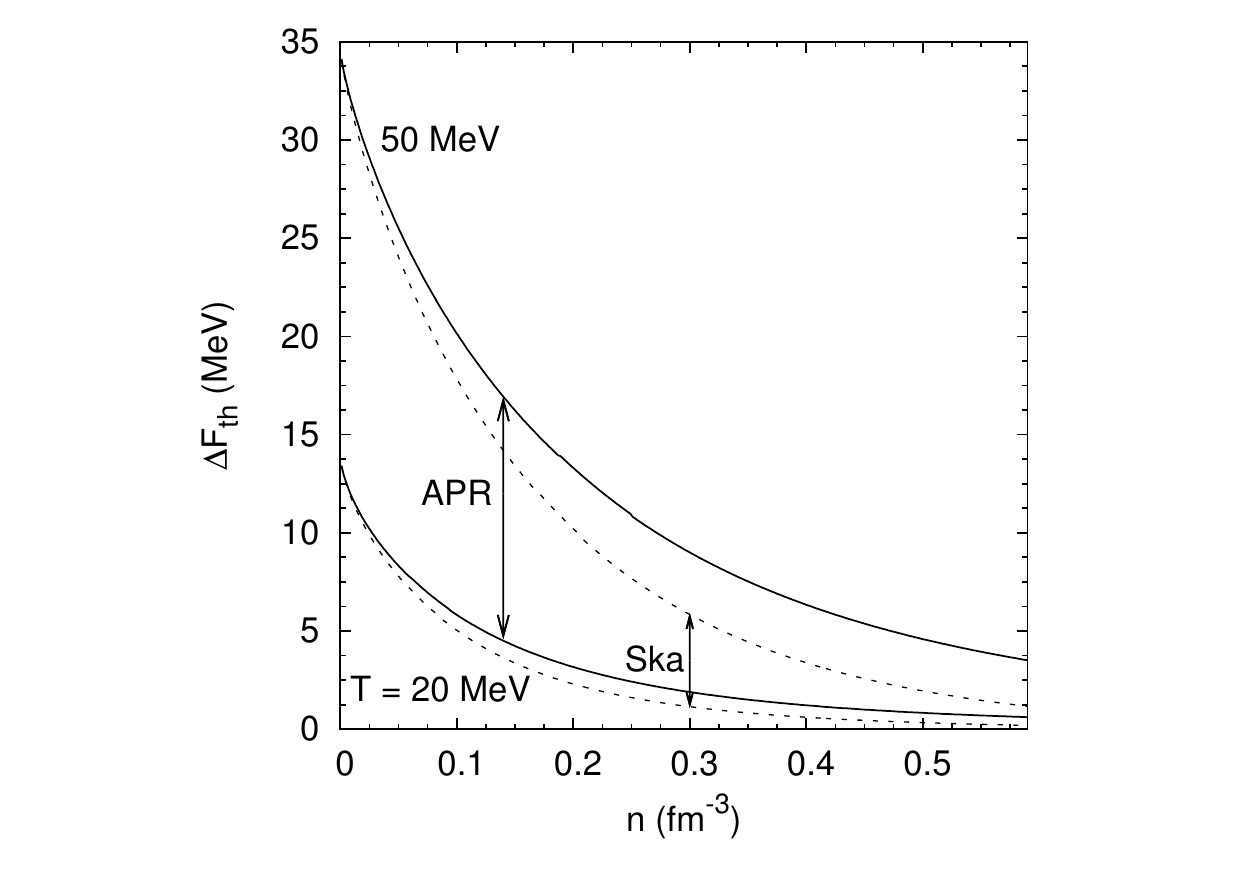}
\caption{Difference between the pure neutron matter and nuclear matter free energies (Eq. (\ref{fden})) 
at the indicated temperatures for the APR and Ska models.  }
\label{APRSka_Efree_Th}
\end{figure}

In Fig. \ref{APRSka_Efree_Th}, the difference between the pure neutron matter and nuclear matter free energies $\Delta F_{th} = F(n,T,x=0) - F(n,T,x=0.5)$ is shown for the two models at $T=20$ and 50 MeV, respectively. For both temperatures shown, the APR model has a larger $\Delta F_{th}$ than that of the Ska model. This feature can be understood in terms of the larger thermal energies of the APR model relative to those of the Ska model at the same density and temperature which dominate over  the 
opposing effects of entropy.

\begin{figure}[!h]
\includegraphics[width=9cm]{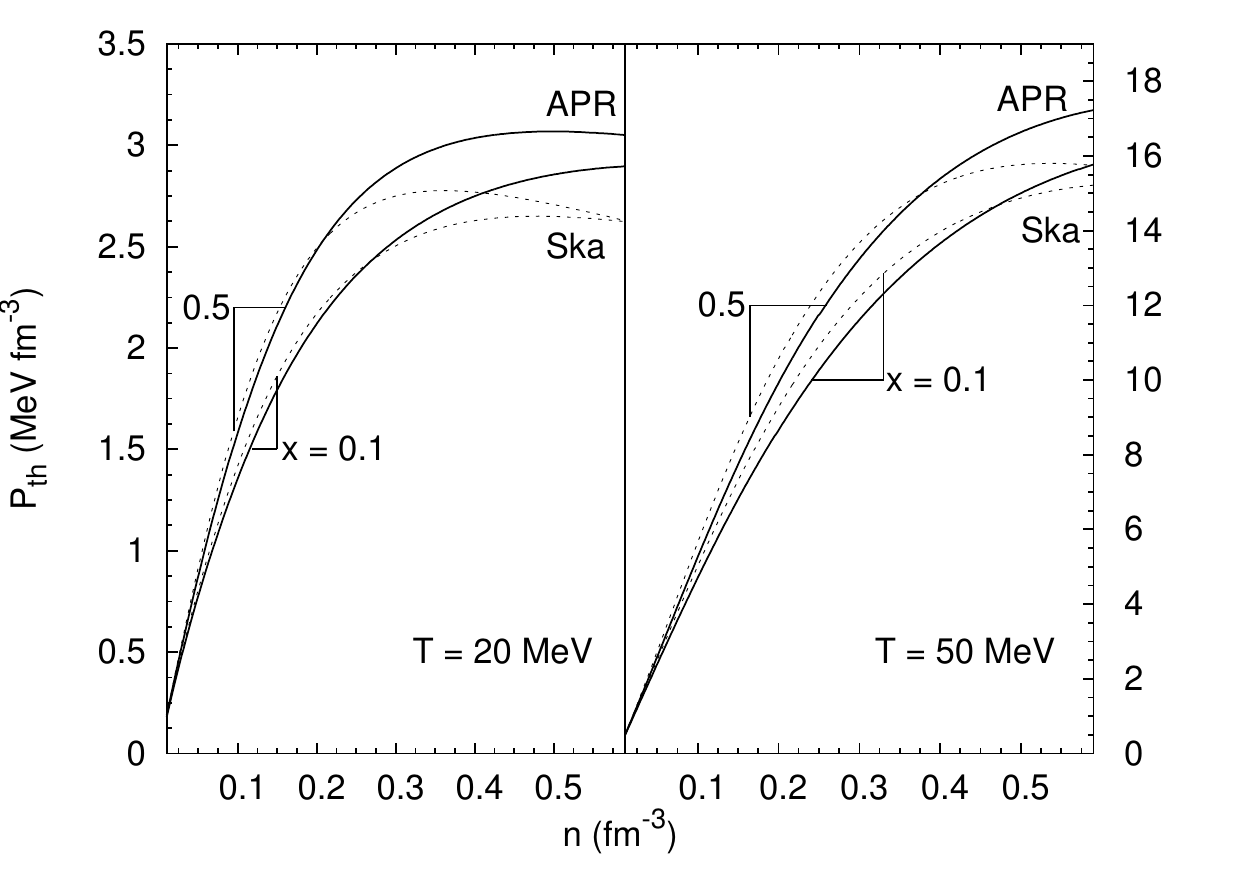}
\caption{Thermal pressure vs baryon density (Eq. (\ref{pth})) at the indicated proton fractions and temperatures.}
\label{APRSKA_Pth}
\end{figure}

The thermal pressures (from Eq. (\ref{pth})) for the two models are shown in Fig. \ref{APRSKA_Pth} for $x=0.5$ and 0.1, and $T=20$ and 50 MeV as functions of density. Both models display the same trend of rising almost linearly  with density until around 1.5 $n_0$ before beginning to saturate at higher densities.  This trend is independent of proton fraction and temperature,
however; the stiffness in pressure is more pronounced for the higher temperature and lower proton fraction. The agreement between the results of the two models becomes progressively worse as the density increases. As with the thermal energy in Fig.  
 \ref{APRSKA_Eth}, these results are a consequence of the increasing degeneracy with increasing density and the behavior of the effective masses in the two models as our discussion in the next sub-section will reveal.   

\begin{figure*}[!ht]
\centering
\begin{minipage}[b]{0.49\linewidth}
\centering
\includegraphics[width=9.5cm]{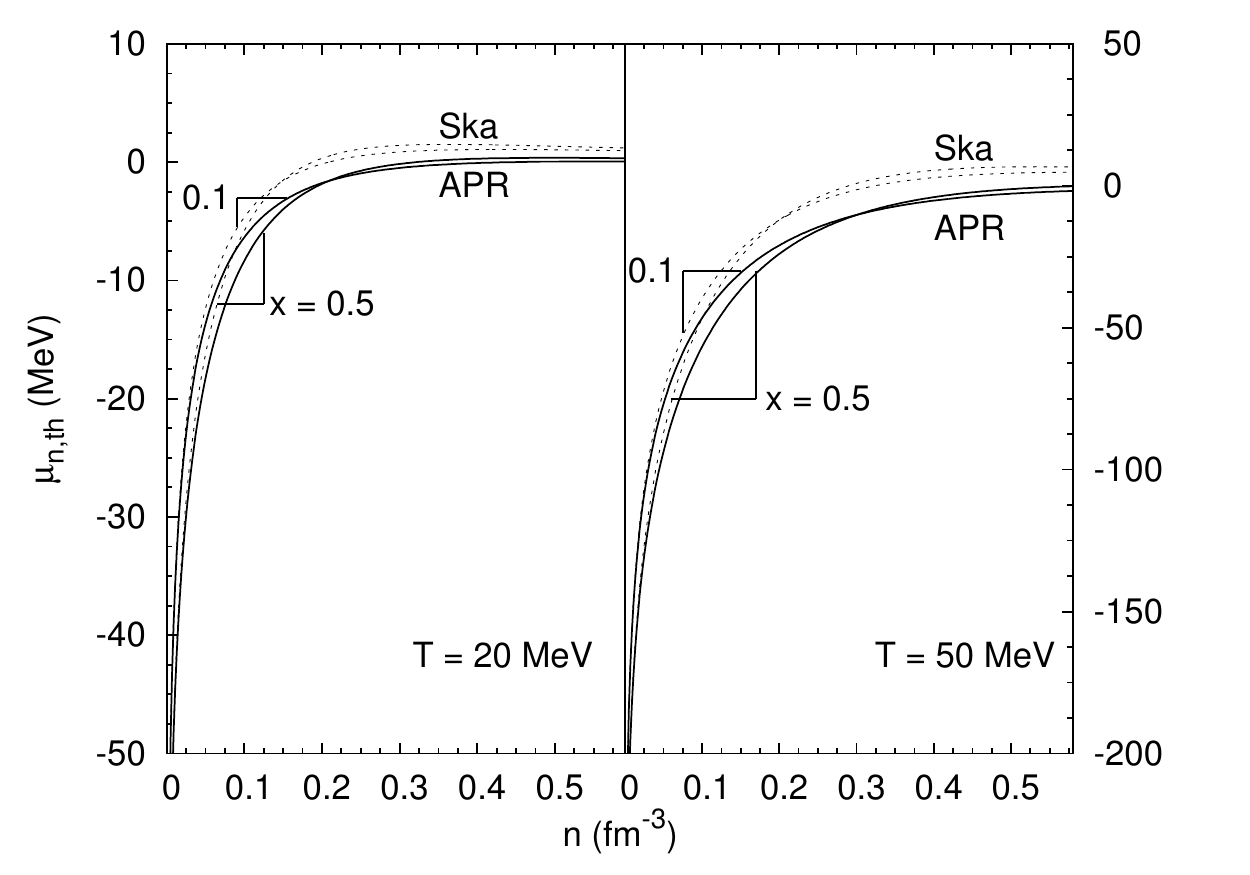}
\end{minipage}
\begin{minipage}[b]{0.49\linewidth}
\centering
\includegraphics[width=9.5cm]{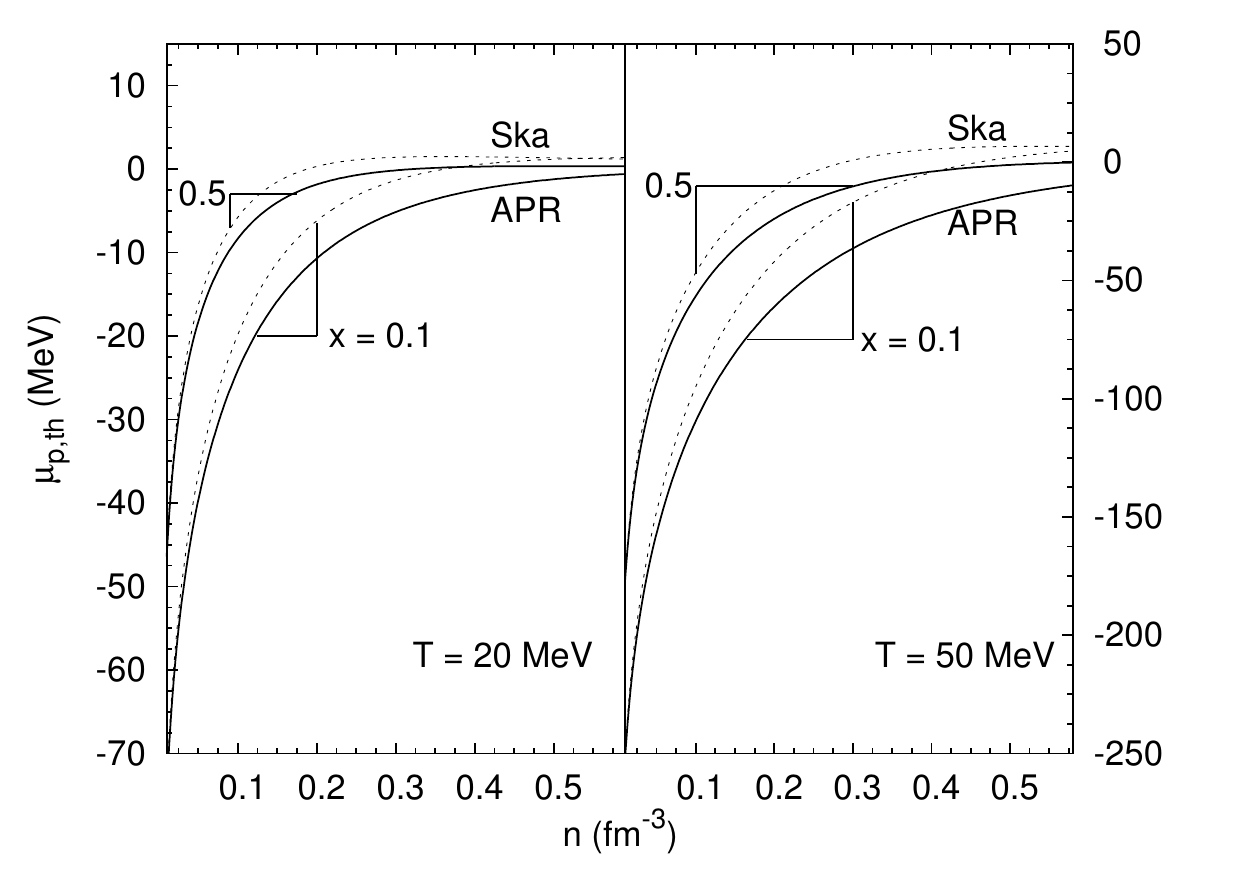}
\end{minipage}
\vskip -0.5cm
\caption{Thermal neutron (left) and proton (right) chemical potentials vs baryon density (Eq. (\ref{muth})). } 
\label{APRSKA_MuTH}
\end{figure*}

The neutron and proton thermal chemical potentials (from Eq. (\ref{muth})) plotted as functions of baryon density are presented in the left and right panels 
of Fig. \ref{APRSKA_MuTH}, respectively. 
Chemical potentials of fermions inclusive of their zero temperature parts decrease with temperature at a fixed density, hence the negative values of their thermal counterparts. 
We observe larger neutron and proton thermal chemical potentials from the Ska model when
compared with the APR model for all but the lowest baryon densities and at both temperatures. The difference between the two models is greatest at intermidiate
densities (between $n_0$ and $2n_0$) and at high temperatures. In the case of the neutron thermal chemical potential there is little difference
between isospin symmetric ($x=0.5$) and neutron rich matter ($x=0.1$). This is not the case for the proton chemical potential which displays a much greater difference as isospin asymmetry increases. 

\begin{figure}[!h]
\includegraphics[width=9cm]{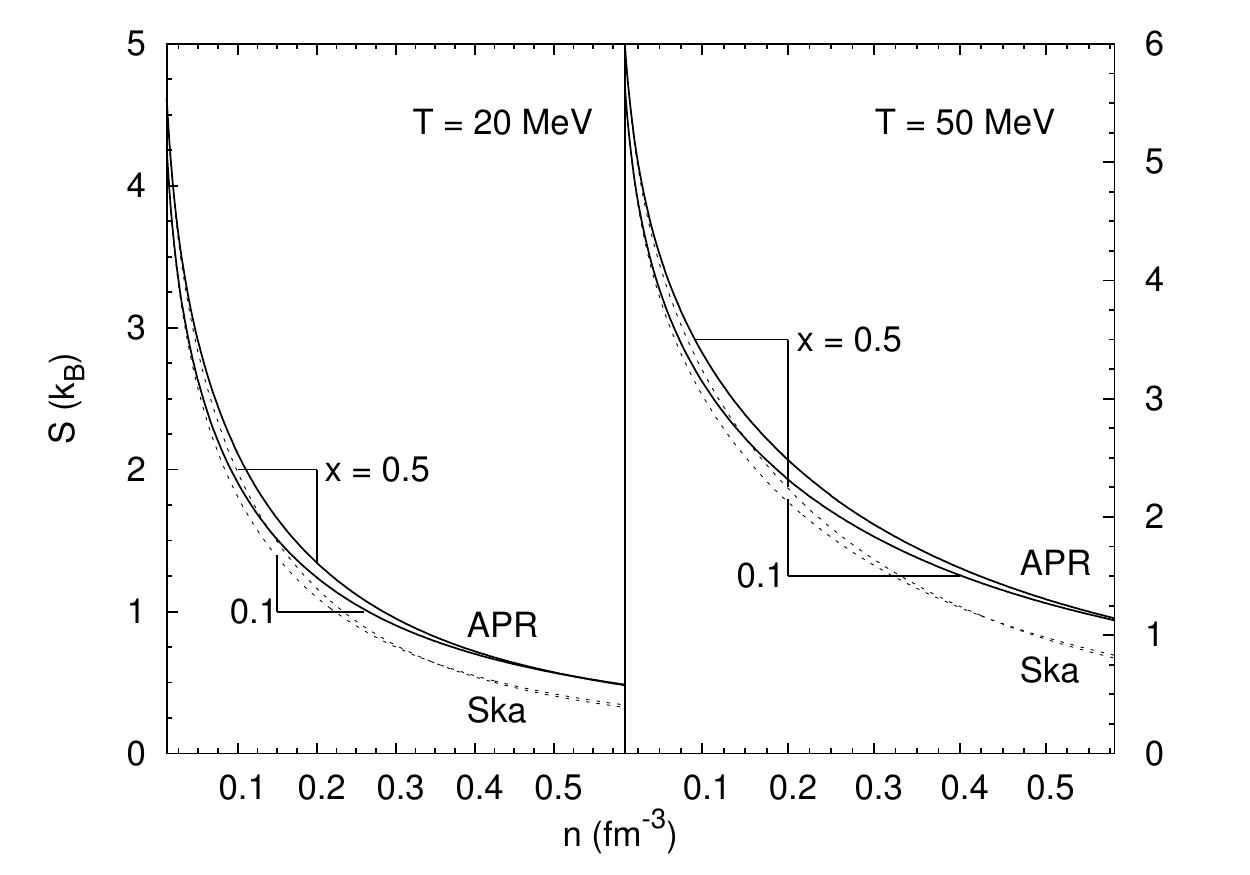}
\caption{Entropy per baryon in units of $k_B$ vs baryon density (Eq. (\ref{entr})).}
\label{APRSKA_SoA}
\end{figure}

In Fig. \ref{APRSKA_SoA}, we present our results for the entropy per baryon for the APR and Ska models. Our results show that
the APR model provides a larger entropy per baryon than the Ska model for all baryon densities, proton fractions and temperatures. The magnitude
of the observed difference is independent of proton fraction $x$ and increases with baryon density $n$ and temperature $T$. For extremely low densities, 
($n \ll n_0$) the difference in entropy per baryon between the models is negligible as interactions play a minor role in a nearly ideal gas for this quantity. 

\begin{figure}[!h]
\includegraphics[width=9cm]{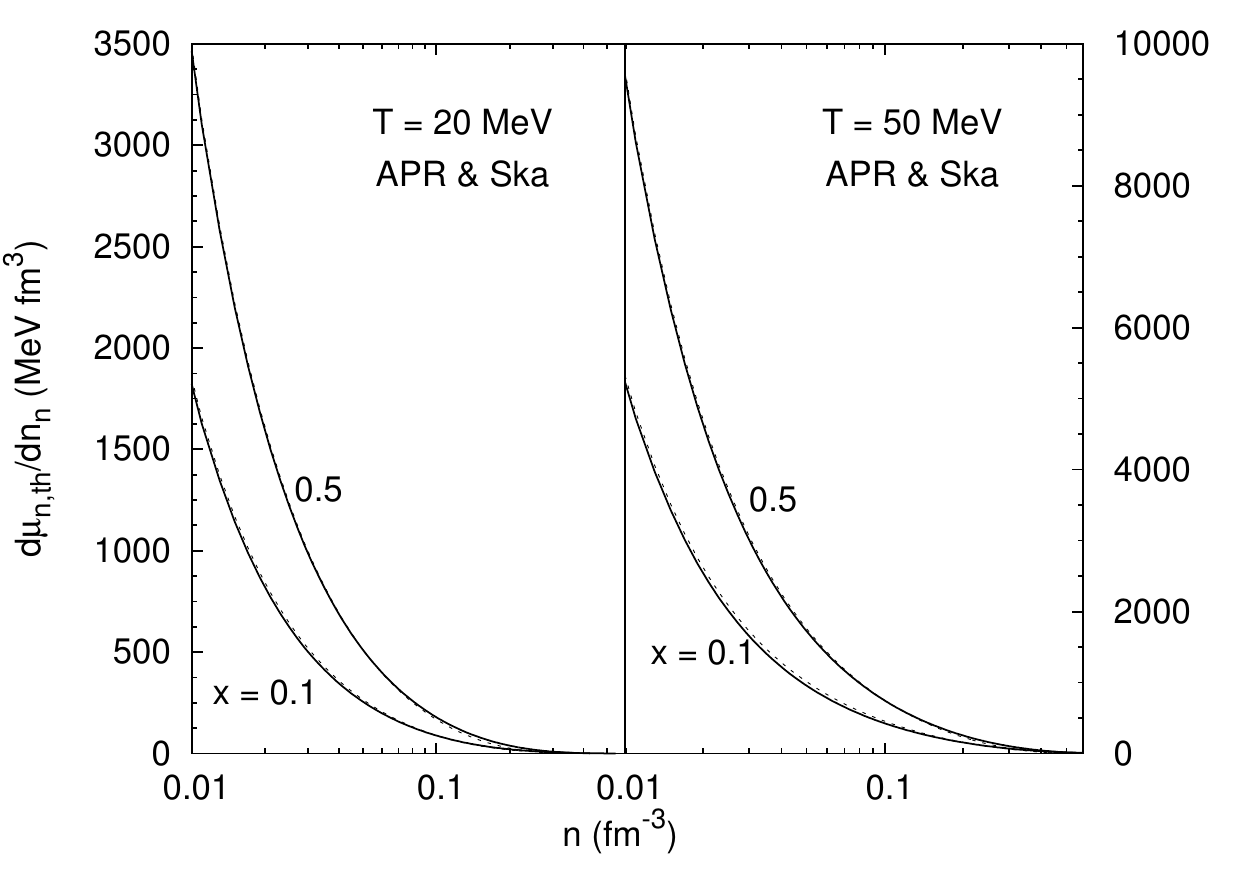}
\caption{Neutron-neutron inverse susceptibility vs baryon density (Eqs. (\ref{xi})-(\ref{xii})) for the APR and Ska models at the indicated proton fractions $x$. The two models are visually indistinguishable
at both temperatures and proton fractions.}
\label{APRSKA_Xnn}
\end{figure}

\begin{figure}[!h]
\includegraphics[width=9cm]{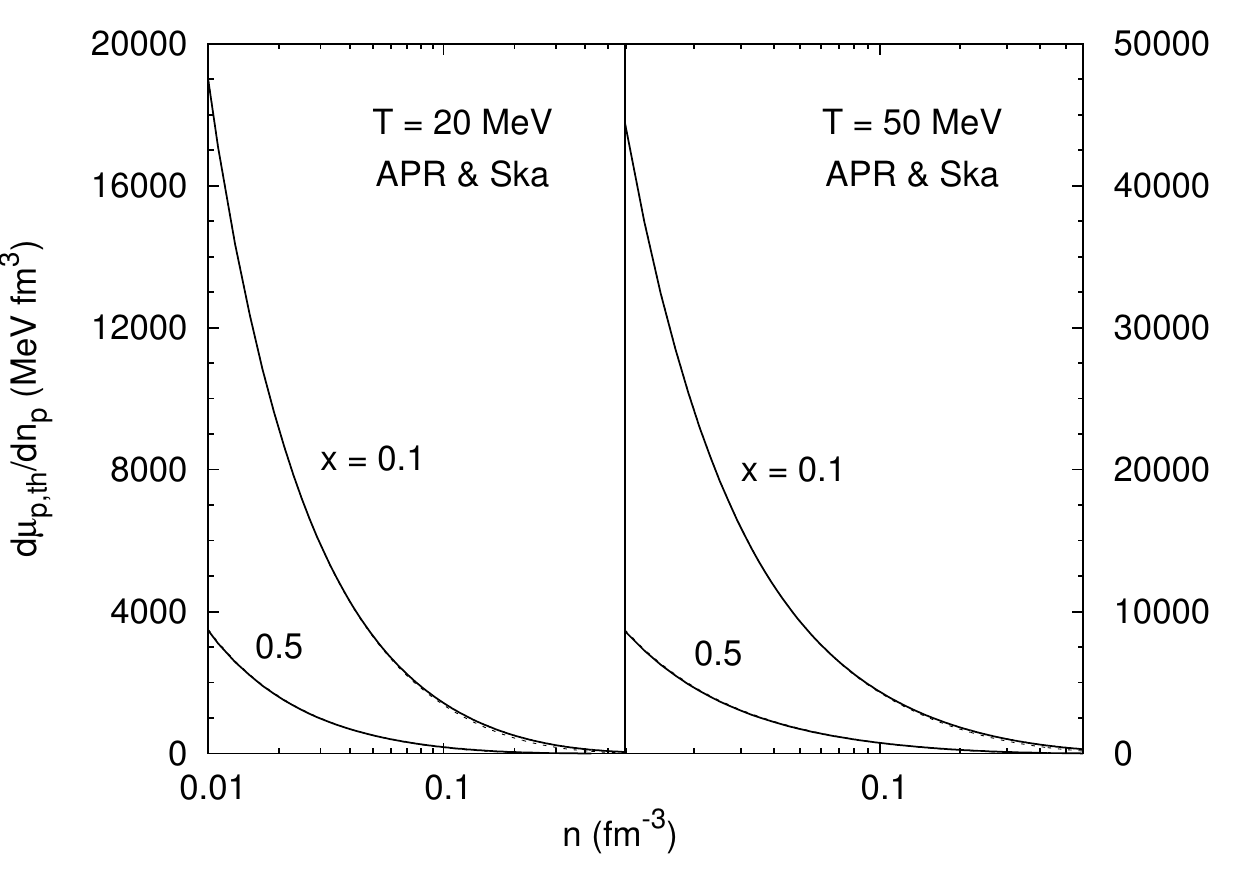}
\caption{Proton-proton inverse susceptibility vs baryon density (Eqs. (\ref{xi})-(\ref{xii})). Just as in 
the case of $\chi_{nn}^{-1}$, the two models are indistinguishable at both temperatures and proton fractions.}
\label{APRSKA_Xpp}
\end{figure}

\begin{figure}[!h]
\includegraphics[width=9cm]{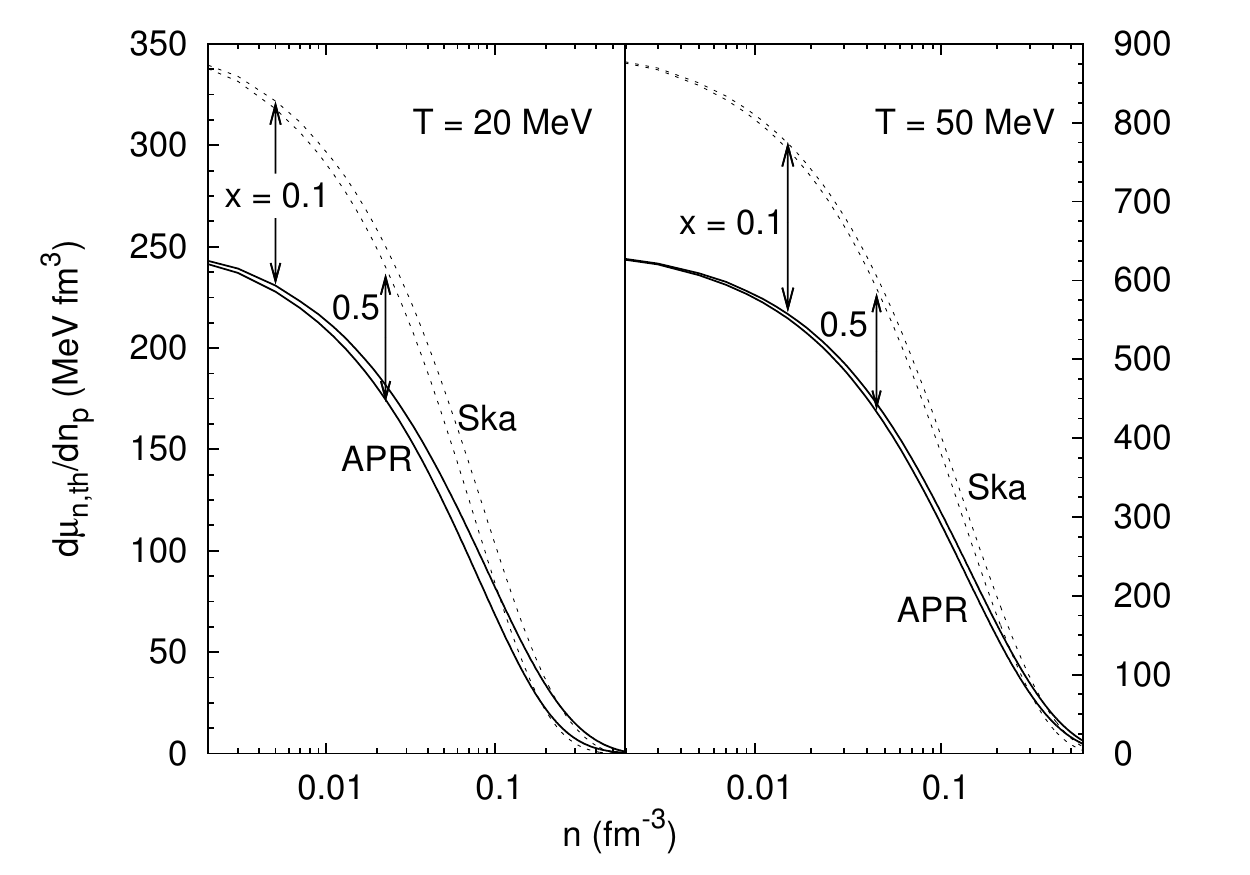}
\caption{Neutron-proton susceptibility vs baryon density (Eqs. (\ref{xi}) and (\ref{xij})). Because 
$d\mu_n/dn_p = d\mu_p/dn_n$, only one of the cross derivatives is shown. Unlike 
$\chi_{nn}^{-1}$ and $\chi_{pp}^{-1}$, $\chi_{np}^{-1}$ exhibits strong model dependence at low densities.}
\label{APRSKA_Xnp}
\end{figure}

In Figs.  \ref{APRSKA_Xnn} through \ref{APRSKA_Xnp} we present results from Eqs. (\ref{xi})-(\ref{xii})
 of the thermal inverse susceptibilities for the APR and Ska models. The neutron-neutron and proton-proton thermal inverse susceptibilities (Figs. \ref{APRSKA_Xnn} and \ref{APRSKA_Xpp}, respectively) show no significant
difference between the two models at all baryon densities, proton fractions and temperatures. The neutron-proton thermal inverse susceptibility (Fig. \ref{APRSKA_Xnp}) shows a significant difference between the two models at densities less than $n_0$. The magnitude of this discrepency is independent of proton fraction and only mildly dependent on temperature. This difference can be attributed to the effective masses as it is explicitly shown in Eqs. (\ref{xiind}) and (\ref{xijnd}) (the non-degenerate limit is appropriate for small densities). The leading terms in $\chi_{ii}$ go as $T/n_i$ thus APR and Ska are similar because the effective mass enters only as a correction. On the other hand, $\chi_{ij}$ differ significantly since their leading 
terms are proportional to $(T/m_i^*)~(dm_i^*/dn_j)$ and thererefore their behavior is primarily influenced by the effective mass. 

\begin{figure*}[!ht]
\centering
\begin{minipage}[b]{0.49\linewidth}
\centering
\includegraphics[width=9.5cm]{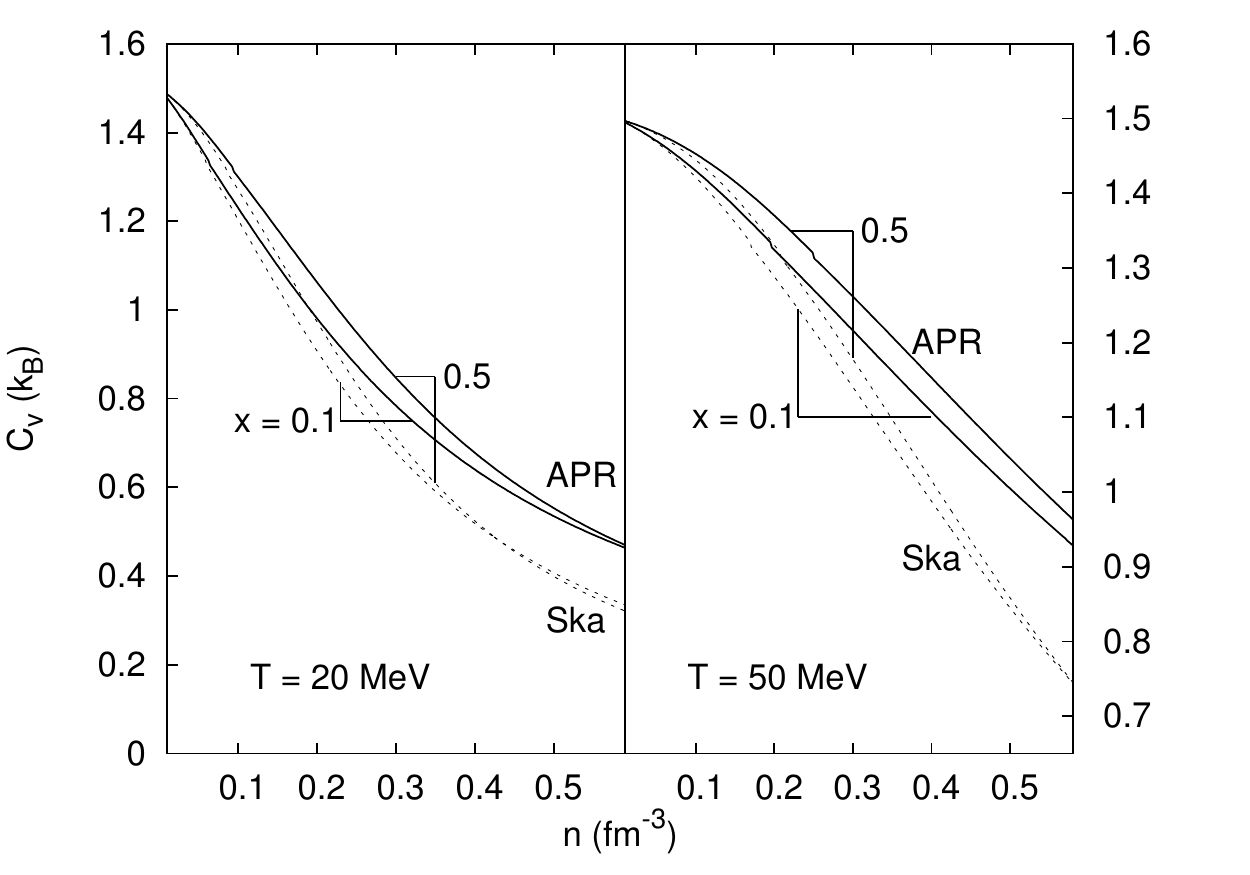}
\end{minipage}
\begin{minipage}[b]{0.49\linewidth}
\centering
\includegraphics[width=9.5cm]{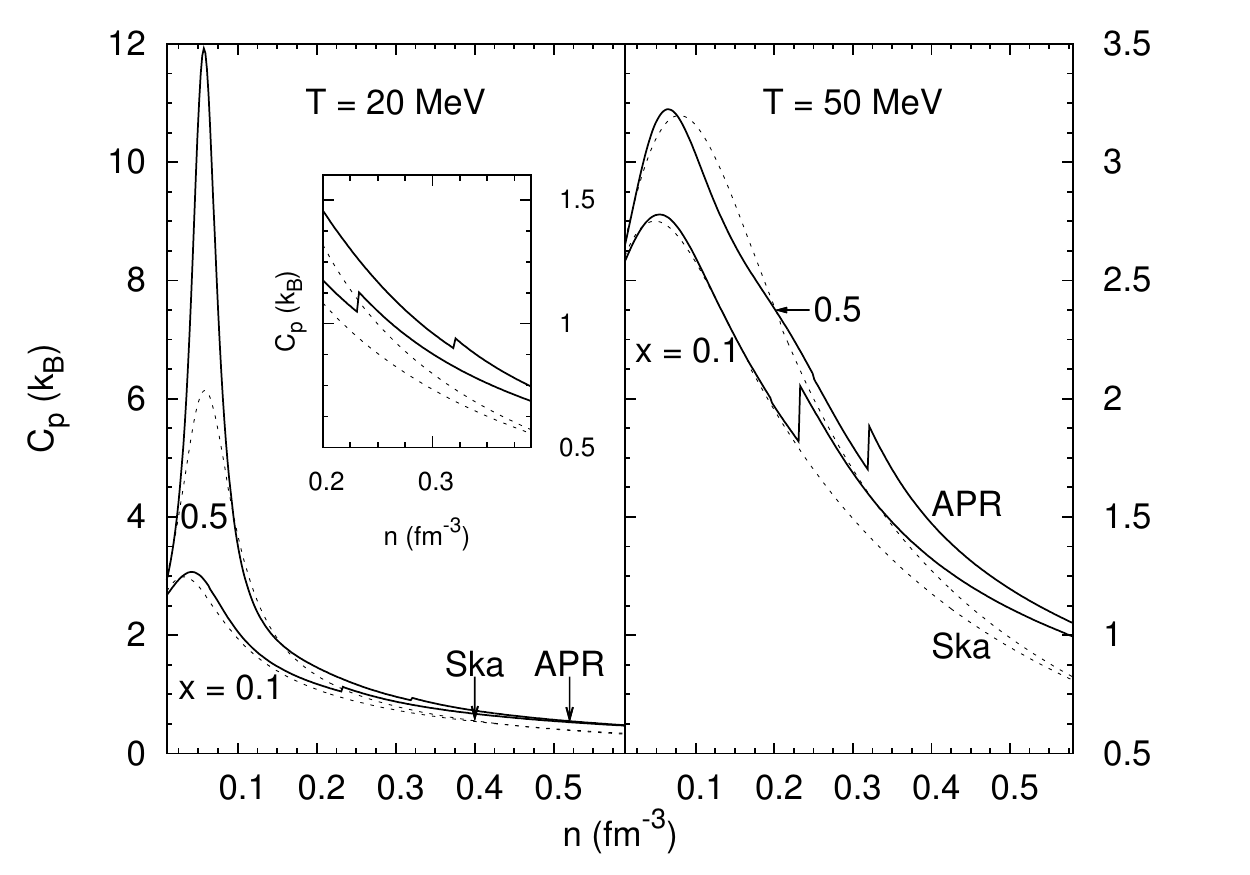}
\end{minipage}
\vskip -0.5cm
\caption{Left panels: Specific heat constant volume, $C_V$ (from Eq. (\ref{cv})) vs baryon density. Right panels: Specific heat at constant pressure, $C_P$ (from Eq. (\ref{cp}))  vs baryon density.  } 
\label{APRSKA_CvCp}
\end{figure*}

In Fig. \ref{APRSKA_CvCp} results for the specific heats at constant volume and at constant pressure, $C_V$ and $C_P$ (from Eqs. (\ref{cv}) and (\ref{cp})) are shown 
as  functions of baryon density for the APR and Ska models at temperatures of $20$ and $50$ MeV, respectively.  Beginning with the value of 1.5 characteristic of a dilute ideal gas, $C_V$ steadily decreases with increasing density as degeneracy begins to set in.  As the EOS of the Ska model is stiffer than that of the APR model at high densities, the fall off of $C_V$ with density is correspondingly more rapid. For both models, $C_V$ exhibits little dependence on proton fraction for both temperatures shown.
Results of $C_P$, shown in the right panel of this figure,  exhibit characteristic maxima that indicate the occurrence of a liquid-gas phase transition at low densities.  
 At $n=n_{c}$ and $T=T_c$, ${dP}/{dn} \rightarrow 0$ (see Fig. \ref{APRSKA_IsoTherm} in which $P$ vs $n$ for the two models are shown at various temperatures) which causes $C_P$ (which is inversely proportional to $dP/dn$) to diverge. For isospin symmetric matter at $T=20$ MeV, the maximum in $C_P$ is greater for the APR model than that for the Ska model. This feature can be understood in terms of  $T=20$ MeV being closer to the $T_c=17.91$ MeV of the APR model than to the  $T_c=15.12$ MeV for the Ska model.   As for $C_V$, there is little dependence on proton fraction for $C_P$. Note that an abrupt  
jump in $C_P$ also occurs for the APR model at the densities for which a transition from the LDP to the HDP takes place due to the onset of pion condensation (see the inset in the first of the right panel figure for its presence also at $T=20$ MeV.)

\subsection{Limiting cases}

In this section, we study the limiting cases when degenerate (low $T$, high $n$ such that $T/E_{F_i} \ll 1$) and non-degenerate 
(high $T$, low $n$ such that $T/E_{F_i} \gg1$) conditions prevail. In these limits, compact analytical expressions for all thermodynamic 
variables can be obtained. From a comparison of the exact, but, numerical, results with their analytical counterparts, the density and temperature ranges
in which supernova matter is degenerate, partially degenerate or non-degenerate can be established. In addition, such a comparison also provides 
a consistency check on our numerical calculations of the thermal variables. 
Because of the 
varying concentrations of neutrons and protons (and leptons, considered in a later section) encountered, one or the other species may well lie
in different regimes of degeneracy. 

\subsection*{Degenerate limit}

In this case, 
we make use of Landau's Fermi Liquid Theory (FLT) ~\cite{ll9, flt}, which allows for a model-independent discussion of the 
various thermodynamical functions. The temperature dependence of these functions is governed by the nature of the single particle 
spectrum. For the APR and Skyrme Hamiltonians, this dependence is characterized by a density dependent effective mass.

In FLT, 
the entropy density $s$ and the number 
density $n$ maintain the same functional forms as those of a free Fermi gas. For a single-component gas,
\begin{eqnarray}
s & = & - \sum_{k,\sigma}\left[n_{k\sigma}~\mbox{ln}~n_{k\sigma}+(1-n_{k\sigma})~\mbox{ln}~(1-n_{k\sigma})\right] 
\label{flts}  \\
n & = & \sum_{k,\sigma}n_{k\sigma} \quad {\rm and} \quad  
n_{k\sigma}  = \frac{1}{e^{(\epsilon_{k\sigma}-\mu)/T}+1} \,,
\label{fltn}
\end{eqnarray}
where $k$ is the wave number, and $\sigma$ stands for spin degrees of freedom, respectively. Note that the quasiparticle energy 
$\epsilon_k$ is itself a function of the distribution function $n_k$. The distribution of particles close to the zero temperature Fermi energy 
$E_F$ determines the general behavior (degenerate versus non-degenerate) of the system. 

The low temperature expansion of $s$ is standard and to order $T$ yields
\ba
s = \frac {\pi^2}{3} N(0)T = \frac {\pi^2}{k_Fv_F} nT \,,
\label{sFLT}
\ea
where $N(0)$ is the density of states at the Fermi surface:
\ba
N(0) = \sum_{\vec k}\delta (\epsilon_{k\sigma}-\mu) = \frac {3n}{k_Fv_F} \,.
\ea
The quantity $v_F$ is the Fermi velocity:
\be
v_F  =  \left.\frac{\partial \epsilon_{ks}^o}{\partial k}\right|_{k=k_F}
                       = \frac{k_F}{m^*} 
\ee
The above equation serves as a definition of the quasiparticle effective mass $m^*$.
Including the 2 spin degrees of freedom, $n=k_F^3/(3\pi^2)$ so that $N(0)=m^*k_F/\pi^2$.
The entropy density in Eq.~(\ref{sFLT}) is often written as 
\be
s = 2anT = \frac {\pi^2}{2} n \left[\frac {T}{T_F} \right]\,.
\ee
Above, the level density parameter $a$ and the Fermi temperature $T_F$ are 
\ba
a &=& \frac {\pi^2N(0)}{6n} = \frac {\pi^2}{2k_Fv_F} = \frac {\pi^2}{4T_F} \nonumber \\
T_F &=& \frac {1}{2}k_Fv_F = \frac {k_F^2}{2m^*} \,.
\ea
In normal circumstances, the leading correction to $s$ above is of order $(T/T_F)^2$ unless there exist soft 
collective modes which give rise to a $(T/T_F)^3 {\rm ln}~(T/T_F)$ behavior ~\cite{flt}.

The generalization to a multi-component gas is straightforward. The sums in Eq. (\ref{flts}) and 
Eq. (\ref{fltn})  
go over particle species so that the end result for the entropy density reads as
\begin{eqnarray}
s & = & \frac{\pi^2}{3}T\sum_i N_i(0) = 2T\sum_ia_in_i  \label{sdeg} \\ 
\mbox{where}~~~ a_i & =&  \frac{\pi^2}{2k_{Fi}v_{Fi}} = 
              \frac{\pi^2}{2}\frac{m_i^*}{k_{Fi}^2}  \label{levelden}
\end{eqnarray}

The rest of the thermal variables follow from thermodynamics, particularly the Maxwell relations.
The thermal energy is obtained from
\begin{eqnarray}
\int dE  &=&  \int TdS    
         =  \frac{2}{n}\sum_i a_in_i\int TdT  \nonumber \\ 
\Rightarrow E_{th}  &=&  \frac{T^2}{n}\sum_i a_in_i
\label{edeg}
\end{eqnarray}
The thermal pressure arises from
\begin{eqnarray}
\int dp & = & \int_0^T\left(s-n\frac{ds}{dn}\right)dT     \nonumber \\
        & = & \sum_i\left[a_in_i-n\frac{d(a_in_i)}{dn}\right]T^2.  \nonumber
\end{eqnarray}
Using $a_i=\frac{\pi^2}{2}\frac{m_i^*}{(3\pi^2n_i)^{2/3}}$, we get
\begin{eqnarray}
n\frac{d(a_in_i)}{dn}  =  a_in_i-\frac{2a_in}{3}
                                       \left(1-\frac{3}{2}\frac{n}{m_i^*}\frac{dm_i^*}{dn}\right) 
\nonumber \\
\end{eqnarray}
This allows us to write the thermal pressure as 
\begin{eqnarray}
 P_{th} 
                   &  = & \frac{2T^2}{3}\sum_ia_i n_i Q_i,  \label{pdeg} 
\end{eqnarray}
where $Q_i$ is given by  Eq. (\ref{qi}).
The thermal chemical potentials are obtained from
\begin{eqnarray}
\int d\mu_i  &=&  -\int \frac{ds}{dn_i}dT 
         =   -\frac{d}{dn_i}\left(\sum_ja_jn_j\right)T^2  \nonumber \\
\Rightarrow \mu_{ith} & = & -T^2\left[\frac{a_i}{3}+\sum_j\frac{n_ja_j}{m_j^*}\frac{dm_j^*}{dn_i}\right].
\label{mudeg}
\end{eqnarray}
Thus, the susceptibilities are
\ba
\frac{d\mu_{i,th}}{dn_i} &=& -\frac{T^2}{3}\left(-\frac{2}{3}\frac{a_i}{n_i}
     +2\frac{a_i}{m_i^*}\frac{dm_i^*}{dn_i}\right.  \nonumber \\
     &&\left.+ 3\frac{n_ia_i}{m_i^*}\frac{d^2m_i^*}{dn_i^2}
     + 3\frac{n_ja_j}{m_j^*}\frac{d^2m_j^*}{dn_i^2}\right)   \label{xiideg} \\
\frac{d\mu_{i,th}}{dn_j} &=& -\frac{T^2}{3}\left(\frac{a_i}{m_i^*}\frac{dm_i^*}{dn_j}
     +\frac{a_j}{m_j^*}\frac{dm_j^*}{dn_i}\right. \nonumber \\
     && \left.+ 3\frac{n_ia_i}{m_i^*}\frac{d^2m_i^*}{dn_idn_j}
     + 3\frac{n_ja_j}{m_j^*}\frac{d^2m_j^*}{dn_idn_j}\right) \label{xijdeg}
\ea

The free energy is given by
\be
F_{th} = E_{th} - TS = -E_{th} = -T^2\sum_i a_i Y_i \label{fdeg}
\ee
from which we get the symmetry energy
\ba
S_{2,th} &=& \frac{T^2a}{9}\left[1+\frac{3}{2m^*}\frac{dm^*}{dx}
     -\frac{9}{4m^{*2}}\left(\frac{dm^*}{dx}\right)^2\right]  \label{s2deg}  \\
a &=& \frac{\pi^2}{2}\frac{m^*}{\hbar^2}\frac{1}{\left(\frac{3\pi^2n}{2}\right)^{2/3}} 
\ea
where $m^*$ and $dm^*/dx$ are given by Eqs.(\ref{msym}) and (\ref{dmsym}) respectively. 

From the relation for the thermal energy, the specific heat at constant volume is 
\begin{equation}
C_V = \frac{2T}{n}\sum_i a_i n_i = S = \frac{2E_{th}}{T}.
\label{cvdeg}
\end{equation}
In the degenerate limit, to lowest order in temperature
\be
C_P = C_V .
\label{cpdeg}
\ee

\subsection*{Non-degenerate limit}

In the ND limit, the degeneracy (and hence the fugacity) is small, so that the FD functions can be 
expanded in a Taylor series about $z=0$:
\begin{equation}
F_{\alpha i} \simeq \Gamma(\alpha+1)\left(z_i-\frac{z_i^2}{2^{\alpha+1}}+\ldots\right) \,.
\end{equation}
Then the $F_{1/2}$ series is perturbatively inverted to get the fugacity in terms of the number 
density and the temperature:
\begin{eqnarray}
z_i & = &\frac{n_i\lambda_i^3}{\gamma} + \frac{1}{2^{3/2}}\left(\frac{n_i\lambda_i^3}{\gamma}\right)^2, \\
\mbox{where} ~~~\lambda_i & = &\left(\frac{2\pi\hbar^2}{m_i^*T}\right)^{1/2}  \\
\mbox{and} ~~~~~\gamma & = & 2 ~~~~\mbox{(the spin orientations)}.  \nonumber
\end{eqnarray}
Subsequently, these are used in the other FD integrals so that they, too, are expressed as explicit 
functions of the number density and the temperature: 
\begin{eqnarray}
F_{3/2i} & = & \frac{3\pi^{1/2}}{4}\frac{n_i\lambda_i^3}{\gamma}
              \left[1+\frac{1}{2^{5/2}}\frac{n_i\lambda_i^3}{\gamma}\right]  \\
F_{1/2i} & = & \frac{\pi^{1/2}}{2}\frac{n_i\lambda_i^3}{\gamma}   \\
F_{-1/2i} & = & \pi^{1/2}\frac{n_i\lambda_i^3}{\gamma}
              \left[1-\frac{1}{2^{3/2}}\frac{n_i\lambda_i^3}{\gamma}\right]
\end{eqnarray}
Finally, we insert these into equations (\ref{eath})-(\ref{muth}) from which we get:
\ba
E_{th}  &=&  \frac{1}{n}\sum_i\left\{\frac{3}{2}Tn_i
             \left[1+\frac{n_i}{4}\left(\frac{\pi\hbar^2}{m_i^*T}\right)^{3/2}\right]
             -\frac{3}{5}\mathcal{T}_{Fi}n_i\right\} \nonumber \\ \label{end} \\  
P_{th}  &=&  \sum_i\left\{TQ_in_i
             \left[1+\frac{n_i}{4}\left(\frac{\pi\hbar^2}{m_i^*T}\right)^{3/2}\right]
             -\frac{2}{5}\mathcal{T}_{Fi}n_i\right\} \nonumber \\ \label{pnd} \\
S &=& \frac{1}{n}\sum_in_i\left\{\frac{5}{2}-\mbox{ln}\left[\left(\frac{2\pi\hbar^2}{m_i^*T}\right)^{3/2}
                                                        \frac{n_i}{2}\right]  \right. \nonumber \\
     &+&\left.\ \frac{n_i}{8}\left(\frac{\pi\hbar^2}{m_i^*T}\right)^{3/2}\right\} \label{snd} \\
\mu_{ith} &=& -T \left\{-\mbox{ln}\left[\left(\frac{2\pi\hbar^2}{m_i^*T}\right)^{3/2}\frac{n_i}{2}\right]
              -\frac{n_i}{2}\left(\frac{\pi\hbar^2}{m_i^*T}\right)^{3/2}\right.  \nonumber \\
          &&  +\frac{3}{2}\frac{n_i}{m_i^*}\frac{dm_i^*}{dn_i}     
               \left[1+\frac{n_i}{4}\left(\frac{\pi\hbar^2}{m_i^*T}\right)^{3/2}\right]  \nonumber \\
          &&  +\frac{3}{2}\frac{n_j}{m_j^*}\frac{dm_j^*}{dn_i}     
      \left.\left[1+\frac{n_j}{4}\left(\frac{\pi\hbar^2}{m_j^*T}\right)^{3/2}\right]\right\}  \nonumber \\
          &&  -\mathcal{T}_{Fi}\left[1-\frac{3}{5}\frac{n_i}{m_i^*}\frac{dm_i^*}{dn_i}\right]
             +\frac{3}{5}\frac{n_j}{m_j^*}\frac{dm_j^*}{dn_i}\mathcal{T}_{Fj}  \label{mund} \,.
\ea
Thus
\ba
F_{th} &=& \sum_i\left\{TY_i\left[-1+\ln\left[\left(\frac{2\pi\hbar^2}{m_i^*T}\right)^{3/2}
                         \frac{n_i}{2}\right]\right.\right. \nonumber \\
       && \left.\left. +\frac{n_i}{4}\left(\frac{\pi\hbar^2}{m_i^*T}\right)^{3/2}\right]
           -\frac{3}{5}\mathcal{T}_{Fi}Y_i\right\}    \label{fnd}\\
S_{2,th} &=&  \frac{T}{8}\left\{8\left(1+\frac{3}{4m^*}\frac{dm^*}{dx}\right)
                 \left[1+\frac{n}{8}\left(\frac{\pi\hbar^2}{m^*T}\right)^{3/2}\right]\right. \nonumber \\
          && ~~~~~ -4\left[1+\frac{3}{8m^{*2}}\left(\frac{dm^*}{dx}\right)^2\right]  \nonumber \\
          && ~~~~~\left. +\frac{3n}{4m^*}\left(\frac{\pi\hbar^2}{m^*T}\right)^{3/2}\frac{dm^*}{dx}
               \left(1+\frac{1}{8m^*}\frac{dm^*}{dx}\right)\right\}  \nonumber \\
          && -\frac{\mathcal{T}_F}{3}\left(1+\frac{3}{2m^*}\frac{dm^*}{dx}\right)  \label{s2nd}  \\
\mathcal{T}_F &=& \left(\frac{3\pi^2n}{2}\right)^{2/3}\frac{\hbar^2}{2m^*}      
\ea
\ba
\frac{d\mu_i}{dn_i} &=& \frac{T}{n_i}\left(1-\frac{3n_i}{m_i^*}\frac{dm_i^*}{dn_i}\right)
           \left[1+\frac{n_i}{2}\left(\frac{\pi \hbar^2}{m_i^*T}\right)^{3/2}
                 +\frac{2}{3}\frac{\mathcal{T}_{Fi}}{T}\right]  \nonumber \\
             &+& O\left(\left(\frac{dm^*}{dn}\right)^2, \frac{d^2m^*}{dn^2}\right) \label{xiind} \\
\frac{d\mu_i}{dn_j} &=& 
           -T\left\{\frac{3}{2m_n^*}\frac{dm_i^*}{dn_j}
           \left[1+\frac{n_i}{2}\left(\frac{\pi \hbar^2}{m_i^*T}\right)^{3/2}
                 +\frac{2}{3}\frac{\mathcal{T}_{Fi}}{T}\right]\right.  \nonumber \\
         &+&  \left.\frac{3}{2m_j^*}\frac{dm_j^*}{dn_i}
           \left[1+\frac{n_j}{2}\left(\frac{\pi \hbar^2}{m_j^*T}\right)^{3/2}
                 +\frac{2}{3}\frac{\mathcal{T}_{Fj}}{T}\right]\right\} \nonumber \\  
         &+& O\left(\left(\frac{dm^*}{dn}\right)^2, \frac{d^2m^*}{dn^2}\right) \label{xijnd}  \\
C_V & = &   \frac{1}{n}\sum_i\left\{\frac{3}{2}n_i
             \left[1-\frac{n_i}{8}\left(\frac{\pi\hbar^2}{m_i^*T}\right)^{3/2}\right]
             \right\} .
\label{cvnd}
\ea
The second derivatives and the squares of the first derivatives of the effective mass are neglected 
because they represent higher order corrections. 

For $C_P$, we need the temperature and density derivatives of pressure 
in the non-degenerate limit, for which we use Eq.(\ref{cp}) in conjuction with
\be
P = \sum_i\left[TQ_in_i\left\{1+\frac{n_i}{4}\left(\frac{\pi\hbar^2}{m_i^*T}\right)^{3/2}\right\}\right]+P_d
\ee
to get
\ba
\left.\frac{\partial P}{\partial T}\right|_n &=&
   \sum_i\left[Q_in_i\left\{1-\frac{n_i}{8}\left(\frac{\pi\hbar^2}{m_i^*T}\right)^{3/2}\right\}\right]  
\label{dpdtnd} \\
\left.\frac{\partial P}{\partial n}\right|_T &=&
  \sum_i\left[T\left\{1+\frac{n_i}{4}\left(\frac{\pi\hbar^2}{m_i^*T}\right)^{3/2}\right\}
                        \left(\frac{\partial Q_i}{\partial n}n_i + Q_iY_i\right)\right] \nonumber \\
&+& \sum_i\left[TQ_i^2n_i\frac{Y_i}{4}\left(\frac{\pi\hbar^2}{m_i^*T}\right)^{3/2}\right] + \frac{dP_d}{dn} \,.
\label{dpdnnd}
\ea

\subsection*{Results}

This section is devoted to comparisons of results from the analytical formulas obtained in the previous section for the limiting cases with those from the exact calculations presented earlier. In addition to providing us with physical insights about the general trends observed, these comparisons will allow us to delineate the range of densities for which matter with varying isospin asymmetry and temperature can be regarded as either degenerate or non-degenerate.  We will restrict our comparisons to results from the APR model only as those for the Ska model yield similar conclusions. 

\begin{figure}[!h]
\includegraphics[width=9cm]{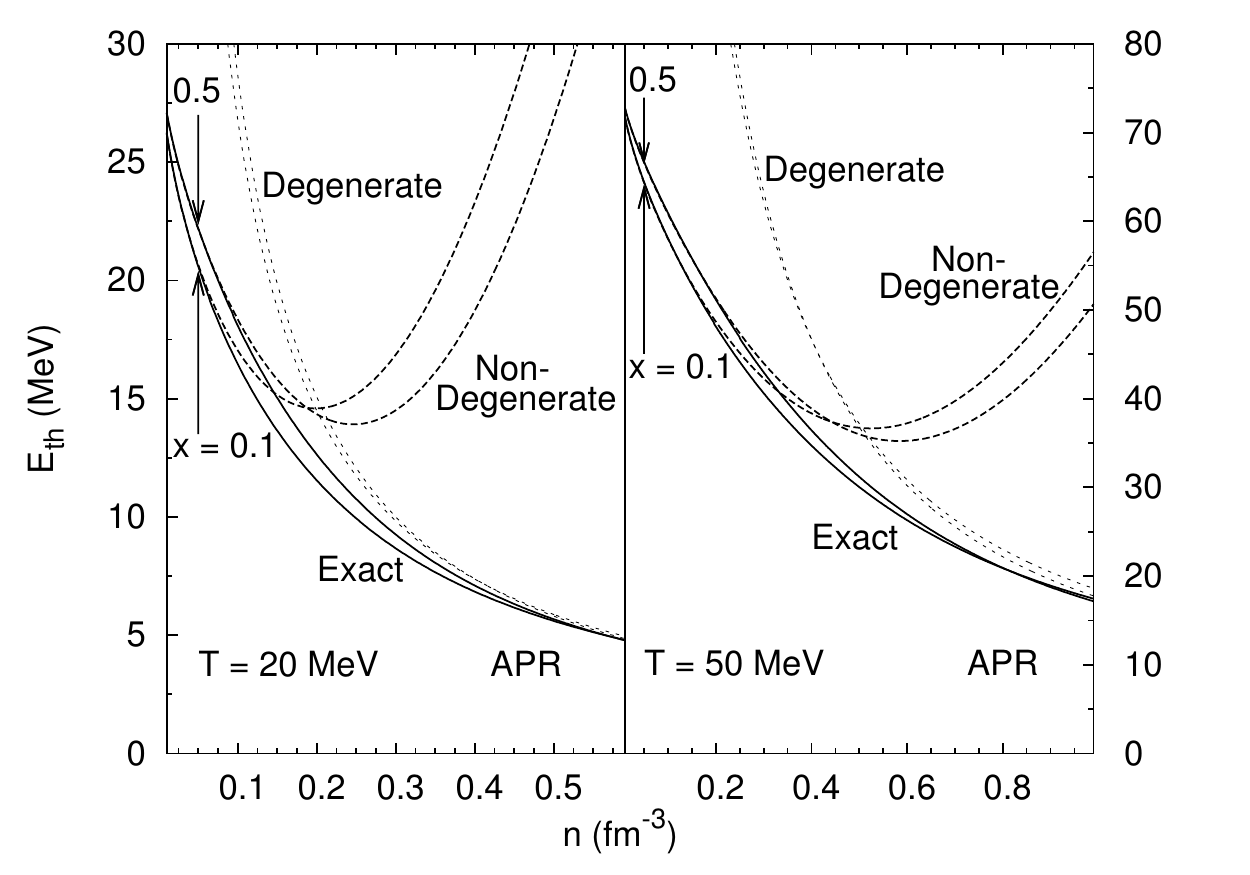}
\caption{Thermal energy per particle (Eq. (\ref{eath})) and limiting cases (Eqs. (\ref{edeg}) and
(\ref{end})) vs baryon density at the indicated temperatures and proton fractions.}
\label{APR_Eth}
\end{figure}
In Fig. \ref{APR_Eth}, we show the thermal energies $E_{th}$ as a function of baryon density $n$ for $T=20$ MeV (left panel) and 50 MeV (right panel) for proton fractions of $x=0.5$ and 0.1, respectively. The $T^2$ dependence implied by the degenerate approximation in  Eq. (\ref{edeg}) is borne out by the the exact results at high densities. Also, the larger the temperature, the larger is the density at which the degenerate approximation approaches the exact result.  
The effective masses introduce an additional density dependence to the $\sim n^{-2/3}$ behavior characteristic of a free gas of degenerate fermions for which $E_{th}$  would be larger than that with momentum dependent interactions. 
Note that in the degenerate limit, both the approximate and exact results are nearly $x$- independent.  
With increasing temperature, the non-degenerate approximation in Eq. (\ref{end})) reproduces the exact results the agreement extending up to nuclear density and even slightly beyond.  As for a free Boltzmann gas, the thermal energy is predominantly linear in $T$ in the non-degenerate limit and is only slightly modified by the density dependence of the effective masses. 
The $\sim n^{-2/3}$ fall off with density arises from the last term in  Eq. (\ref{eath}) (the degeneracy energy of fermions at $T=0$) with sub-dominant corrections from the density dependence of the effective masses.   
Effects of isospin asymmetry are somewhat more pronounced in the non-degenerate case when compared to the degenerate limit. Results for highly asymmetric matter from    
 Eq. (\ref{end})) begin to deviate from the exact results at lower densities than for symmetric matter 
because the two components are in different regimes of degeneracy.  
\begin{figure}[!h]
\includegraphics[width=8.5cm]{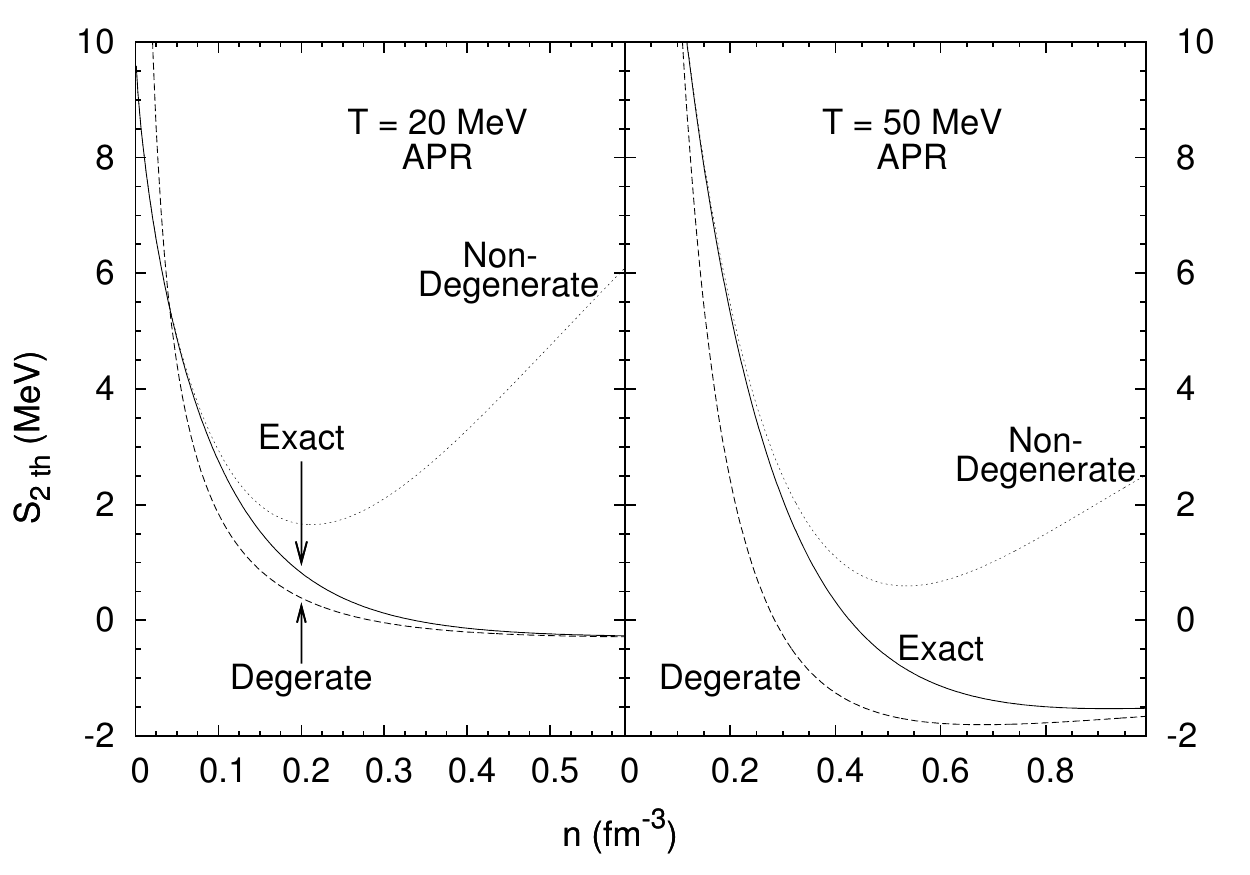}
\caption{Thermal contributions to the symmetry energy, $S_{2,th}$, from Eq. (\ref{s2t}) compared with its limiting cases (Eqs. (\ref{s2deg}) and (\ref{s2nd})) at the indicated temperatures. }
\label{APR_S2_Lim}
\end{figure}

In Fig. \ref{APR_S2_Lim},  thermal contributions to the symmetry energy, $S_{2,th}$ from Eq. (\ref{s2t}) and its limiting cases from Eqs. (\ref{s2deg}) and (\ref{s2nd}) for the APR model are shown as functions of baryon density at temperatures of $20$ and $50$ MeV, respectively. Agreement 
between the degenerate limit and the exact result is obtained around $3n_0$ for $T=20$ MeV and at much larger densities ($n>1~{\rm fm}^{-3}$) for the $T=50$ MeV.   The
non-degenerate limit coincides with the exact result for densities less than $\approx0.5n_0$ for the $20$ MeV temperature. At $T=50$ MeV the non-degenerate limit has a greater range 
of baryon densities for which it agrees with the exact result, reaching up to 1-1.5~$n_0$.   A noteworthy feature in this figure is that both the exact and the degenerate result for $S_{2,th}$ become negative after a certain baryon density. 
Note that for free fermions, $S_{2,th}$ in Eq. (\ref{s2deg}) is strictly positive, pointing to the fact that derivatives of $m^*$ with respect to proton fraction $x$ are at the root of driving $S_{2,th}$ negative.  
In what follows, we examine the rate at which the identity $\Delta F_{th} = \sum_i S_{i,th}$ with $i$ even (odd terms cancelling) is fulfilled.

\begin{figure}[!h]
\includegraphics[width=8.5cm]{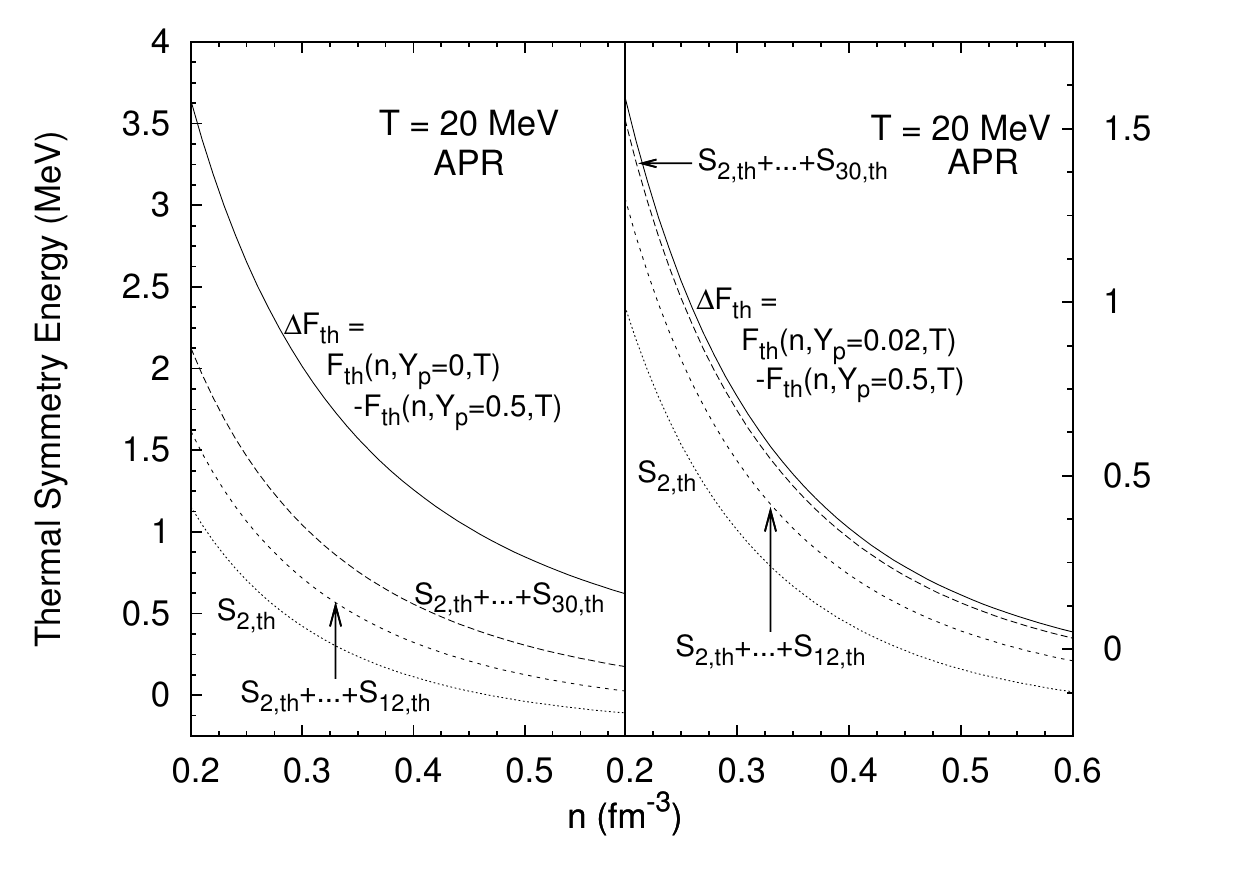}
\caption{Thermal symmetry free energies $S_{i,th}$ and their contributions to $\Delta F_{th} = \sum_i S_{i,th}$ as defined in the insets.}
\label{APR_Asym_S12_Double}
\end{figure}

The left panel of Fig. \ref{APR_Asym_S12_Double} shows the difference of the exact free energies $\Delta F_{th} = F_{th}(n,x=0,T) - 
 F_{th}(n,x=0.5,T)$ at $T=20$ MeV.  Also shown are contributions from various $S_{i,th}$ at the same temperature. To be specific, 
 we consider only the degenerate limit results for $S_{i,th}$ in this comparison. 
 It turns out that only $S_2$ turns negative at a finite baryon density, whereas $S_4,~S_6,\cdots$ which contain higher derivatives of $m^*$ with respect to the proton fraction $x$ are all positive whose magnitudes decrease very slowly.  
 We have calculated up to thirty terms in $S_{i,th}$ and show how their sums compare with $\Delta F_{th}$.   
 It is clear that the convergence to the exact result is relatively poor, in contrast to the 
rapid convergence of symmetry energies at zero temperature (see Fig. \ref{APR_SymE}).  
 The situation is better, although by no means impressive, for  $\Delta F_{th} = F_{th}(n,x=0.02,T) - F_{th}(n,x=0.5,T)$ at $T=20$ MeV. These results indicate that the happenstance of rapid convergence of symmetry energies at zero temperature cannot automatically be taken to hold for their thermal parts as well. It should be noted, however, that the latter represent relatively small corrections to the total symmetry energy where the main contribution is due to the zero temperature component. The asymptotic nature of the Taylor series expansion  of $\Delta F_{th}$ in even powers of $(1-2x)$ at finite temperature is at the origin of such poor convergence for large isospin asymmetries. Exact, albeit numerical, calculations of the Fermi integrals are necessary for high isospin asymmetry.

\begin{figure}[hbt]
\includegraphics[width=9cm]{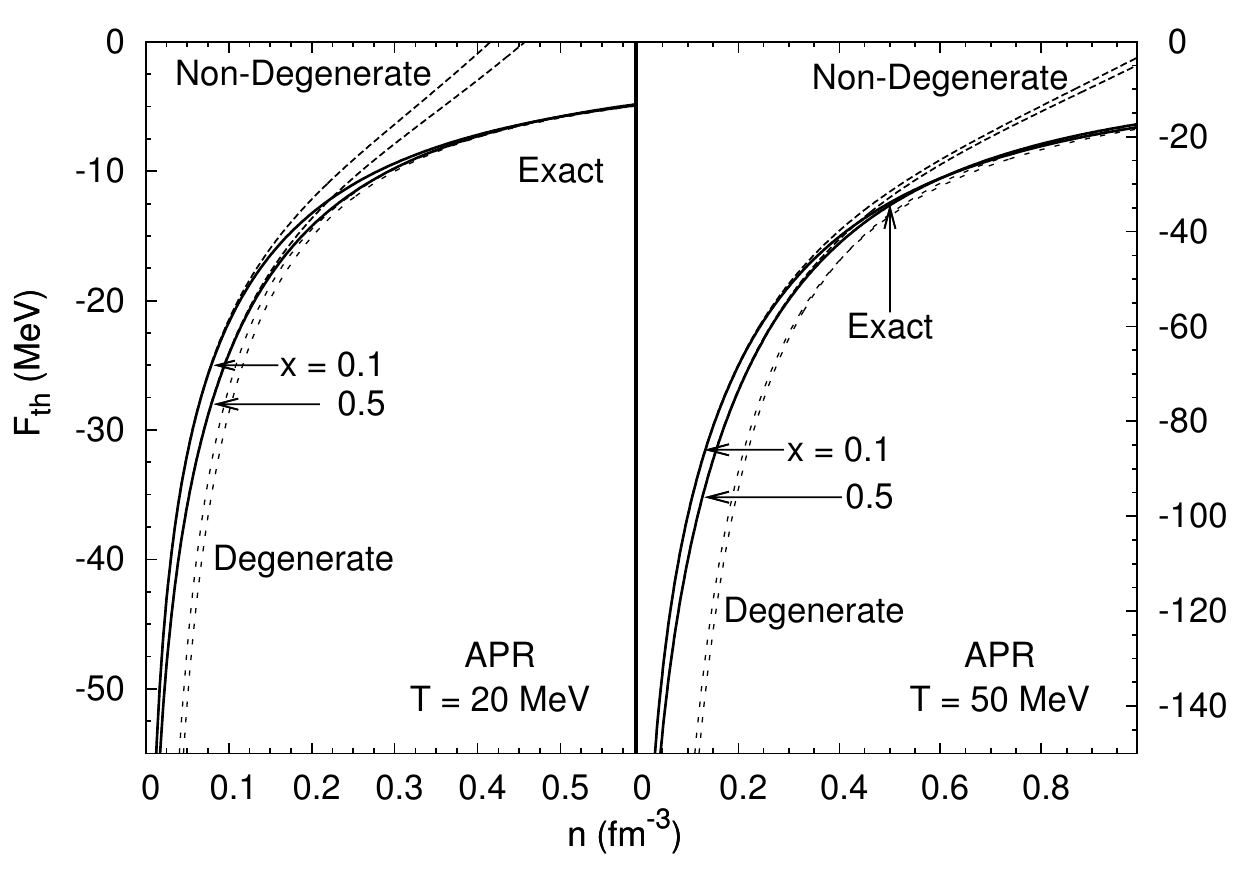}
\vskip -0.5cm
\caption{Thermal free energy (Eq. (\ref{fden})) and its limiting cases (Eqs. (\ref{fdeg}) and (\ref{fnd}) 
vs baryon density at the indicated proton fractions and temperatures. }
\label{APR_MxT_Fth}
\end{figure}

  In Fig. \ref{APR_MxT_Fth},  we show results for the thermal free energy from Eq. (\ref{fden}) and its limiting cases from Eqs. (\ref{fdeg}) and (\ref{fnd}) as functions of baryon density. The degenerate limit and 
the exact result of $F_{th}$ are in agreement for densities greater than $1.5n_0$ for $T=20$ MeV and only for much larger densities ($n\geq 5n_0$) for $T=50$ MeV. The convergence 
between the degenerate limit and the exact result of $F_{th}$ is independent of proton fraction for both temperatures. The non-degenerate limit begins to differ from the exact result 
at around $n_0$ for $T=20$ MeV and about $2n_0$ for $T=50$ MeV. For both temperatures shown,  the convergence between the non-degenerate limit and the exact solution is nearly independent 
of proton fraction.

\begin{figure}[h]
\includegraphics[width=9cm]{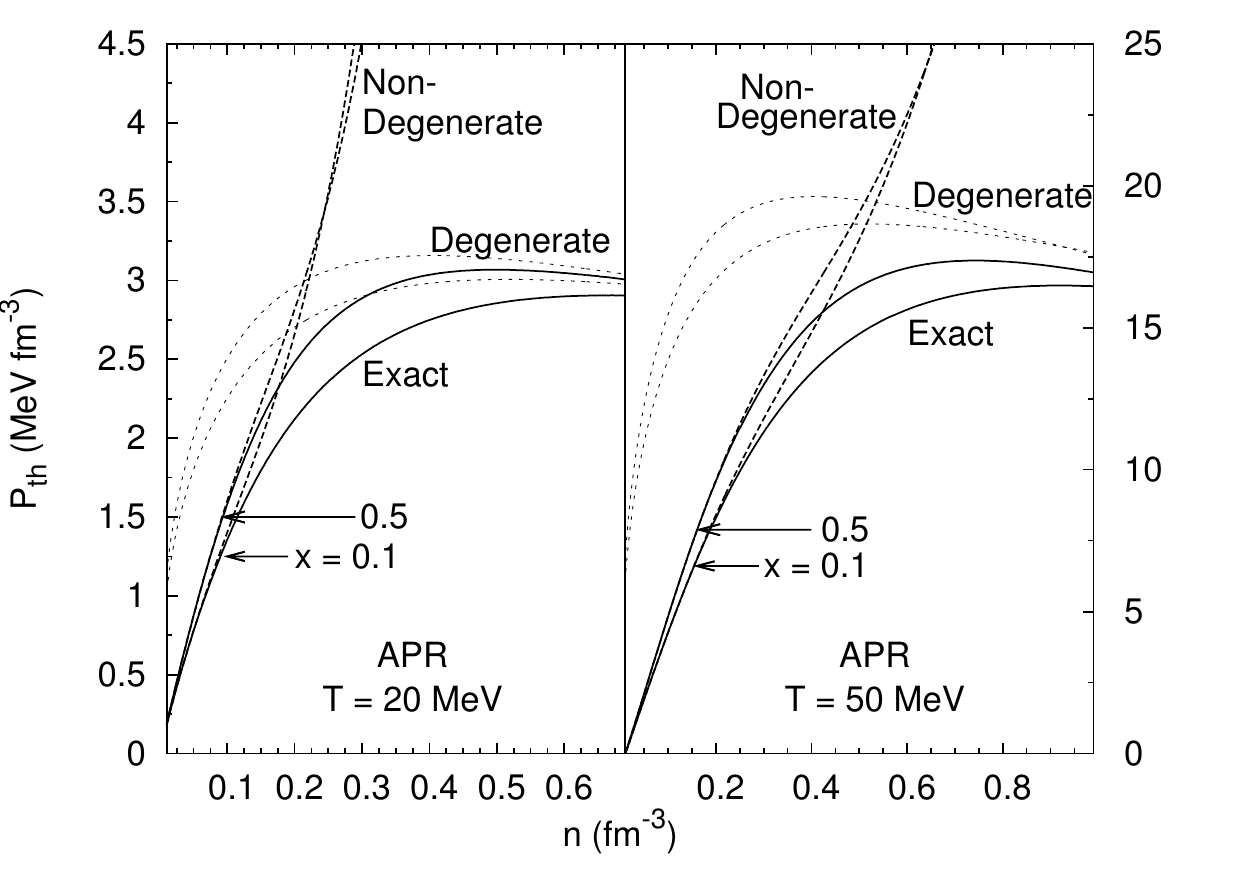}
\caption{Thermal pressure (Eq. (\ref{pth})) and limiting cases (Eqs. (\ref{pdeg}) and (\ref{pnd})) vs baryon density.}
\label{APR_Pth}
\end{figure}

Results for the exact thermal pressures $P_{th}$ (from Eq. (\ref{pth}))  and those of its limiting cases 
(Eqs. (\ref{pdeg}) and (\ref{pnd})  are presented in  Fig. \ref{APR_Pth} for the APR model. 
For both temperatures considered, the initial rise of $P_{th}$ (in the non-degenerate regime) is linear with slope $\sim T$ modulated by the factors $Q_i$ highlighting the role of density dependent effective masses relative to a free fermi gas for which the slope would be $T$. The linear rise is halted as matter begins to become increasingly degenerate when effective mass corrections begin to gain importance.  Quantitative agreement of the exact results with those from the limiting form of the degenerate expression is, however, reached at densities much larger than shown in this figure. Note that isospin asymmetry effects are more pronounced for $P_{th}$ than for $E_{th}$ except at very low and very high densities. 

\begin{figure*}[!ht]
\centering
\begin{minipage}[b]{0.49\linewidth}
\centering
\includegraphics[width=9cm]{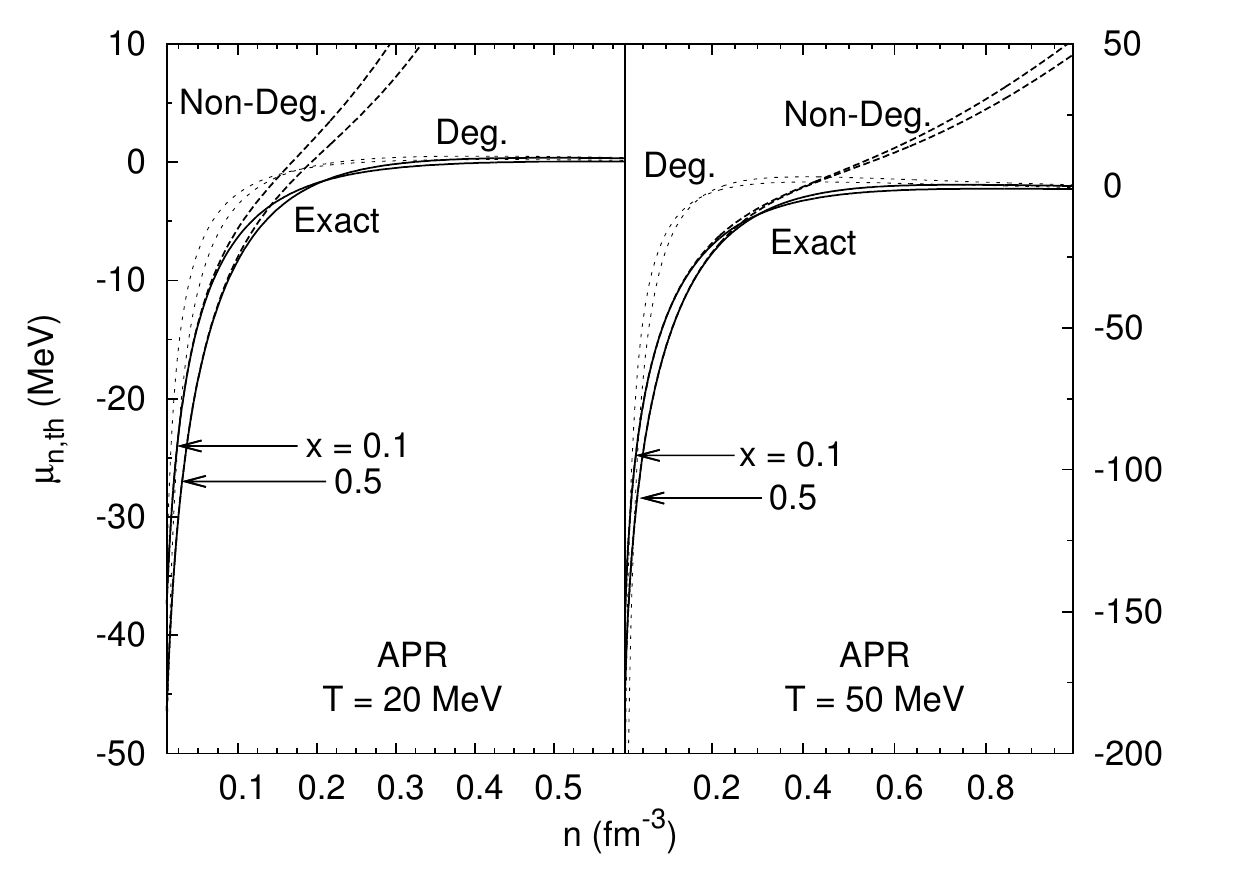}
\end{minipage}
\begin{minipage}[b]{0.49\linewidth}
\centering
\includegraphics[width=9cm]{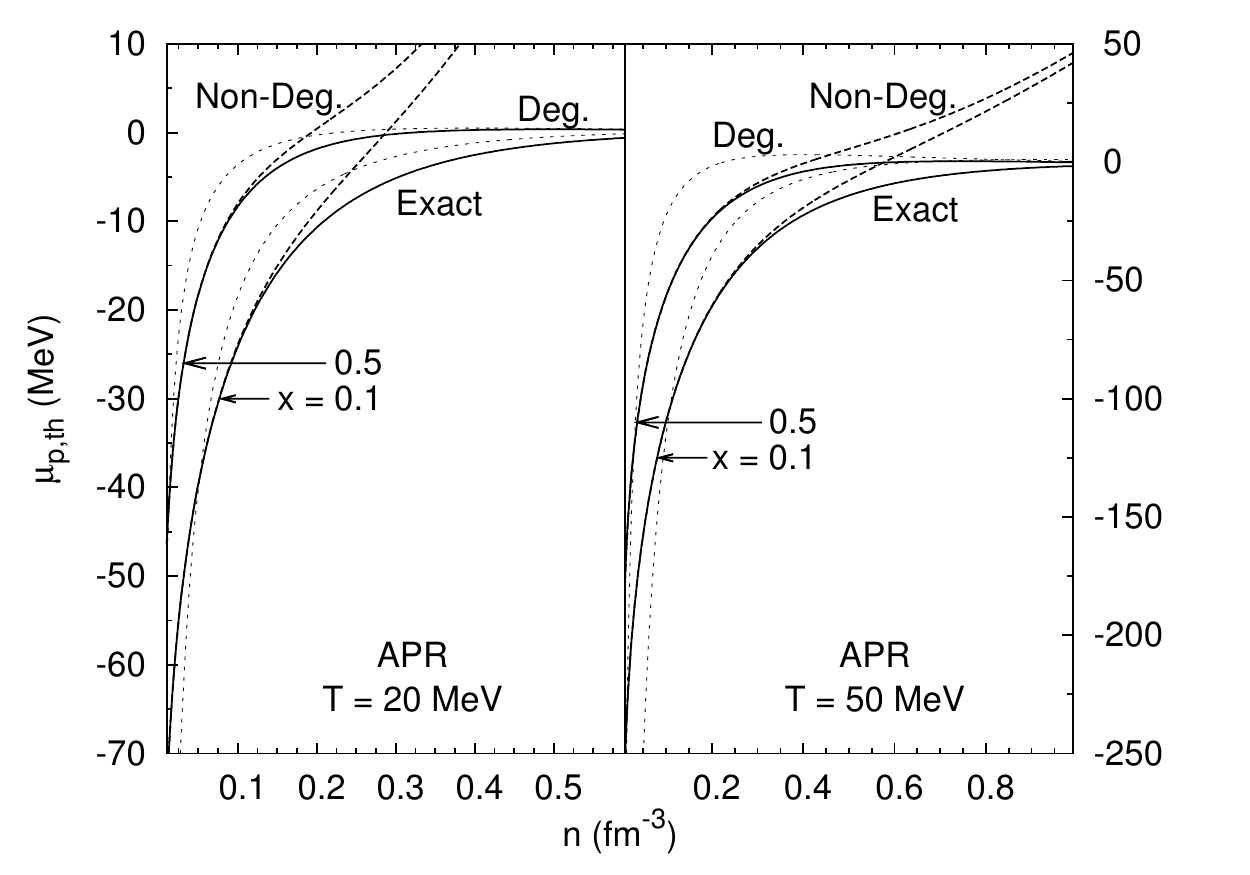}
\end{minipage}
\vskip -0.5cm
\caption{Proton (left) and neutron (right) thermal chemical potentials (Eq. (\ref{muth})) with limits 
(Eqs. (\ref{mudeg}) and (\ref{mund})) vs baryon density.} 
\label{APR_MuTH}
\end{figure*}

Thermal contributions to the neutron and proton chemical potentials $\mu_{n,th}$ and $\mu_{p,th}$ versus baryon density $n$ are shown in Fig. \ref{APR_MuTH} in which comparisons between between results from the exact  (Eq. (\ref{muth})) and limiting cases (Eqs. (\ref{mudeg}) and (\ref{mund}))  are made.  For $\mu_{n,th}$ (left panels), good agreement is found between the 
non-degenerate limit and the exact result for densities up to $n_0$ for $T=20$ MeV and up to $2n_0$ for $T=50$ MeV. 
Results in the degenerate limit rapidly approach the exact results, unlike in the cases of $E_{th}$ and $P_{th}$. 
Note that this level of quantitative agreement, in both non-degenerate and degenerate cases, required derivatives of the density dependent effective masses (Eqs. (\ref{mudeg}) and (\ref{mund})). Isospin asymmetry effects are not very pronounced for $\mu_{n,th}$.

The thermal contribution to the proton chemical potential $\mu_{p,th}$ (right panels) exhibits a greater difference between isospin symmetric and asymmetric matter when compared to $\mu_{n,th}$.
The agreement between the exact results for $\mu_{p,th}$  and those of the limiting cases
is much the same as it was for $\mu_{n,th}$. Both the degenerate and non-degenerate limits agree to 
a greater degree for the higher temperature and for isospin symmetric matter. \\

\begin{figure}[!h]
\includegraphics[width=9cm]{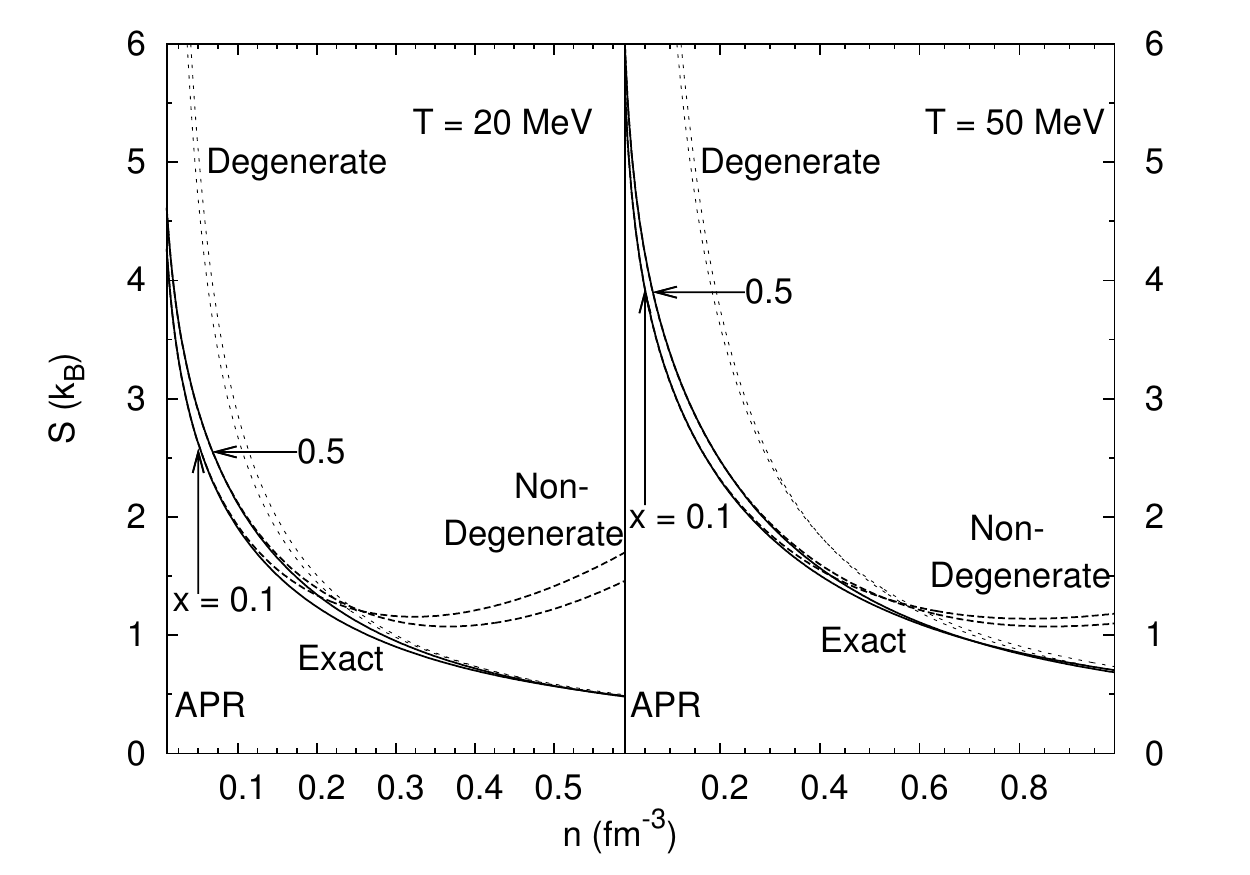}
\caption{Entropy per baryon (Eq. (\ref{entr})) and its limiting cases (Eqs. (\ref{sdeg}) and (\ref{snd})) vs baryon density at the indicated proton fractions and temperatures.}
\label{APR_SoA}
\end{figure}

In Fig. \ref{APR_SoA}, we present the exact results for the entropy per baryon (Eq. (\ref{entr}))  and its limiting cases 
(Eqs. (\ref{sdeg}) and (\ref{snd}))  
  as functions of baryon density $n$. The exact results show
little difference between isospin symmetric and asymmetric matter. 
A comparison of the results in the two panels reveals the range of densities over which the non-degenerate and degenerate approximations reproduce the exact results. 
The agreement between the exact results and those of the limiting cases
is almost independent of proton fraction, although what little difference there is points to isospin symmetric matter having a slightly better agreement.

\begin{figure}[!ht]
\includegraphics[width=10cm]{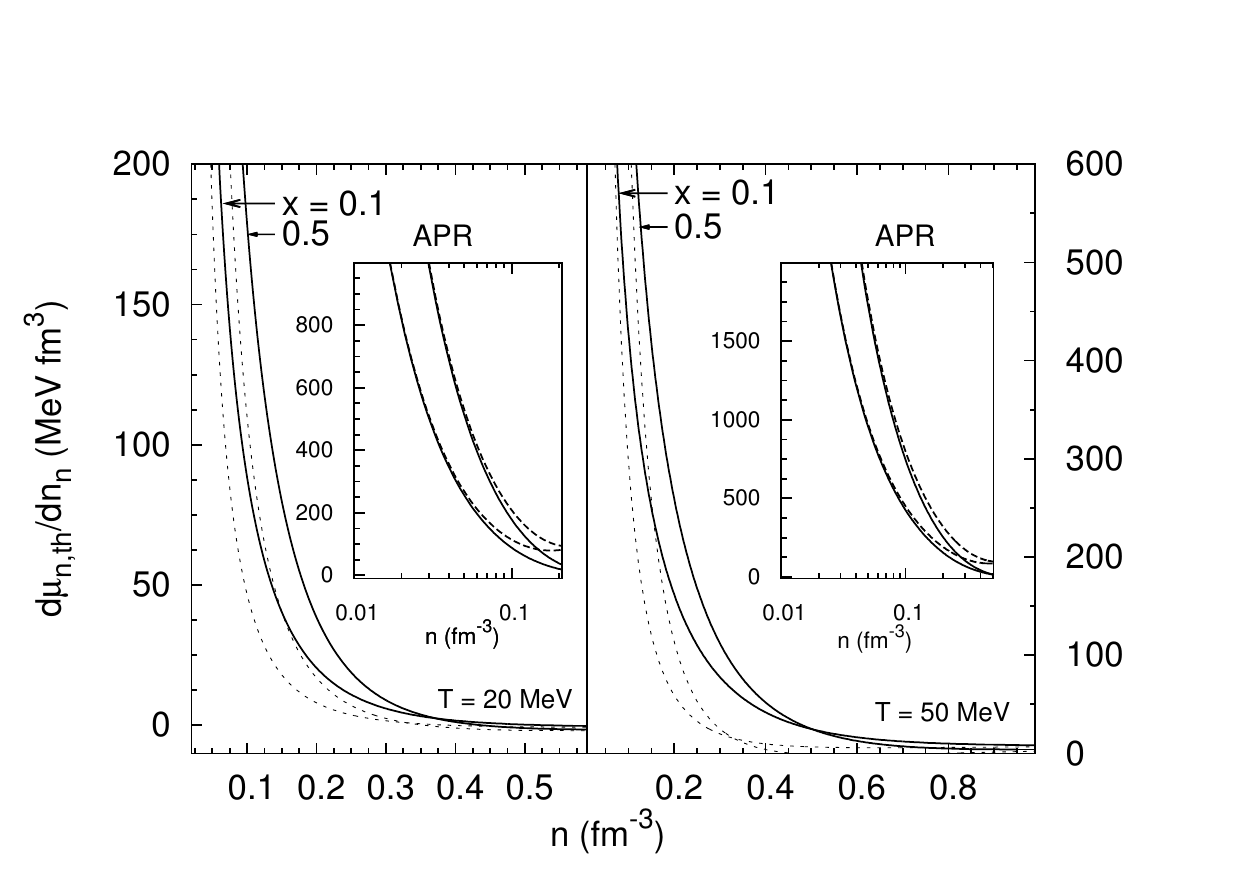}
\caption{Neutron-neutron inverse susceptibility vs baryon density (Eqs. (\ref{xi}) and (\ref{xii})) for the APR model and its limiting cases at the indicated proton fractions $x$. The degenerate limit 
(Eq. (\ref{xiideg})) and the non-degenerate limit (Eq. (\ref{xiind})), see inset, are both shown. }
\label{APR_Xnn}
\end{figure}
\begin{figure}[!hb]
\includegraphics[width=10cm]{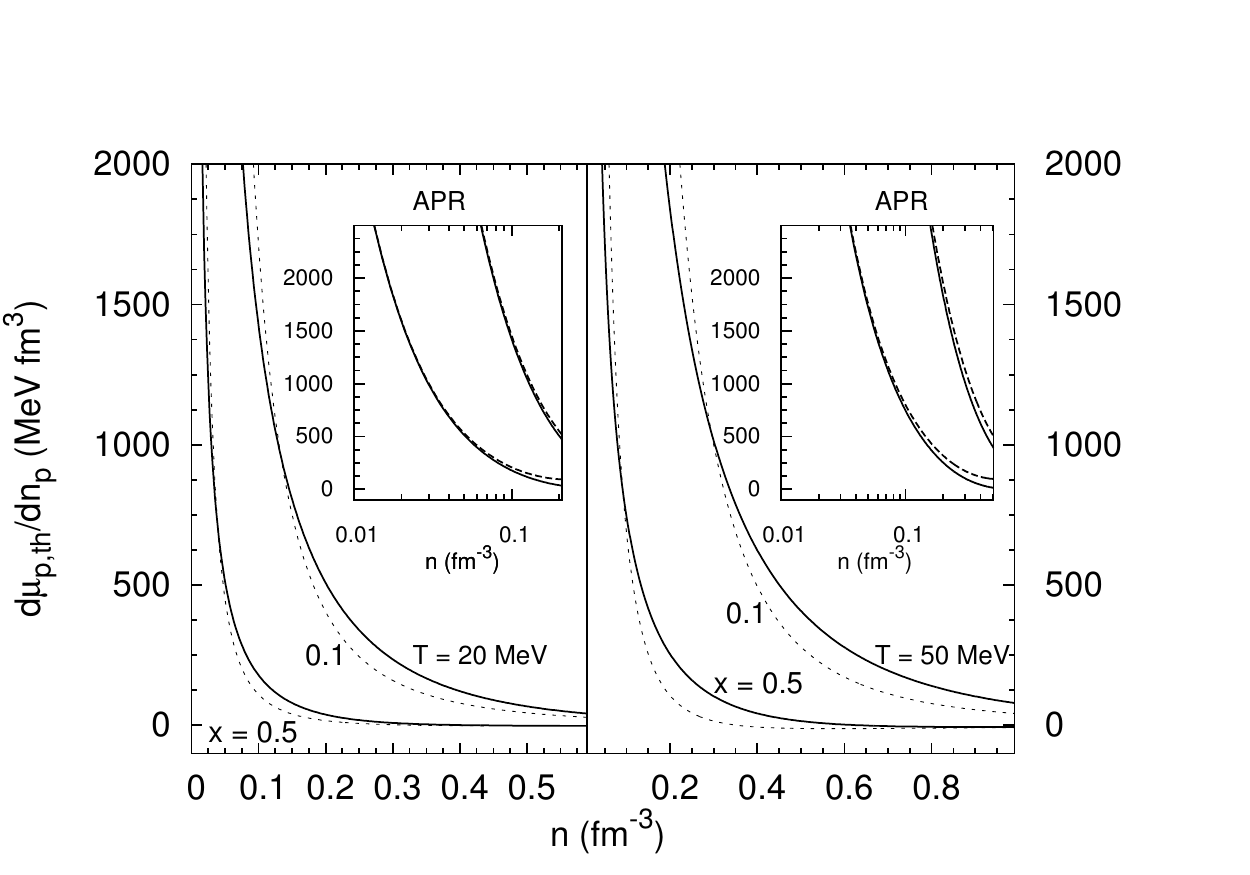}
\caption{Proton-proton inverse susceptibility vs baryon density (Eqs. (\ref{xi}) and (\ref{xii})) and its limiting cases. Both the exact result and its degerate limit (Eq. (\ref{xiideg})) are shown. The inset compares the non-degenerate limit (Eq. (\ref{xiind})) with the exact result.}
\label{APR_Xpp}
\end{figure}

\begin{figure}[!h]
\includegraphics[width=10cm]{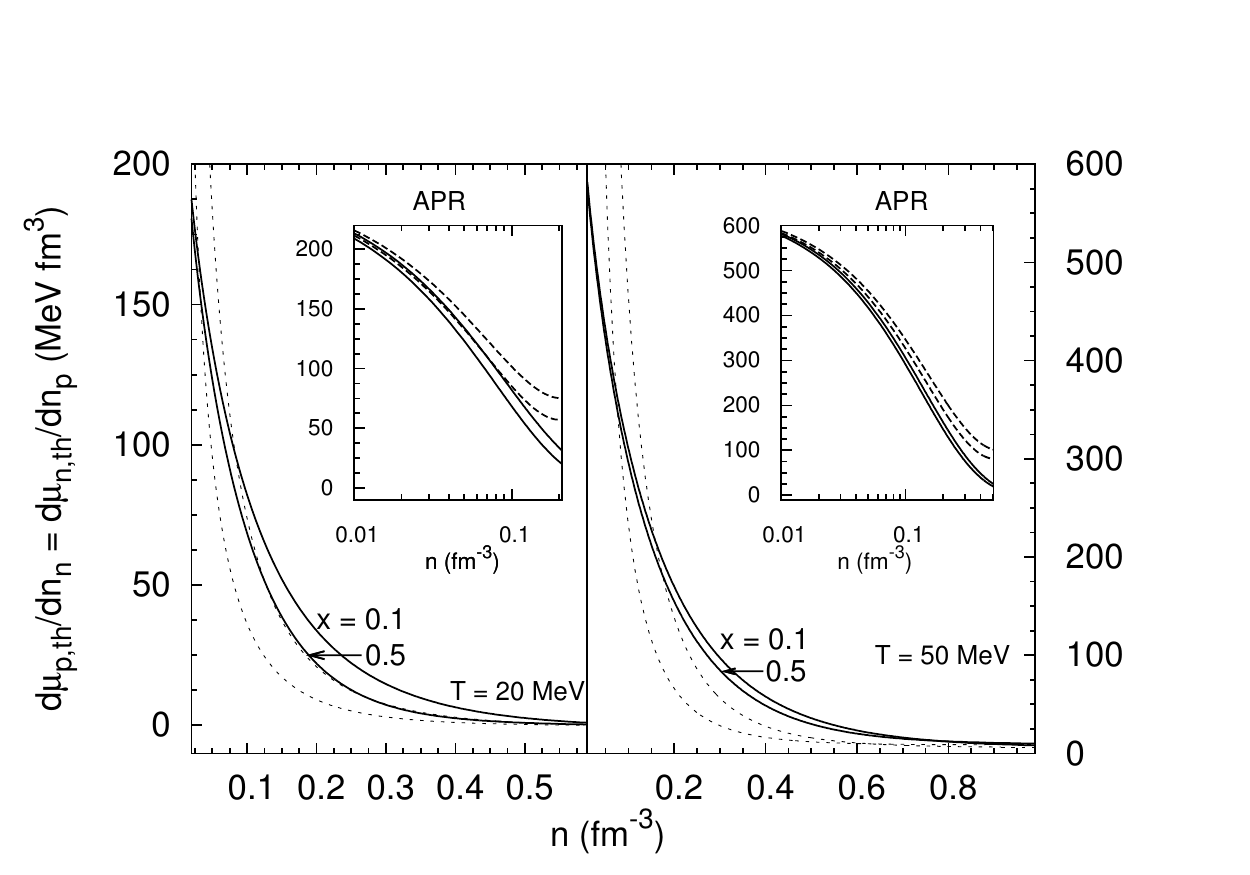}
\caption{Mixed inverse susceptibilities (Eqs. (\ref{xi}) and (\ref{xij}) and the limiting cases 
(Eqs. (\ref{xijdeg}) and (\ref{xijnd})) vs baryon densities. As $d\mu_n/dn_p = d\mu_p/dn_n$, only one of the mixed derivatives is shown.}
\label{APR_Xnp}
\end{figure}

In Figs. \ref{APR_Xnn}, \ref{APR_Xpp}, and \ref{APR_Xnp}, thermal contributions to the inverse susceptibilities $\chi^{-1}_{nn}$, 
$\chi^{-1}_{pp}$, and $\chi^{-1}_{np}$ (Eqs. (\ref{xi}), (\ref{xii})  and (\ref{xij})) are shown together with their limiting cases 
in Eqs. (\ref{xiideg}),  (\ref{xijdeg}) (\ref{xiind}) and (\ref{xijnd})
(the non-degenerate limits are in the insets of all three figures). Note that where expected, the degenerate and non-degenerate approximations provide an accurate description of the exact results. It is intriguing that for densities slightly above the nuclear density, neither of the approximations works very well. 

\begin{figure*}[!ht]
\centering
\begin{minipage}[b]{0.49\linewidth}
\centering
\includegraphics[width=9.5cm]{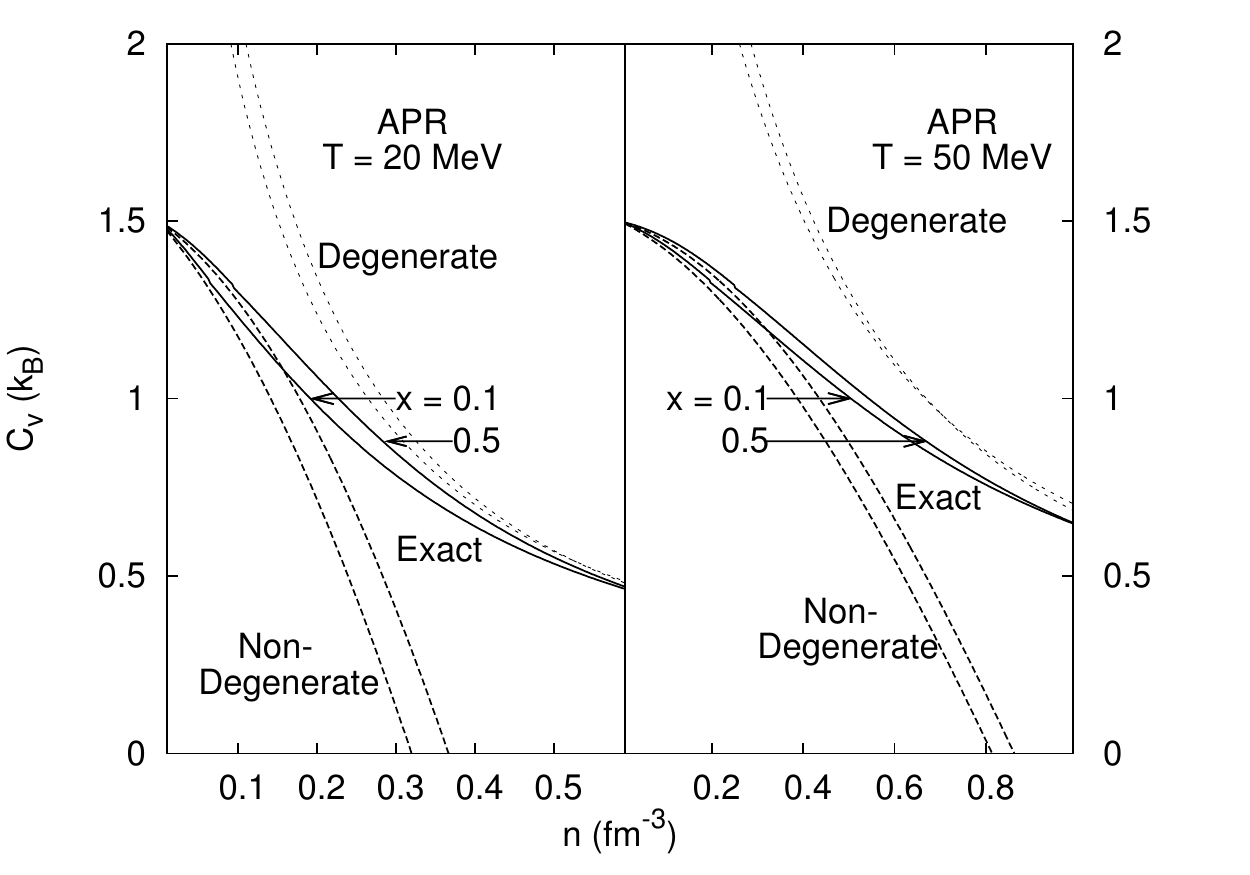}
\end{minipage}
\begin{minipage}[b]{0.49\linewidth}
\centering
\includegraphics[width=9.5cm]{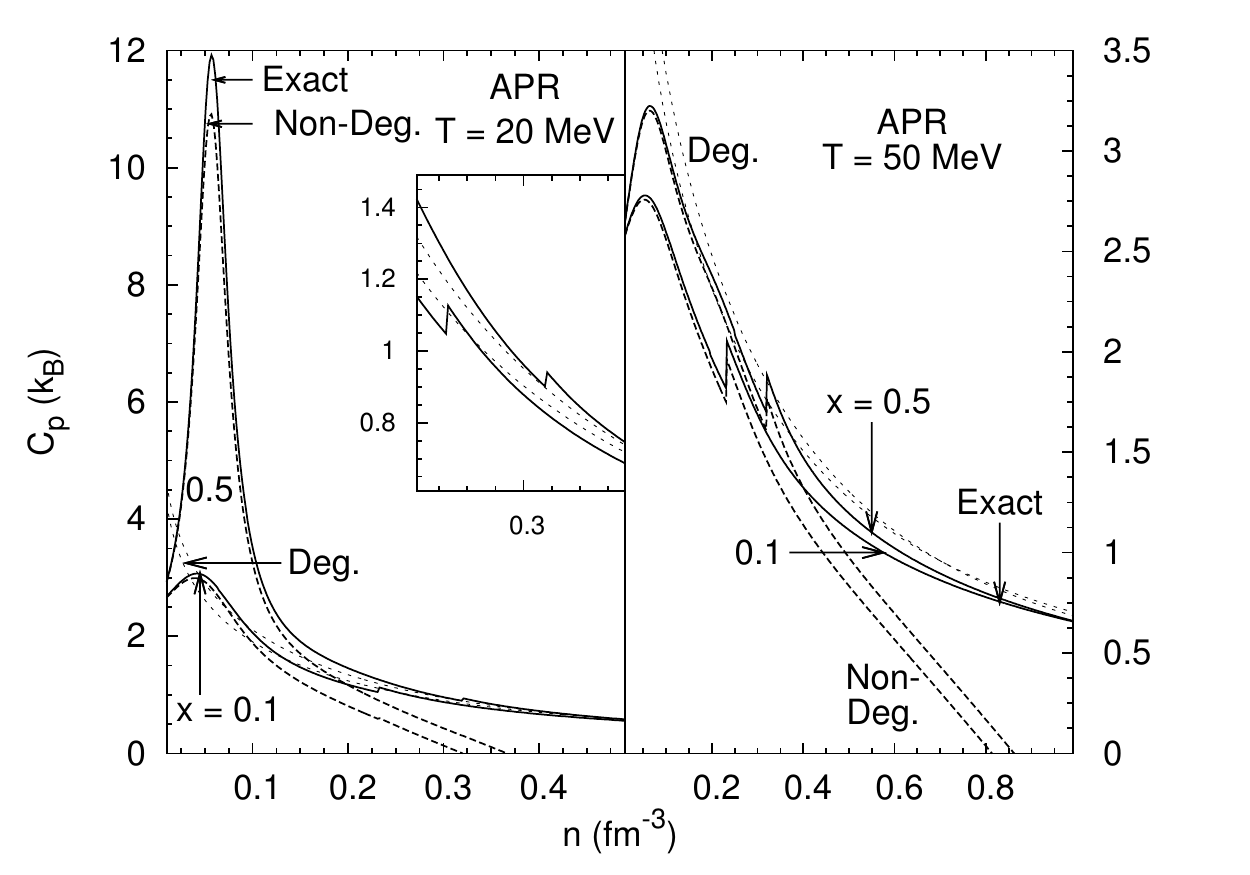}
\end{minipage}
\vskip -0.5cm
\caption{Left: Specific heat at constant volume from Eq. (\ref{cv}) and its limiting cases from Eqs. (\ref{cvdeg}) and (\ref{cvnd}). Right: Specific heat at constant pressure from Eq. (\ref{cp}) 
and its limiting cases (Eqs. (\ref{cpdeg}) and (\ref{cp}) using Eqs. (\ref{cvnd}), (\ref{dpdtnd}), and 
(\ref{dpdnnd})).  }  
\label{APR_CvCp_deg}
\end{figure*}

  In the left panels of Fig. \ref{APR_CvCp_deg}, we present our results of the specific heat at constant volume from Eq. (\ref{cv}) and its limiting cases from Eqs. (\ref{cvdeg}) and (\ref{cvnd}) for the APR model. Results shown are for  
for isospin symmetric ($x=0.5$) and neutron rich matter ($x=0.1$) at temperatures of 20 and 50 MeV, respectively.  The degenerate limit (\ref{cvdeg}) converges with 
the exact result for densities larger than $0.4~{\rm fm}^{-3}$ at $T=20$ MeV and for densities larger than ($1~{\rm fm}^{-3}$) for $T=50$ MeV with little to no dependence on proton fraction.
As expected, the non-degenerate limit holds at low densities, the agreement with the exact result extending to slightly above $n_0$ at the higher temperature. The extent of disagreement   
is somewhat dependent on the proton fraction with neutron rich matter differing from the exact result 
at slightly lower baryon densities than for symmetric matter.

  The right panels of Fig. \ref{APR_CvCp_deg} show the specific heat at constant pressure from Eq. (\ref{cp}) and its limiting cases 
  from Eqs. (\ref{cpdeg}) and (\ref{cp}) using Eqs. (\ref{cvnd}), (\ref{dpdtnd}), and 
(\ref{dpdnnd}) as functions of baryon density. The degenerate limit of 
$C_P$ (Eq. \ref{cpdeg}) provides good agreement with the exact solution at densities greater than  about $n_0$ at $T=20$ MeV. At $T=50$ MeV, the degenerate limit of $C_P$ provides a good
approximation to the exact result at densities greater than $2n_0$, but does not converge until large densities ($n>1~{\rm fm}^{-3}$). The non-degenerate limit of $C_P$ in Eq. (\ref{cp}) using 
Eqs. (\ref{cvnd}), (\ref{dpdtnd}), and (\ref{dpdnnd}) is in agreement with the exact solution up to $n_0$ for $T=20$ MeV and up to almost $2n_0$ for $T=50$ MeV. However, the liquid-gas phase transition pushes the exact 
solution to larger $C_P$ when compared to the effect of this transition on the degenerate limit. Even 
including the effects of the liquid-gas phase transition, the agreement between the 
non-degenerate limit and the exact solution is very good. The rate of converence between the two 
limits and the exact solution is independent of proton fraction.

\subsection*{Results for leptons}

Here we present results of our calculations for the contribution from leptons to the energy $E_e$ per baryon and the electron chemical potential $\mu_e$   as functions
of baryon density $n$. Other state variables follow in a straightforward manner and are summarized in Appendix C. 
We present the exact results obtained using the scheme in Ref. \cite{jel} (Eqs. (\ref{eejel}) and (\ref{muejel}) labelled JEL in figures) and those of the relativistic approach with mass corrections  (Eqs. (\ref{eerel}) and (\ref{muerel})  labelled Rel in figures).  
Comparisons are made  both at $T=0$ and 50 MeV,  and in isospin symmetric and neutron rich matter. 

\begin{figure}[!h]
\includegraphics[width=9cm]{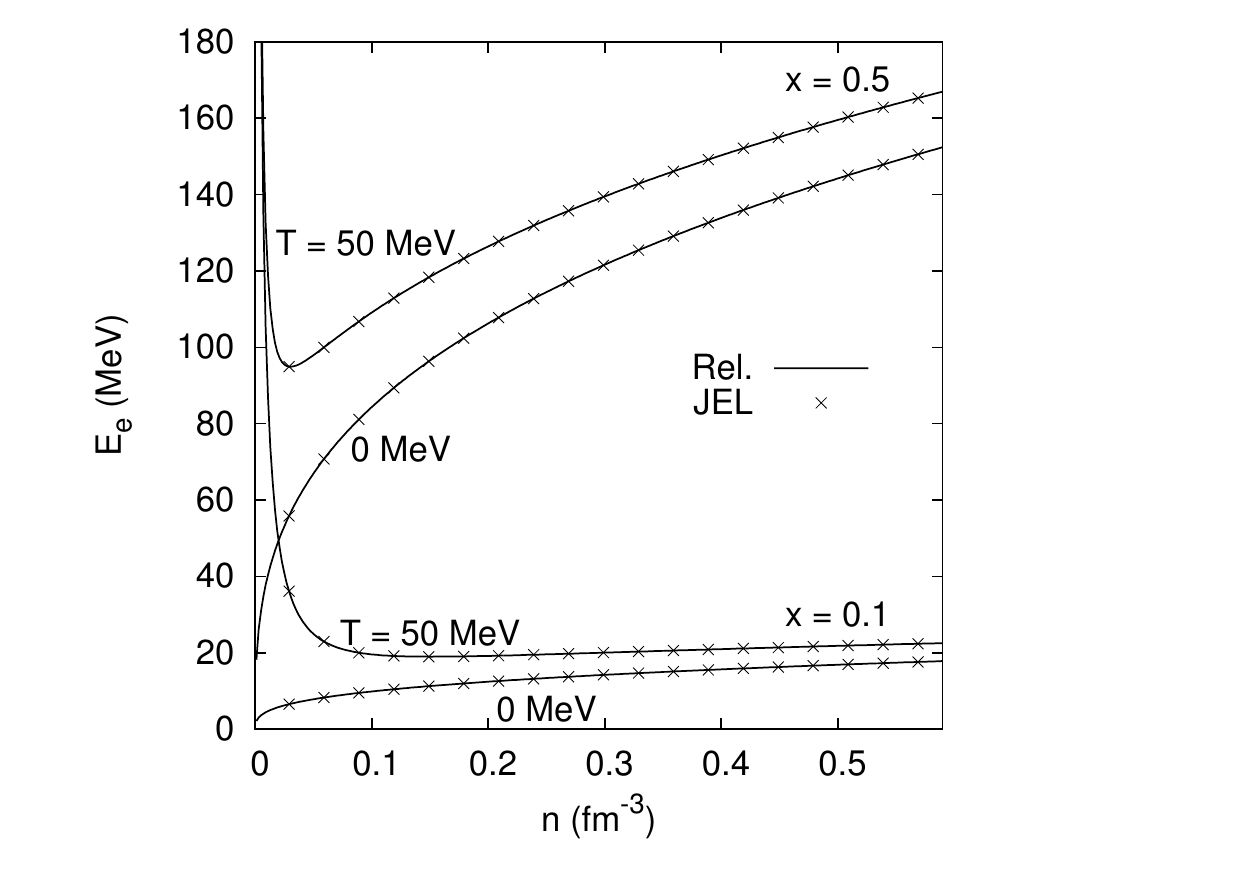}
\caption{Contribution to energy per particle from leptons vs baryon density at the indicated 
temperatures and proton fractions. The solid lines are obtained using the approximate analytical expression Eq. (\ref{eerel}) and the crosses correspond to a full numerical calculation using Eq. (\ref{eejel}). }
\label{LEP_E}
\end{figure}

In Fig. \ref{LEP_E}, we display the energy per baryon  $E_e$ of electrons and positrons as 
a function of baryon density $n$. The two approaches (JEL and Rel) are in complete agreement at all $n$ for both temperatures and for isospin symmetric and asymmetric matter. 
Isospin symmetric matter provides a larger contribution to the energy of leptons than neutron rich matter. This is expected as the system is 
charge neutral, thus the quantity of leptons is dependent on the number of protons. For both temperatures considered, the contribution from positrons is negligible. 

\begin{figure}[!h]
\includegraphics[width=9cm]{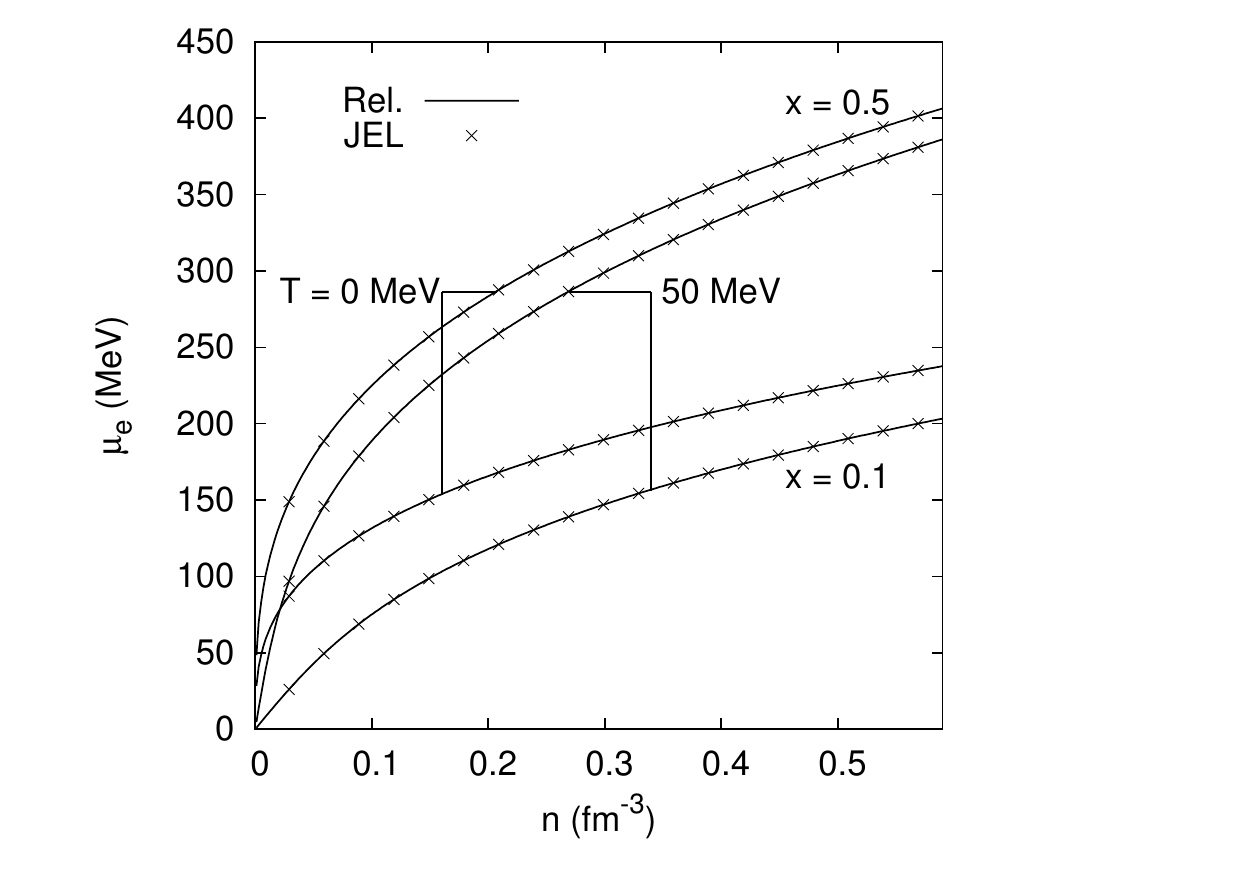}
\caption{Electron chemical potential vs baryon density at the indicated temperatures and proton fractions. 
The solid lines are obtained using the approximate analytical expression Eq. (\ref{muerel}) and the crosses correspond to a full numerical calculation using Eq. (\ref{muejel}). The positron chemical potential has the same magnitude but opposite sign. }
\label{LEP_Mu}
\end{figure}

The electron chemical potential $\mu_e$ is shown as a function of baryon density $n$ in Fig. \ref{LEP_Mu}. As
was the case with the contribution to energy from leptons, the two approaches (JEL and Rel.) are in complete agreement for all baryon densities
at both temperatures, and for both isospin symmetric and asymmetric matter. 

\section{EQUATION OF STATE WITH A PION CONDENSATE}

We have seen in earlier sections that the APR Hamiltonian density incorporates a phase transition involving a neutral pion condensate and that  at the transition density several of the state variables exhibited a jump. In this section, we discuss how an equation of state that satisfies the physical requirements of stability is constructed in the presence of this first-order phase transition.   

Mechanical stability requires that the inequality
\be
\frac{dP}{dn} \geq 0
\ee
is always satisfied. 
However, in the case of APR model, the transition from the LDP to the HDP is accompanied by a decrease in pressure 
pointing to a negative incompressibility. We deal with this unphysical incompressibility by means of a 
Maxwell construction which takes advantage of the thermodynamic equilibrium conditions 
\ba
P_L(n_L) &=& P_H(n_H)  \label{co1}\\
\mu_L(n_L) &=& \mu_H(n_H) \label{co2}
\ea
to establish the mixed-phase region such that 
\be
\frac{dP}{dn} = 0 \,.
\ee
The entropy density is discontinuous across the region (even though it contains none of the terms in the 
Hamiltonian that drive the phase change) thus generating a latent heat
\be
l = T\left[s_H(n_H) - s_L(n_L)\right]
\ee
which signifies a first-order transition. 

The numerical implementation of the coexistence conditions in Eqs. (\ref{co1})-(\ref{co2}) is accomplished as in 
Ref. \cite{LLPR} where the average chemical potential (as electrons contribute similarly in both phases)
\be
\mu = Y_n\mu_n +Y_p(\mu_p+\mu_e) \label{muave}
\ee
and the function
\be
Q = n_t\mu - P
\ee
are expanded in a Taylor series about $n_t$ (the density at which transition from the LDP to HDP occurs) to first and second order respectively, yielding
\ba
\mu(n) &=& \mu(n_t) + (n-n_t)\left.\frac{d\mu}{dn}\right|_{n_t}  \\
Q(n) &=& Q(n_t) + \frac{(n-n_t)^2}{2}\left.\frac{d\mu}{dn}\right|_{n_t} \,.
\ea
Then the LDP and the HDP counterparts are set equal, as stipulated by equilibrium, forming a system of 
two equations the solution of which gives the densities that define the boundary of the coexistence 
region
\ba
n_L &=& n_t + \frac{\mu_H(n_t)-\mu_L(n_t)}
              {\mu_L'(n_t)^{1/2}\left[\mu_L'(n_t)^{1/2}+\mu_H'(n_t)^{1/2}\right]}  \label{nl}\\
n_H &=& n_t + \frac{\mu_L(n_t)-\mu_H(n_t)}
              {\mu_H'(n_t)^{1/2}\left[\mu_L'(n_t)^{1/2}+\mu_H'(n_t)^{1/2}\right]} \label{nh}
\ea
The primes ($\prime$) denote derivatives with respect to the number density $n$.

These results serve as initial guesses which are further improved by adopting an iterative procedure. We 
define the functions 
\ba
f(n_L,n_H) &=& P_L(n_L) - P_H(n_H)  \\
g(n_L,n_H) &=& \mu_L(n_L) - \mu_H(n_H) 
\ea
and expand to first order in Taylor series about the $m^{th}$ iterative solution
\ba
f(n_L^{m+1},n_H^{m+1}) &=& f(n_L^m,n_H^m) 
          + (n_L^{m+1} - n_L^m)\left.\frac{\partial f}{\partial n_L}\right|_{n_L^m}   \nonumber \\
         &+& (n_H^{m+1} - n_H^m)\left.\frac{\partial f}{\partial n_H}\right|_{n_H^m}
\label{iter1} \\
g(n_L^{m+1},n_H^{m+1}) &=& g(n_L^m,n_H^m) 
          + (n_L^{m+1} - n_L^m)\left.\frac{\partial g}{\partial n_L}\right|_{n_L^m}   \nonumber \\
         &+& (n_H^{m+1} - n_H^m)\left.\frac{\partial g}{\partial n_H}\right|_{n_H^m}.
\label{iter2}
\ea
Equations (\ref{iter1}) and (\ref{iter2}) are independent of each other and can thus be used to 
determine $n_L$ and $n_H$. If we assume that  $n_L^{m+1}$ and $n_H^{m+1}$ are the ``true'' solutions 
of the system (i.e. $f(n_L^{m+1},n_H^{m+1}) = g(n_L^{m+1},n_H^{m+1}) = 0$), then 
\ba
n_L^{m+1} &=& n_L^m + \frac{f(n_L^m,n_H^m)\left.\frac{\partial g}{\partial n_H}\right|_{n_H^m}
                    -g(n_L^m,n_H^m)\left.\frac{\partial f}{\partial n_H}\right|_{n_H^m}}
       {\left.\frac{\partial f}{\partial n_H}\right|_{n_H^m}
          \left.\frac{\partial g}{\partial n_L}\right|_{n_L^m}
        -\left.\frac{\partial f}{\partial n_L}\right|_{n_L^m} 
             \left.\frac{\partial g}{\partial n_H}\right|_{n_H^m}}  \nonumber \\ \\
n_H^{m+1} &=& n_H^m + \frac{f(n_L^m,n_H^m)\left.\frac{\partial g}{\partial n_L}\right|_{n_L^m}
                    -g(n_L^m,n_H^m)\left.\frac{\partial f}{\partial n_L}\right|_{n_L^m}}
       {\left.\frac{\partial f}{\partial n_L}\right|_{n_L^m}
          \left.\frac{\partial g}{\partial n_H}\right|_{n_H^m}
        -\left.\frac{\partial f}{\partial n_H}\right|_{n_H^m} 
             \left.\frac{\partial g}{\partial n_L}\right|_{n_L^m}}  \nonumber \\ 
\ea
This process is repeated until the difference $n^{m+1} - n^m$ is less than some prescribed value.

\subsection*{Results}

\begin{figure}[hbt]
\begin{center}
\includegraphics[width=10cm]{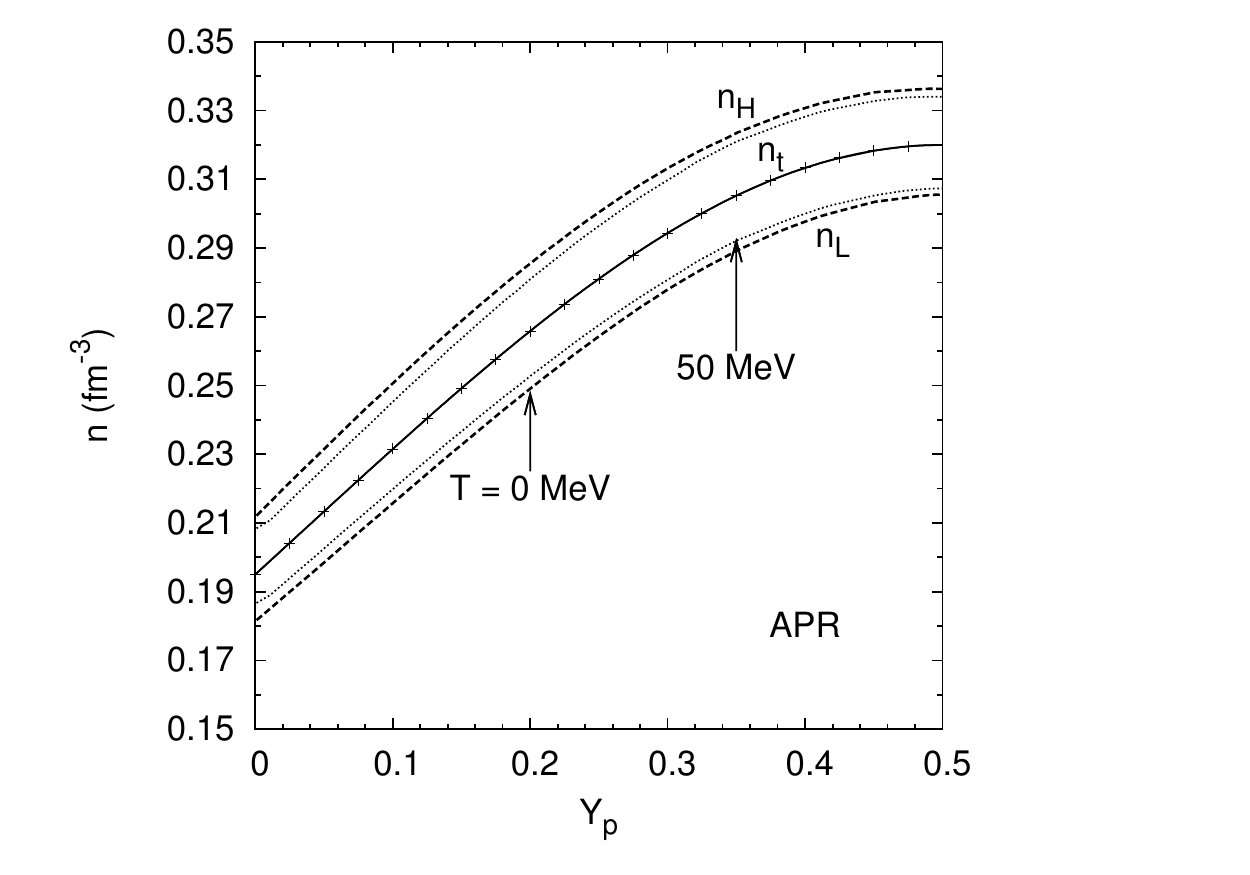}
\end{center}
\vskip -1cm
\caption{The curve labeled $n_t$ shows the trajectory in the $n-Y_p$ plane along which the transition from the LDP to the HDP occurs according to 
Eq.~(\ref{transition}). Results for $n_t$ are a reproduction of those in Fig. 7 of Ref.~\cite{apr}.  
The crosses show results from our polynomial fit in Eq. (\ref{polfit}). Curves labeled $n_L$ and $n_H$ indicate the mixed-phase boundary at zero and 50 MeV temperatures, respectively, determined by a Maxwell construction as described in the text.} 
\label{HDPline_wfit}
\end{figure}

  The transition densities between the LDP and HDP phases from Eq. (\ref{polfit}) are shown by the solid curve (and crosses) in Fig. \ref{HDPline_wfit} 
as a function of proton fraction at zero and 50 MeV, respectively.  In addition, results from the determination of the mixed 
phase region (curves labeled $n_L$ and $n_H$) using a Maxwell construction are presented as a function of proton fraction. The range of baryon densities 
in the mixed phase region has only slight dependence on the proton fraction and temperature. As the neutral pion condensate is mainly driven by density effects in the APR model, effects of temperature in the range considered are small .  

\begin{figure*}[!ht]
\centering
\begin{minipage}[b]{0.49\linewidth}
\centering
\includegraphics[width=9cm]{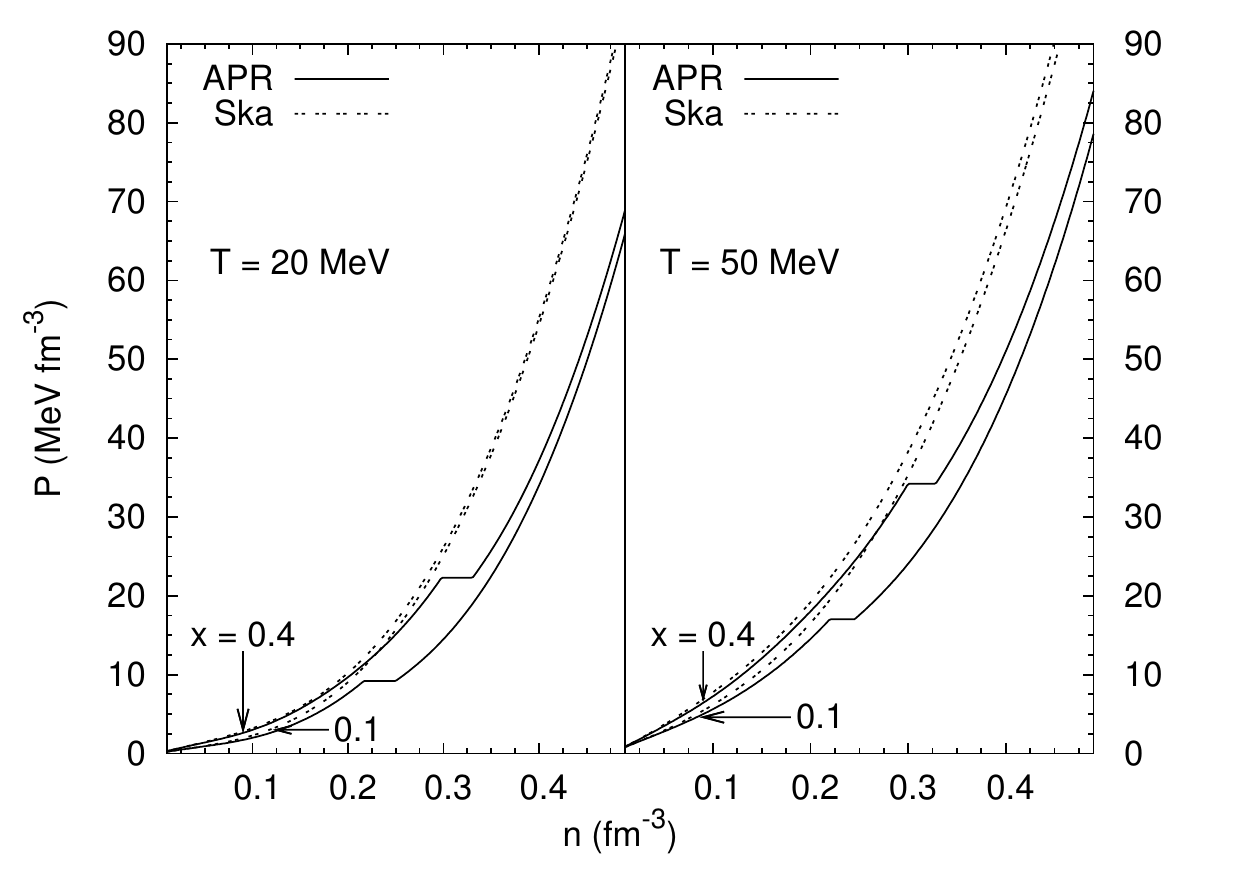}
\end{minipage}
\begin{minipage}[b]{0.49\linewidth}
\centering
\includegraphics[width=9cm]{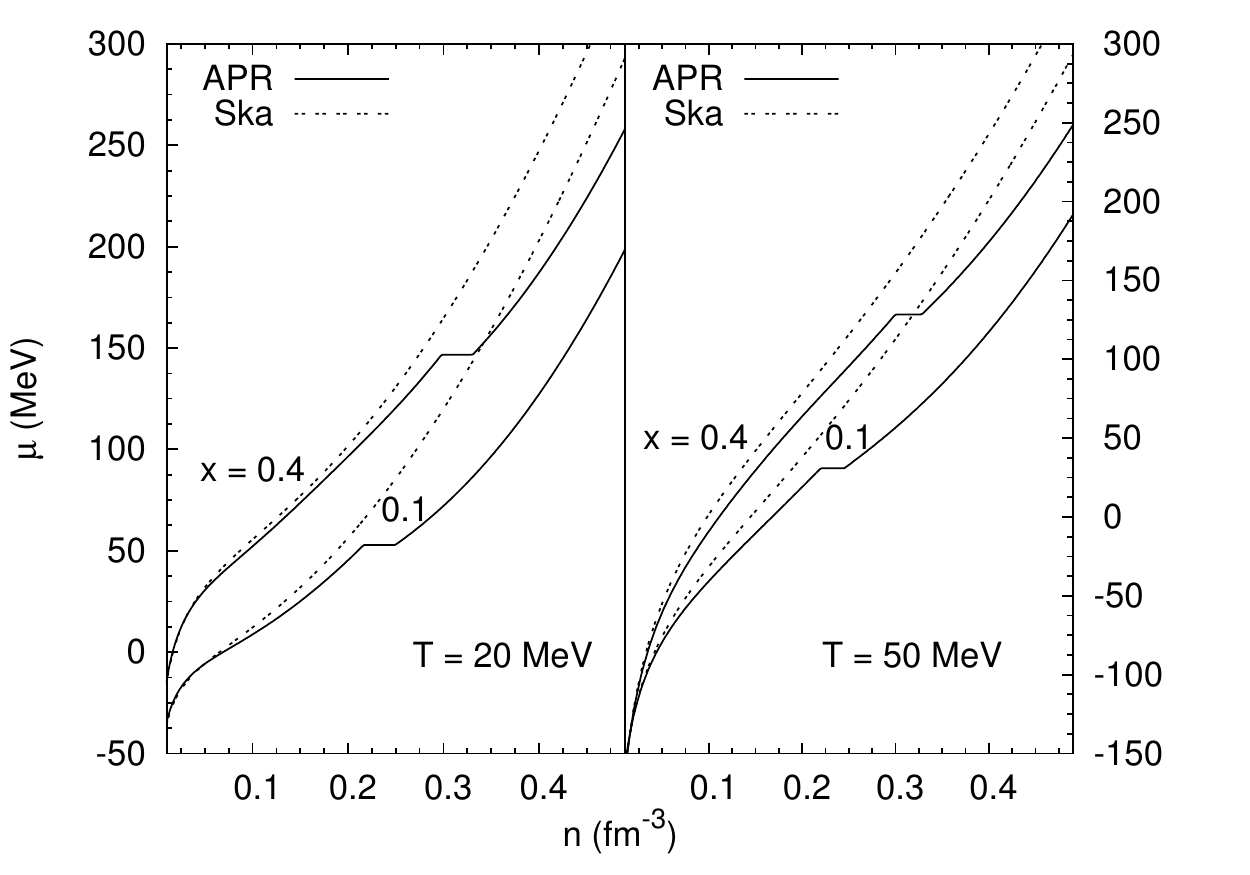}
\end{minipage}
\vskip -0.5cm
\caption{Pressure (left) (Eq. (\ref{pth})) and average chemical potential (right) (Eq. (\ref{muave})) for the APR (solid) and Ska (dashed) models at the indicated proton fractions and temperatures. The flat portions of the APR curves are due to the Maxwell construction for the mixed-phase region, the boundaries of which are given by Eqs. (\ref{nl})-(\ref{nh}).}
\label{APR_MxT_PMu}
\end{figure*}

In Fig. \ref{APR_MxT_PMu}, we show the total pressure (left panels) and the average chemical potential (right panels) as functions of baryon density using a Maxwell construction. Results of our calculations are shown for 
$Y_p=0.1, 0.3,~{\rm and}~ 0.5$, and at $T=20$ and 50 MeV, respectively. The mixed phase region exists in the horizontal portions of the pressure and chemical potential curves. For both P and $\mu$, the abrupt transitions into and out of the mixed phase regions after Maxwell construction are evident. 
\begin{figure}[!h]
\includegraphics[width=9cm]{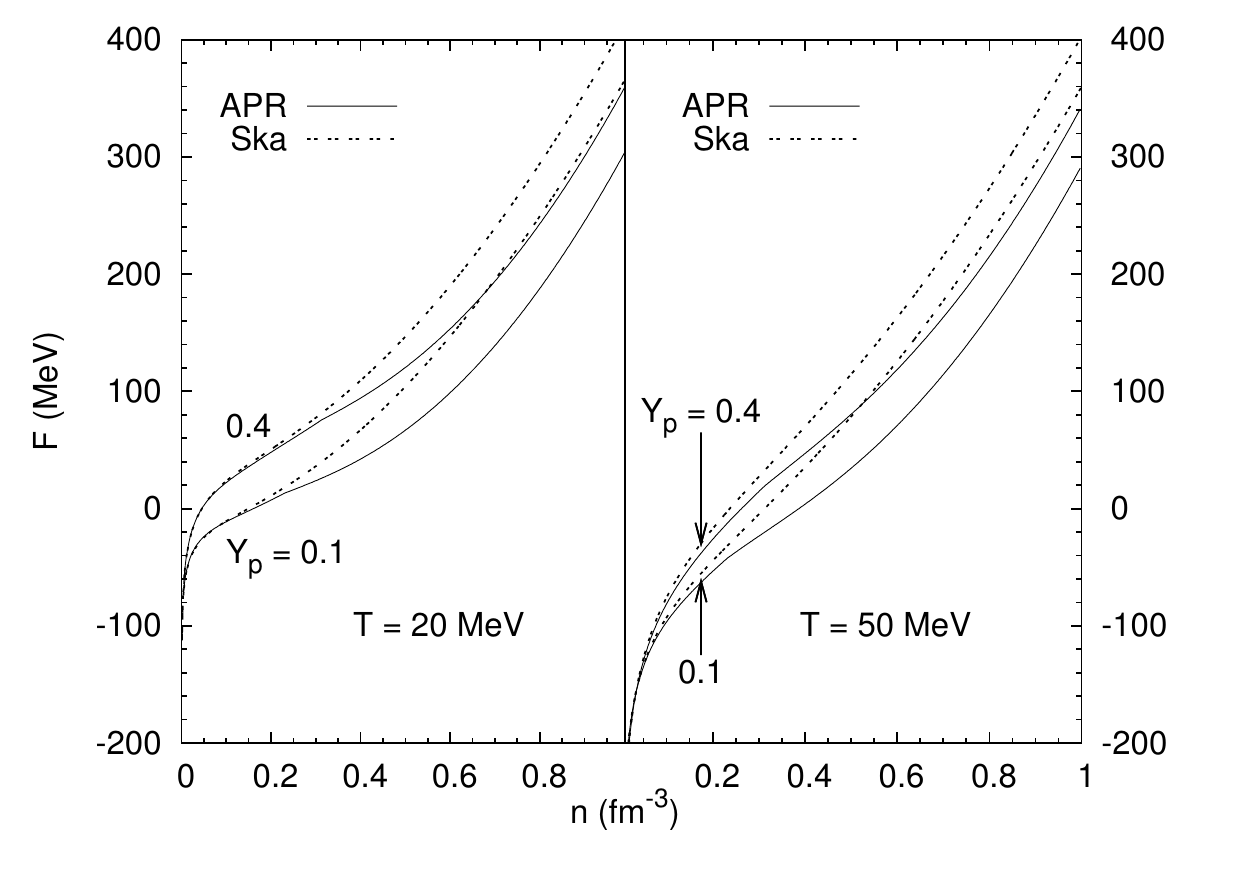}
\caption{Free energy (Eq. (\ref{totEfree})) vs baryon density for the APR (solid) and Ska (dashed) models. Results for $Y_p=0.1~{\rm and}~ 0.4$ at $T=20$ MeV (left) and 50 MeV (right) are presented. The onset of pion condensation appears as a cusp at the appropriate densities.}
\label{APRska_EfreeMeV}
\end{figure}

A comparison between the free energies of APR and Ska is presented in Fig. \ref{APRska_EfreeMeV}. The two models are in close agreement up to $n \sim 0.2$ fm$^{-3}$ but for higher densities, APR is softer due to 
pion condensation. 
\begin{figure}[!h]
\includegraphics[width=9cm]{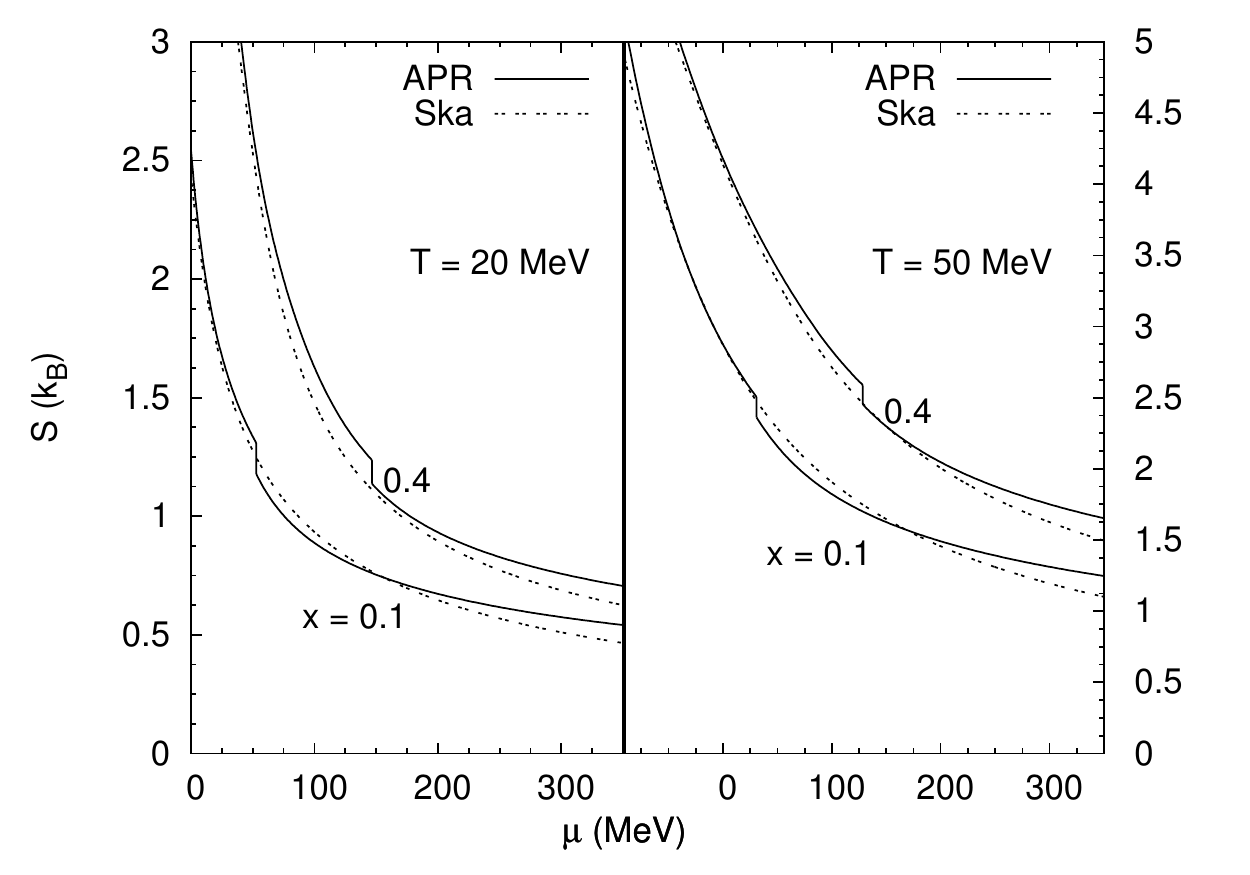}
\caption{Total entropy of baryons (Eq. (\ref{entr})) and leptons (Eq. (\ref{serel})) vs average chemical potential (Eq. (\ref{muave})) at the temperatures and proton fractions shown after Maxwell construction.}
\label{APR_sMu}
\end{figure}

In Fig. \ref{APR_sMu}, the total entropy as a fuction of the average chemical potential is shown for 
representative proton fractions at temperatures of 20 and 50 MeV, respectively. The  vertical portions in these curves show the entropy jumps across the mixed phase region after Maxwell construction.

\begin{figure*}[!ht]
\centering
\begin{minipage}[b]{0.49\linewidth}
\centering
\includegraphics[width=9cm]{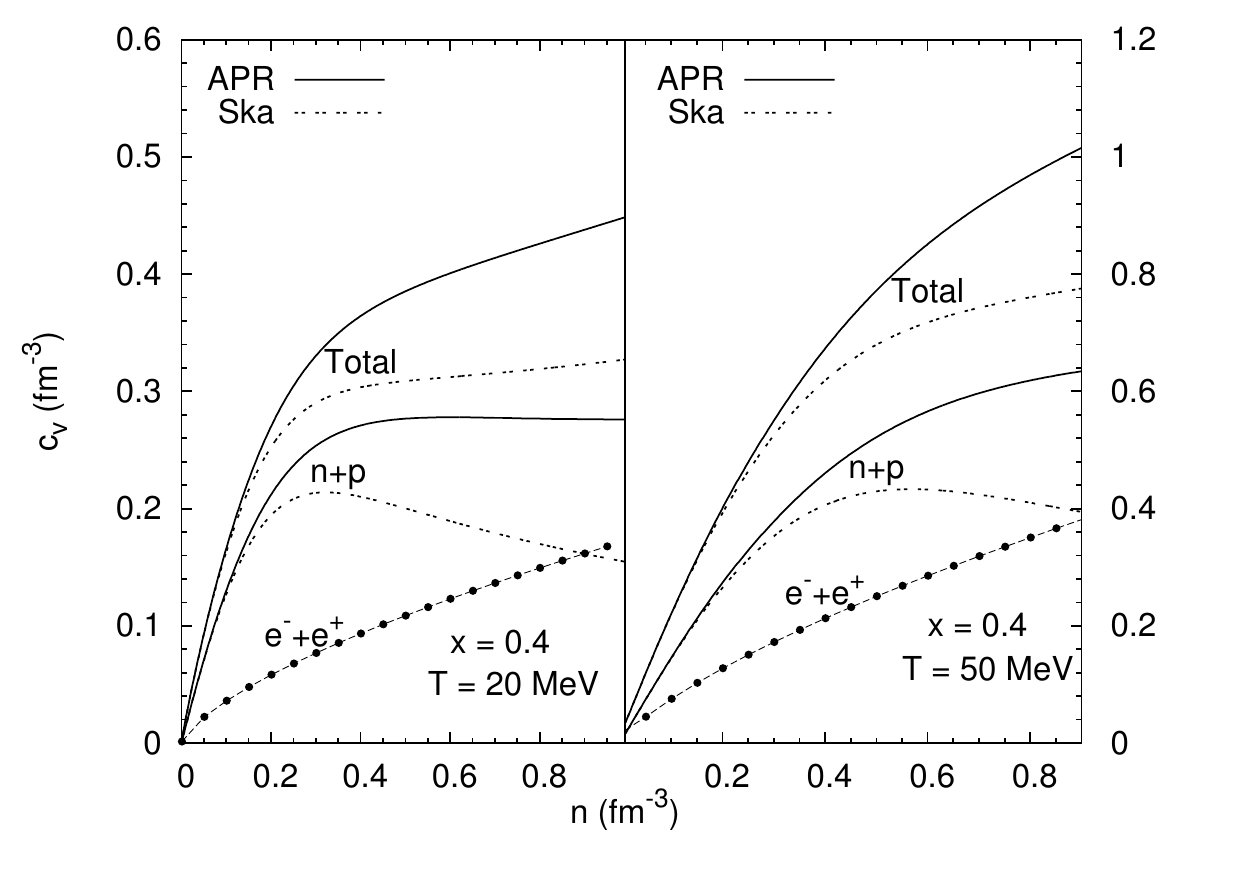}
\end{minipage}
\begin{minipage}[b]{0.49\linewidth}
\centering
\includegraphics[width=9cm]{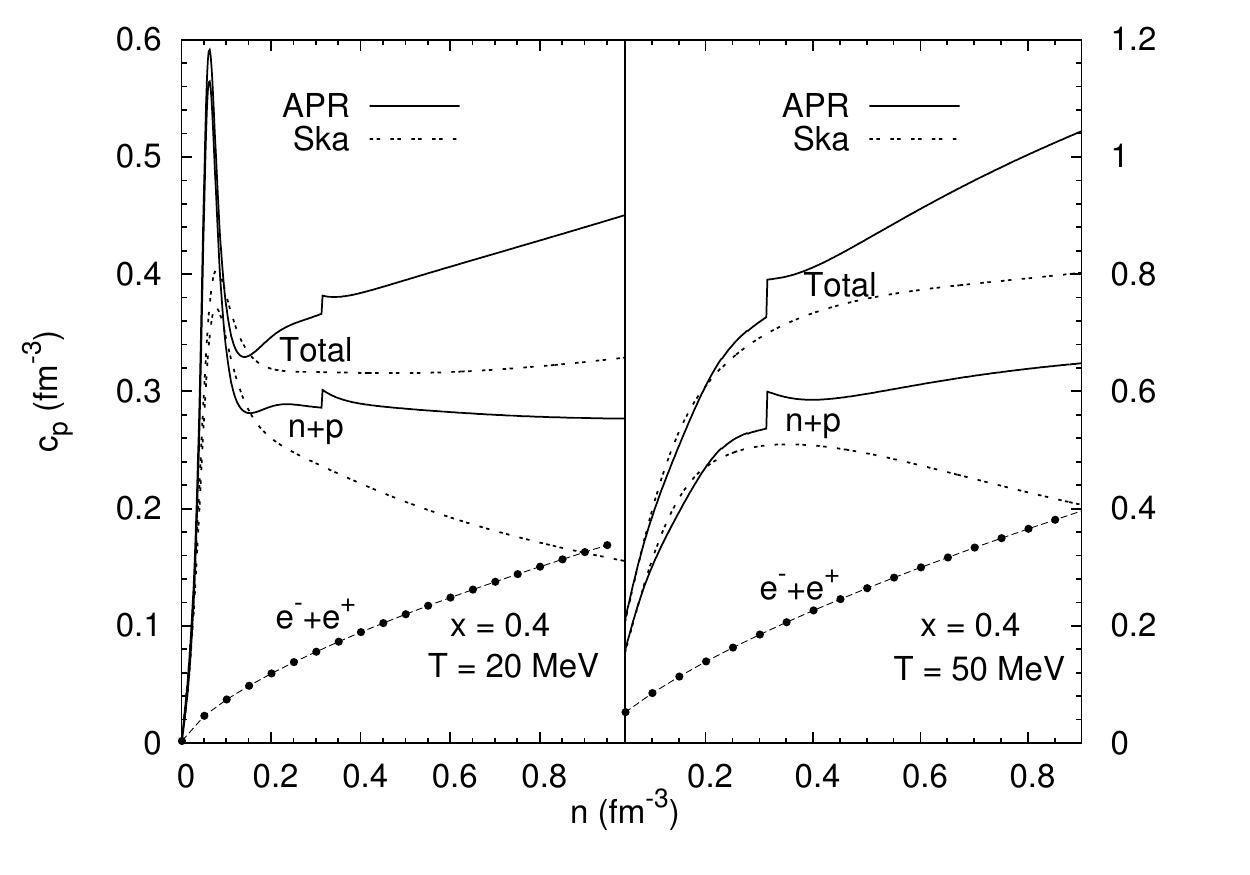}
\end{minipage}
\vskip -0.5cm
\caption{Contributions from nucleonic and leptonic  constituents for the specific heat densities 
at constant volume (left panels) and constant pressure (right panels). The nucleonic contributions  
are from Eqs. (\ref{cv})-(\ref{cp})) for the APR (solid) and Ska (dashed) models and leptonic 
contributions (dotted) are from (Eqs. (\ref{cve})-(\ref{cpe})). }
\label{APRskaJEL_CvCp_cont}
\end{figure*}
\begin{figure}[!ht]
\includegraphics[width=9cm]{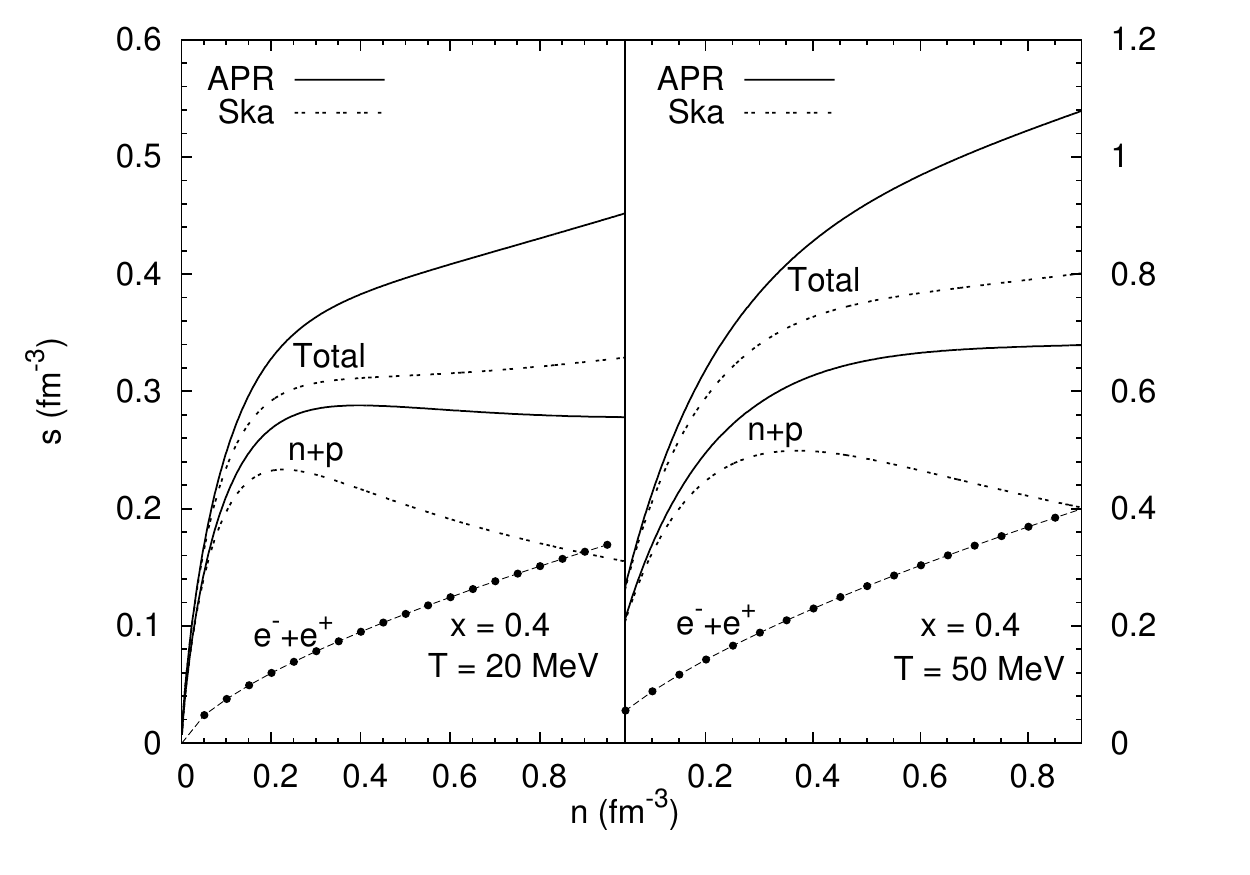}
\caption{Nucleonic  (Eq. (\ref{entr})) and leptonic  (dotted) (Eq. (\ref{se})) contributions for the total entropy density 
for the APR (solid) and Ska (dashed) models at the indicated proton fractions and temperatures. }
\label{APR_sCont}
\end{figure}

In Fig. \ref{APRskaJEL_CvCp_cont}, we present the individual contributions of nucleons and leptons to the total specific heat densities at 
constant volume and pressure. The contribution from leptons was obtained using the JEL scheme (see Appendix D) 
while the nucleonic contribution was calculated by adapting the general results of section V 
(Eqs.(\ref{cv})-(\ref{cp})) to APR and Ska . The two models are in agreement for densities up to $n_0$, 
whereas for larger densities, the specific heat densities of APR are higher (both $c_V$ and $c_P$).  Except for the highest densities shown in these figures, the dominant contributions arise from nucleons. 

The individual contributions of nucleons and leptons to the total entropy density for the APR and Ska models are displayed in Fig. \ref{APR_sCont}.  Note that in the degenerate limit $s\simeq c_V \simeq c_P$. As with the specific heat densities, the largest contributions are from nucleons for densities of relevance in core-collapse supernovae. 

\begin{figure}[!h]
\includegraphics[width=9cm]{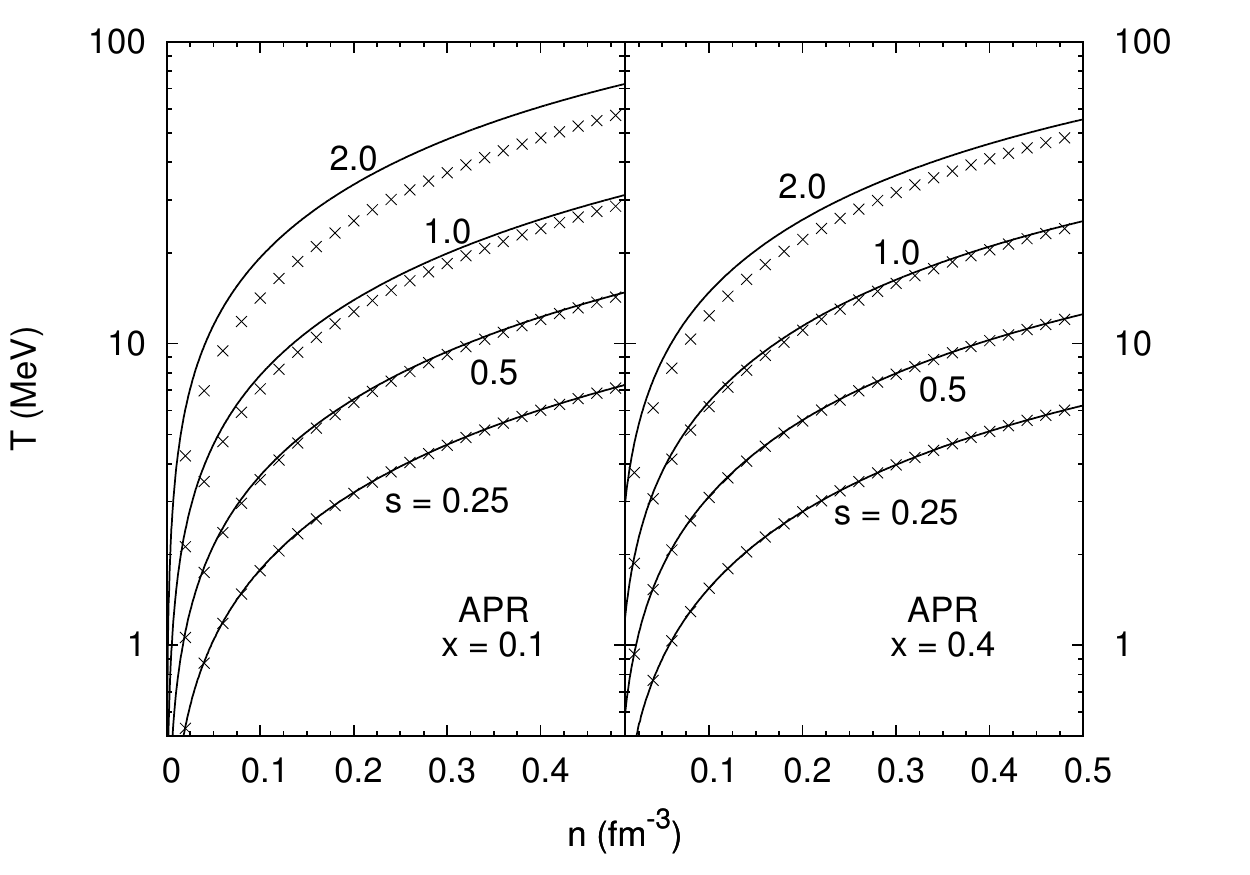}
\caption{Curves of constant entropy in the $(T,n)$-plane for the APR model.  Solid curves show results from exact numerical calculations and the crosses show results from the degenerate limit expression in  Eq. (\ref{T1}) at the indicated proton fractions.}
\label{APR_Tvn_Isent}
\end{figure}
\begin{figure}[!h]
\includegraphics[width=9cm]{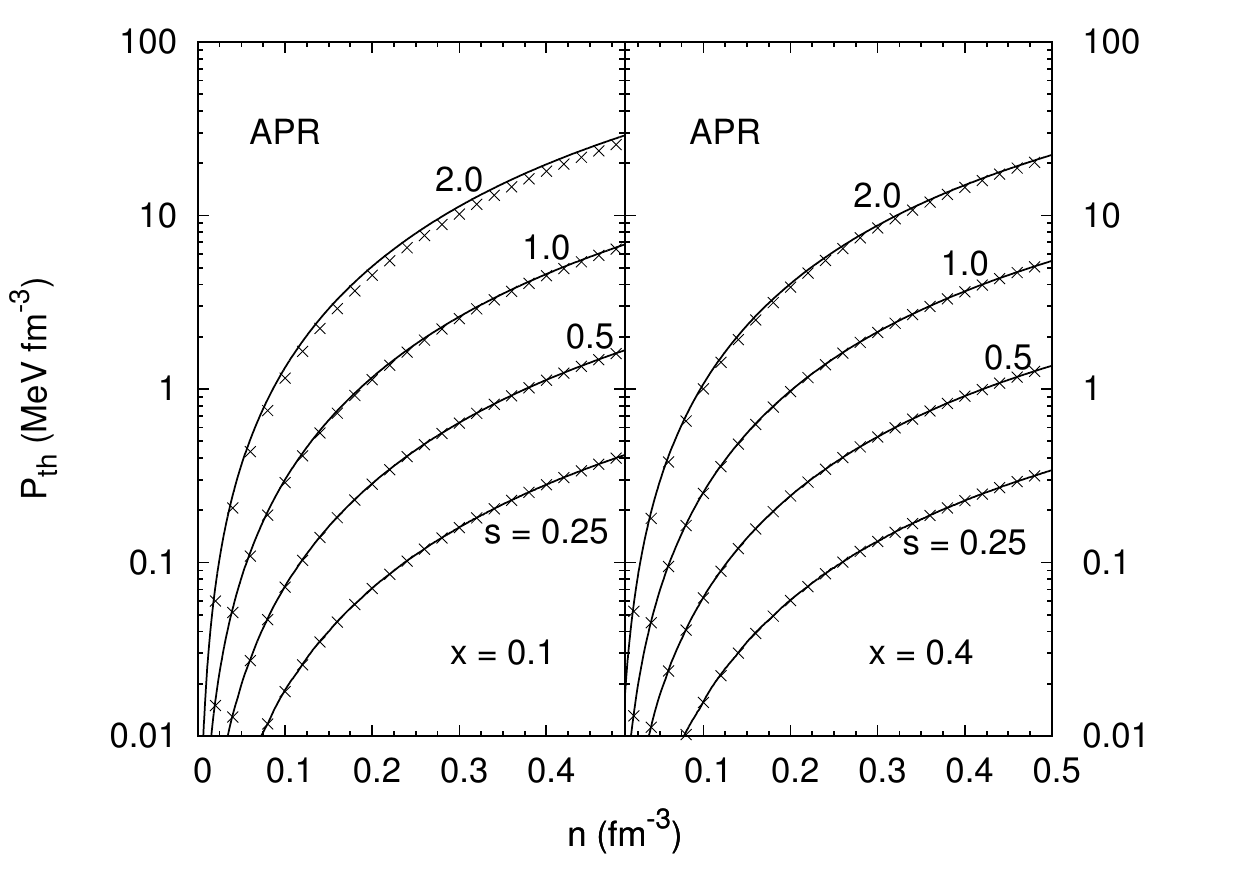}
\caption{Isentropes in the $(P_{th},n)$-plane for the APR model at the indicared proton fractions. Solid curves are from the exact numerical calculations. Results (crosses) from the  degenerate limit expression are from Eq. (\ref{PS}). }
\label{APR_PTHvn_Isent}
\end{figure}
\begin{figure}[!h]
\includegraphics[width=9cm]{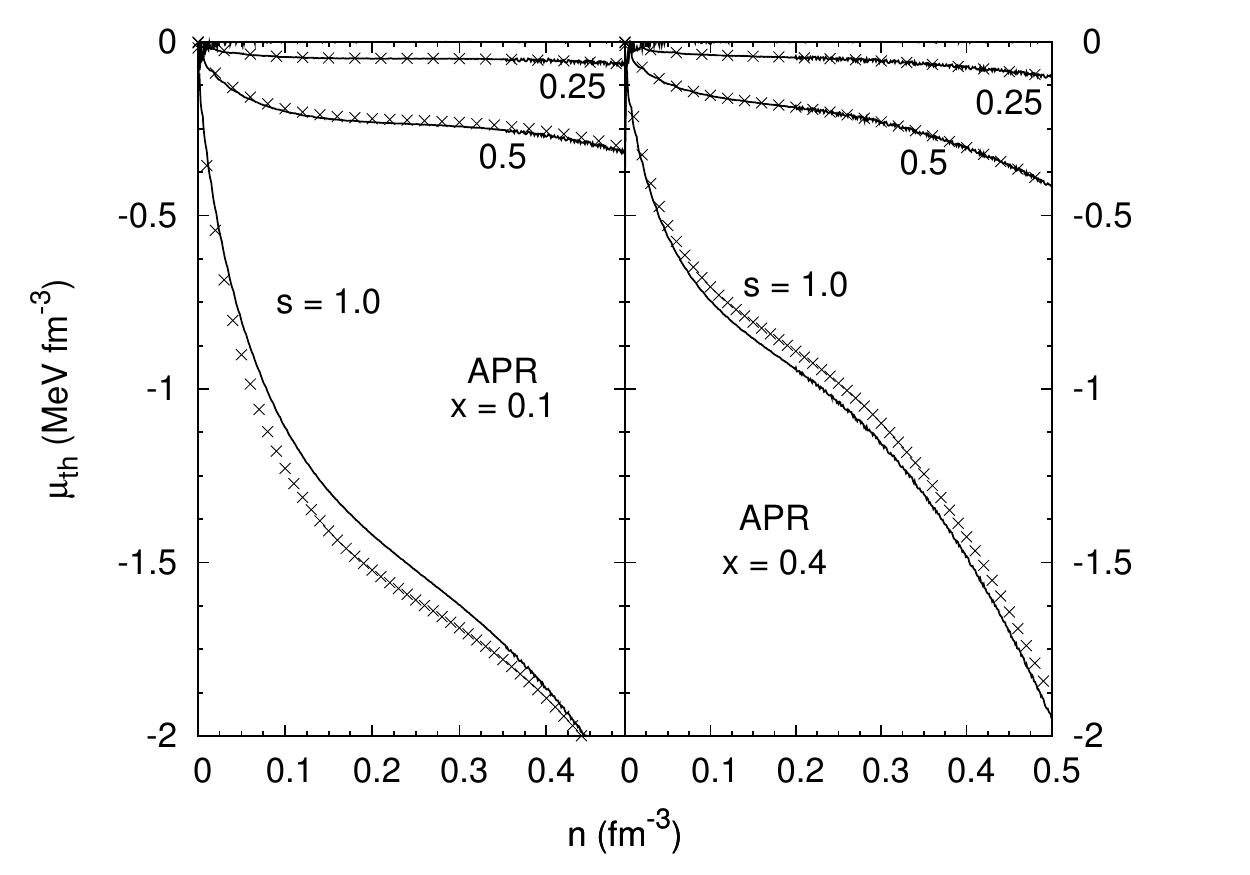}
\caption{Isentropes in the $(\mu_{th},n)$-plane for the APR model. Solid curves  are results from the exact numerical calculations and the crosses are from expressions in the degenerate limit in Eqs. (\ref{muav2})-(\ref{T2})) at the indicated proton fractions.}
\label{APR_MUTHvn_Isent}
\end{figure}

Thermal variables for constant entropy, that is isentropes, often provide valuable guidance to the hydrodynamic evolution of a system, as in ideal hydrodynamics (meaning without viscous terms) the entropy density current is conserved. Ever since the observation by Bethe et al., \cite{BBAL79}, who pointed out that the entropy in supernova evolution is low, a great deal of qualitative understanding has been gained by studying isentropes for the various thermodynamical variables. In view of this, we present some isentropes in what follows. 

Isentropes of the APR model in the $T$-$n$ plane are shown in Fig. \ref{APR_Tvn_Isent}. The crosses in this figure show results from the degenerate limit expression
\be
T = \frac{S}{2[a_nY_n+(a_p+a_e)Y_p]}
\label{T1}
\ee
with excellent agreement for $S\le1$. The level density parameters $a_n$ and $a_p$  above are as in 
Eq. (\ref{levelden}), whereas that for the electrons is $a_e = (\pi^2/2)(E_{Fe}/k_{Fe}^2)$ 
as electrons are relativistic for near nuclear and supra-nuclear densities. 
We have verified that a similarly excellent agreement is obtained for the Ska model (results not shown).

Isentropes of the APR model in the $P_{th}$-$n$ plane are shown in Fig. \ref{APR_PTHvn_Isent} in which the exact numerical results are compared with those in the degenerate limit \cite{pr280}:
\be
P_{th} = \frac{2n}{3\pi^2}~S^2~\frac{\sum_i \frac{Y_i}{T_{Fi}}Q_i}
               {\left(\sum_i\frac{Y_i}{T_{Fi}}\right)^2}    ~;~~~ i=n,p,e.
\label{PS}
\ee
We observe nearly identical results for $S\le2$. For nucleons, $Q_i$ are those from Eq. (\ref{qi}). 
For electrons, $Q_e = 1/2$ and $T_{Fe} = k_{Fe}^2/(2E_{Fe})=\pi^2/(4a_e)$.

Isentropes of the APR model in the $\mu_{th}$-$n$ plane are shown in Fig. \ref{APR_MUTHvn_Isent}. 
To compare the exact results with those from the 
degenerate limit results, it was necessary to expand the expressions for the entropy and
the nucleon thermal chemical potentials to ${\cal O}(T^3)$ and ${\cal O}(T^4)$ respectively:
\ba
S &=& 2T\sum_{i=n,p,e}a_iY_i  
     -\frac{16T^3}{5\pi^2}\sum_{i=n,p,e}a_i^3Y_i    \label{S2nd}\\
\mu_{i=n,p} &=& -T^2\left[\frac{a_i}{3}+\frac{a_in_i}{m_i^*}\frac{dm_i^*}{dn_i}
                                 +\frac{a_jn_j}{m_j^*}\frac{dm_j^*}{dn_i}\right]   \nonumber \\
            &+&  \frac{4T^4}{5\pi^2}\left[-a_i^3 +\frac{3a_i^3n_i}{m_i^*}\frac{dm_i^*}{dn_i}
                        +\frac{3a_j^3n_j}{m_j^*}\frac{dm_j^*}{dn_i}\right]  ~~;~~ i \ne j  \nonumber \\
\label{mi2nd}     
\ea
Then, the average thermal chemical potential is given by
\ba
\mu_{av,th} &=& -T^2\left[\frac{Y_na_n}{3}\left(1+\frac{3n}{m_n^*}\frac{dm_n^*}{dn}\right)\right. \nonumber \\
            &+& \left. \frac{Y_pa_p}{3}\left(1+\frac{3n}{m_p^*}\frac{dm_p^*}{dn}\right)
                          +\frac{2a_eY_p}{3}\right]  \nonumber  \\
        &+&   \frac{4T^4}{5\pi^2}
                         \left[-Y_na_n^3\left(1-\frac{3n}{m_n^*}\frac{dm_n^*}{dn}\right)\right. \nonumber \\
        &-& \left. Y_pa_p^3\left(1-\frac{3n}{m_p^*}\frac{dm_p^*}{dn}\right)\right].
\label{muav2}
\ea
The temperature used in the above expression is obtained from Eq. (\ref{S2nd}) by perturbative inversion:
\be
T = \frac{S}{2\sum a_iY_i}\left[1+\frac{2S^2}{5\pi^2}\frac{\sum a_i^3Y_i}{(\sum a_iY_i)^3}\right]
              ~~;~~ i=n,p,e   
\label{T2}
\ee
At this level of approximation (made necessary by the weak density dependence of the chemical 
potential in the degenerate limit), we get fairly good consistency between the exact and the approximate 
results for $S\le1$.

\section{Conclusions}

Our primary objective in this work has been to build an equation of state 
of supernova matter in the bulk homogeneous phase based on the zero-temperature 
APR Hamiltonian density which has been devised to reproduce the results of 
the microscopic potential model calculations of Akmal and Pandharipande for 
nucleonic matter with varying isospin asymmetry. One of the main features of 
the APR model is that it incorporates a neutral pion condensate at 
supra-nuclear densities found in the calculations of AP for all values of 
proton fraction. Consequently, its high density behavior is somewhat soft in 
its pressure variation, yet it is able to support a neutron star in excess of 
2 M$_{\odot}$ required by recent observations. Our principal contribution in 
this work is the extension of the APR model to finite temperature for use in 
numerical simulations of core-collapse supernovae. In order to provide a 
contrast, we have also calculated the finite temperature properties of a 
model (termed Ska) using an energy density functional stemming from Skyrme 
effective forces. The methods developed in this work are applicable and 
easily adapted to investigate thermal properties of other Skyrme-like 
energy density functionals.

We have studied the behavior of the state variables energy $E$, pressure $P$, 
the neutron and proton chemical potentials $\mu_n$ and $\mu_p$, entropy per 
baryon $S$, the free energy $F$, and the response functions such as the 
compressibility $K$, the inverse susceptibilities $\chi_{ij}$, and specific 
heats $C_V$ and $C_P$ of the APR and the Ska models as functions of the 
temperature $T$, the baryon density $n$, and the proton-to-baryon fraction $x$. 
The two EOS's are quantitatively similar for densities up to $\sim$1.5 $n_0$, 
but differ significantly at higher densities. The cross susceptibilities 
$\chi_{np}$, $\chi_{pn}$ and the ratio $P_c/(n_cT_c)$ evaluated at the critical 
density $n_c$ of the liquid-gas phase transition are the only exceptions to 
the above general observation.

We have also calculated several properties of isospin-symmetric matter at 
the saturation density and compared with experimental results, although 
the latter, in some cases, are associated with large uncertainties. 
Considerable attention has been paid to the symmetry energy $S_2$ as a 
function of the density and the temperature. Our results reveal a weak 
dependence on the temperature which leads to the conclusion that $S_2$ is 
determined mainly by the density dependent effective mass. It is also 
evident herein that, in the case of matter with a phase transition, the
quantities $S_2$, and $F_{\rm sym}=F(x=0)-F(x=1/2)$  are fundamentally different.

We also find that the density jump across the coexistence region of the 
LDP to the HDP transition of the APR model depends weakly on the 
temperature, the proton fraction, and the leptonic contributions.

That thermal effects are, in general, less pronounced in degenerate matter 
is expected as this is the regime where $T/T_F \ll 1$; i.e., temperature 
effects are overwhelmed by density effects. However, when looking at the 
thermal part of any given thermodynamic quantity, the aforementioned 
density effects are entirely determined by the effective masses. As we 
have seen, the density dependence of the effective mass for nucleons
interacting via Skyrme or Skyrme-like forces is responsible for several 
degenerate limit effects not encountered in a free gas. In particular, 
as a function of density the thermal pressure $P_{th}$ flattens (whereas in 
a free gas it increases monotonically), $\mu_{i,th}$ become positive (strictly 
negative in the free case), and $S_{2,th}$ becomes negative (always positive 
for a free gas). The results of Eqs. (\ref{pdeg}),(\ref{mudeg}), and 
(\ref{s2deg}) in which terms involving the derivatives of the effective mass 
with respect to the density encode effects of momentum dependent interactions 
and modify the expressions from what would have been their free forms. The role 
of the effective mass in the non-degenerate limit, although present, is minimal for most of the 
state variables. Intriguingly, our results indicate that, for the temperatures (up to 30 MeV) and proton 
fractions (0.38-0.42) of most relevance to supernova evolution, densities in the vicinity of the
nuclear saturation density can be considered neither degenerate nor non-degenerate. The quantitative results
presented in this work (particularly, the neutron and proton chemical potentials) can be used to advantage to
determine the rates of electroweak reactions such as electron capture and neutrino-matter interactions in hot dense matter. 

Based on the APR model, work on the inhomogeneous phase at subnuclear densities where nuclei 
coexist with leptons, nucleons, nuclei, and light nuclear clusters as well as pasta-like configurations is 
in progress and will be reported in a separate work.

\section*{ACKNOWLEDGEMENTS}

We have benefitted greatly from the unpublished Ph.D. thesis of Matthew Carmell from Stony Brook University. Computational help from Kenneth Moore at Ohio University during the initial stages of this work is gratefully acknowledged.   
This work was supported in part by the 
US DOE under Grants  No. DE-FG02-87ER-40317 (for C.C. and J.M.L) and 
No. DE-FG02-93ER-40756 (for B.M. and M.P.).

\appendix

\section{SINGLE-PARTICLE SPECTRA}
Here we provide a derivation of the expression in Eq. (\ref{spectra}) which
is a direct consequence of the fact that the expectation value of the Hamiltonian is stationary 
with respect to variations of its eigenstates ~\cite{atomic, vb}:
\begin{eqnarray}
\frac{\delta}{\delta\phi_k}\left(E-\sum_k \epsilon_k \int |\phi_k(\vec r)|^2d^3r\right) & = & 0,
\label{variation}
\end{eqnarray}
where $\epsilon_k$ is the eigenvalue corresponding to the eigenstate $\phi_k$, $E=\langle H\rangle$ 
and $k$ is the set of all relevant quantum numbers.

For a many-body Hamiltonian, $\phi_k$ are the single particle states making up the Slater determinant, 
and therefore the set of all $\epsilon_k$ is the single-particle energy spectrum of the Hamiltonian.

Consider now a nucleonic Hamiltonian density \\ $\mathcal{H} = \mathcal{H}(\tau_i,n_i)$, where
\begin{eqnarray}
\tau_i(\vec r) & = & \sum_{k,s} |\nabla\phi_k(\vec r,s,i)|^2  \\
n_i(\vec r) & = & \sum_{k,s} |\phi_k(\vec r,s,i)|^2
\end{eqnarray}
are the kinetic energy density and number density respectively, of the nucleon species with isospin $i$.

The variation of the number density with respect to $\phi$ is 
\begin{eqnarray}
\delta n_i & = & \sum_{k,s}[\delta\phi^*(\vec r,s,i)\phi(\vec r,s,i)+\phi^*(\vec r,s,i)\delta\phi(\vec r,s,i)].
\nonumber \\
\label{delni}
\end{eqnarray}

Imposing time-translational invariance leads to  
\begin{eqnarray}
\phi(\vec r,s,i) & = & i^{2s}\phi^*(\vec r,-s,i)  \\
\mbox{and}~~~ \delta\phi(\vec r,s,i) & = & i^{2s}\delta\phi^*(\vec r,-s,i).
\label{delphi}
\end{eqnarray}
Therefore,
\begin{eqnarray}
\delta n_i & = & \sum_{k,s}[\delta\phi^*\phi+(-1)\phi(-s)\times(-1)\delta\phi^*(-s)] \nonumber \\
           & = & \sum_{k,s}[\delta\phi^*\phi+\delta\phi^*(-s)\phi(-s)] \nonumber \\
           & = & 2\sum_{k,s}\delta\phi^*\phi,
\end{eqnarray}
as the sum is over all spins.Similarly,
\begin{eqnarray}
\delta\tau_i & = & 2\sum_{k,s}\nabla\delta\phi_k^*\cdot\nabla\phi_k.  
\end{eqnarray}
Furthermore,
\begin{eqnarray}
E & = & \sum_i\int d^3r~\mathcal{H}(\tau_i,n_i).
\end{eqnarray}
Combining this with (\ref{delni}) and (\ref{delphi}) implies
\begin{eqnarray}
\delta E & = & \sum_i\int d^3r \left[\frac{\partial \mathcal{H}}{\partial \tau_i}\delta\tau_i
                 + \frac{\partial \mathcal{H}}{n_i}\delta n_i\right] \nonumber \\
         & = & \int d^3r \sum_i\left[\frac{\partial \mathcal{H}}{\partial \tau_i}
                                      (2\sum_{k,s}\nabla\phi_k^*\cdot\nabla\phi_k)
                            + \frac{\partial \mathcal{H}}{n_i}(2\sum_{k,s}\delta\phi^*_k\phi_k)\right]
                       \nonumber \\ 
         & = & \int d^3r \sum_{k,s}\left[2\delta\phi_k^*
                 \sum_i\left(-\nabla \frac{\partial \mathcal{H}}{\partial \tau_i}\nabla
                    +\frac{\partial \mathcal{H}}{n_i}\right)\phi_k\right].
\label{dele}
\end{eqnarray}
The minus sign is a consequence of the anti-hermiticity of the $\nabla$ operator: 
$\langle\nabla\phi| = \langle\phi|\nabla^{\dag} = \langle\phi|(-\nabla)$.

\noindent Finally, by inserting (\ref{dele}) into (\ref{variation}) we get
\begin{eqnarray}
0 & = & \int d^3r \sum_{k,s}2\delta\phi_k^*\left[\sum_i\left(
    -\nabla\frac{\partial\mathcal{H}}{\partial\tau_i}\nabla
     +\frac{\partial\mathcal{H}}{\partial n_i}\right)\right]\phi_k \nonumber  \\
  && \hspace{40pt}  -\int d^3r \sum_{k,s}2\delta\phi_k^*\epsilon_k\phi_k   \nonumber \\
  & = & \int d^3r \sum_{k,s}2\delta\phi_k^*\left[\sum_i\left(
    -\nabla\frac{\partial\mathcal{H}}{\partial\tau_i}\nabla+\frac{\partial\mathcal{H}}{\partial n_i}\right)
     -\epsilon_k\right]\phi_k  \nonumber \\
  & \Rightarrow & \sum_i\left(
    -\nabla\frac{\partial\mathcal{H}}{\partial\tau_i}\nabla+\frac{\partial\mathcal{H}}{\partial n_i}\right)
     -\epsilon_k\ = 0   \nonumber \\
  & \Rightarrow & 
    -\nabla\frac{\partial\mathcal{H}}{\partial\tau_i}\nabla+\frac{\partial\mathcal{H}}{\partial n_i}
     -\epsilon_{ki}\ = 0 .
\end{eqnarray}
Thus in momentum space, 
\be
k_i^2\frac{\partial\mathcal{H}}{\partial\tau_i}+\frac{\partial\mathcal{H}}{\partial n_i}= \epsilon_{ki}.
\ee

\section{APR STATE VARIABLES}

In this appendix we summarize results pertaining to the zero temperature state variables of APR. 
Combining the density-dependent parts (see below) of these, with the appropriate thermal expressions from sections VI and VII yields the corresponding expressions at finite temperature. 
It is convenient to write $\mathcal{H}_{APR}$ as the sum of a kinetic part $\mathcal{H}_k$, 
a part consisting of the momentum-dependent interactions $\mathcal{H}_m$, and a density-dependent 
interactions part $\mathcal{H}_d$:
\be
\mathcal{H}_{APR} = \mathcal{H}_k + \mathcal{H}_m + \mathcal{H}_d   
\ee
where
\ba
\mathcal{H}_k &=& \frac{\hbar^2}{2m}(\tau_n+\tau_p)  \\
\mathcal{H}_m &=& (p_3+(1-x)p_5)ne^{-p_4 n}\tau_n \nonumber \\
              &+& (p_3+xp_5)ne^{-p_4 n}\tau_p  \\
\mathcal{H}_d &=&  g_1(n)[1-(1-2x)^2)]+g_2(n)(1-2x)^2 
\ea
Furthermore, the following quantities are necessary:
\ba
\delta g_1 &=& g_{1H}-g_{1L}  \nonumber \\
           &=& -n^2\left[p_{17}(n-p_{19})+p_{21}(n-p_{19})^2 \right] e^{p_{18}(n-p_{19})}  \nonumber \\ \\
\delta g_2 &=& g_{2H}-g_{2L}  \nonumber  \\
          &=& -n^2\left[p_{15}(n-p_{20})+p_{14}(n-p_{20})^2 \right] e^{p_{16}(n-p_{20})}  \nonumber \\ \\
f_{1L} &=& \frac{dg_{1L}}{dn} -\frac{2g_{1L}}{n} \nonumber \\
        &=& -n^2\left[p_2+2p_6n  \right. \nonumber  \\
        && \left. ~~~~~+(p_{11}-2p_9^2p_{10}n-2p_9^2p_{11}n^2)e^{-p_9^2n^2}\right]  \\
f_{1H} &=& f_{1L}+\delta f_1  \\
\delta f_1 &=& \left[2p_{19}(p_{17}-p_{19}p_{21})n \right. \nonumber  \\
           && +\left\{3(2p_{19}p_{21}-p_{17})+p_{18}p_{19}(p_{17}-p_{19}p_{21})\right\}n^2 \nonumber \\
           && +\left\{p_{18}(2p_{19}p_{21}-p_{17})-4p_{21}\right\}n^3  \nonumber \\
           && \left. -p_{18}p_{21}n^4\right]e^{p_{18}(n-p_{19})}   
\ea
\ba   
h_{1L} &=& \frac{df_{1L}}{dn}-\frac{2f_{1L}}{n}  \nonumber  \\
             &=& -n^2\left[2p_6-2p_9^2(p_{10}+3p_{11}n  \right.  \nonumber \\
             && \left. ~~~~~~~~~~~~~~~~~~~~-2p_9^2p_{10}n^2-2p_9^2p_{11}n^3)e^{-p_9^2n^2}\right] 
             \nonumber \\ \\
h_{1H} &=& h_{1L}+\delta h_1  \\
\delta h_1 &=& \left[2p_{19}(p_{17}-p_{19}p_{21}) \right. \nonumber  \\
           && +\left\{6(2p_{19}p_{21}-p_{17})+4p_{18}p_{19}(p_{17}-p_{19}p_{21})\right\}n \nonumber \\
           && +\left\{6p_{18}(2p_{19}p_{21}-p_{17}) \right.   \nonumber \\
           && \left.~~+p_{18}^2p_{19}(p_{17}-p_{19}p_{21})-12p_{21}\right\}n^2  \nonumber \\
           && +\left\{p_{18}^2(2p_{19}p_{21}-p_{17})-8p_{18}p_{21}\right\}n^3 \nonumber \\
           && \left. -p_{18}^2p_{21}n^4\right]e^{p_{18}(n-p_{19})}   
\ea
\ba
w_{1L} &=& \frac{dh_{1L}}{dn}-\frac{2h_{1L}}{n} \nonumber  \\
         &=& -n^2\left(-3p_{11}+6p_9^2p_{10}n+12p_9^2p_{11}n^2 \right.   \nonumber \\
         && \left. -4p_9^4p_{10}n^3-4p_9^4p_{11}n^4\right)2p_9^2e^{-p_9^2n^2}  \\
w_{1H} &=& w_{1L}-\delta w_1 \\
\delta w_1 &=& \left[6\left\{(2p_{19}p_{21}-p_{17})+p_{18}p_{19}(p_{17}-p_{19}p_{21})\right\} \right.
           \nonumber \\
           && +\left\{18p_{18}(2p_{19}p_{21}-p_{17}) \right. \nonumber \\
           && ~~\left.+6p_{18}^2p_{19}(p_{17}-p_{19}p_{21})-24p_{21}\right\}n \nonumber \\
           && +\left\{9p_{18}^2(2p_{19}p_{21}-p_{17}) \right.  \nonumber \\
           && ~~\left. +p_{18}^3p_{19}(p_{17}-p_{19}p_{21})-36p_{18}p_{21}\right\}n^2  \nonumber \\
           && +\left\{p_{18}^3(2p_{19}p_{21}-p_{17})-12p_{18}^2p_{21}\right\}n^3 \nonumber \\
           && \left. -p_{18}^3p_{21}n^4\right]e^{p_{18}(n-p_{19})}   
\ea
\ba 
f_{2L} &=& \frac{dg_{2L}}{dn} -\frac{2g_{2L}}{n} \nonumber \\
       &=& -n^2\left(-\frac{p_{12}}{n^2}+p_8-2p_9^2p_{13}ne^{-p_9^2n^2}\right)  \\
f_{2H} &=& f_{2L}+\delta f_2  \\
\delta f_2 &=& \left[2p_{20}(p_{15}-p_{20}p_{14})n \right. \nonumber  \\
           && +\left\{3(2p_{20}p_{14}-p_{15})+p_{16}p_{20}(p_{15}-p_{20}p_{14})\right\}n^2 \nonumber \\
           && +\left\{p_{16}(2p_{20}p_{14}-p_{15})-4p_{14}\right\}n^3  \nonumber \\
           && \left. -p_{16}p_{14}n^4\right]e^{p_{16}(n-p_{20})}   
\ea
\ba  
h_{2L} &=& \frac{dh_{2L}}{dn} -\frac{2h_{2L}}{n} \nonumber \\
       &=& -n^2\left[\frac{2p_{12}}{n^3}-2p_{19}^2p_{13}(1-2p_9^2n)e^{-p_9^2n^2}\right]  \\
h_{2H} &=& h_{2H}+\delta h_2  \\
\delta h_2 &=& \left[2p_{20}(p_{15}-p_{20}p_{14}) \right. \nonumber  \\
           && +\left\{6(2p_{20}p_{14}-p_{15})+4p_{16}p_{20}(p_{15}-p_{20}p_{14})\right\}n \nonumber \\
           && +\left\{6p_{16}(2p_{20}p_{14}-p_{15}) \right.   \nonumber \\
           && \left.~~+p_{16}^2p_{20}(p_{15}-p_{20}p_{14})-12p_{14}\right\}n^2  \nonumber \\
           && +\left\{p_{16}^2(2p_{20}p_{14}-p_{15})-8p_{16}p_{14}\right\}n^3 \nonumber \\
           && \left. -p_{16}^2p_{14}n^4\right]e^{p_{16}(n-p_{20})}   
\ea
\ba   
w_{2L} &=& \frac{dw_{2L}}{dn} -\frac{2w_{2L}}{n} \nonumber \\
       &=& -n^2\left[-\frac{6p_{12}}{n^4}+4p_9^4p_{13}(1+n-2p_9^2n^2)e^{-p_9^2n^2}\right]  \nonumber \\ \\
w_{2H} &=& w_{2L}+\delta w_2  \\
\delta w_2 &=& \left[6\left\{(2p_{20}p_{14}-p_{15})+p_{16}p_{20}(p_{15}-p_{20}p_{14})\right\} \right. 
           \nonumber  \\
           && +\left\{18p_{16}(2p_{20}p_{14}-p_{15}) \right. \nonumber \\
           && ~~\left.+6p_{16}^2p_{20}(p_{15}-p_{20}p_{14})-24p_{14}\right\}n \nonumber \\
           && +\left\{9p_{16}^2(2p_{20}p_{14}-p_{15}) \right.  \nonumber \\
           && ~~\left. +p_{16}^3p_{20}(p_{15}-p_{20}p_{14})-36p_{16}p_{14}\right\}n^2  \nonumber \\
           && +\left\{p_{16}^3(2p_{20}p_{14}-p_{15})-12p_{16}^2p_{14}\right\}n^3 \nonumber \\
           && \left. -p_{16}^3p_{14}n^4\right]e^{p_{16}(n-p_{20})}      
\ea
The subscripts $L$ and $H$ imply the low density and the high density phase respectively. \\
Expressions for the state variables are collected below.

\subsection*{Energy per particle}
\begin{eqnarray}
\frac{E}{A} &=& \frac{E_k}{A} + \frac{E_m}{A} + \frac{E_d}{A} = \frac{\mathcal{H}}{n}  
\label{EA1}  \\
\frac{E_k}{A} &=& \frac{(3\pi^2)^{5/3}}{5\pi^2}\frac{\hbar^2}{2m}n^{2/3}[(1-x)^{5/3}+x^{5/3}]  \\ 
\nonumber \\
\frac{E_m}{A} &=& \frac{(3\pi^2)^{5/3}}{5\pi^2}\left\{p_3[(1-x)^{5/3}+x^{5/3}]  \right. \nonumber \\
               && ~~~~~~~~~~~~\left.+p_5[(1-x)^{8/3}+x^{8/3}]\right\}n^{5/3}e^{-p_4n} \nonumber \\  \\
\frac{E_d}{A} &=&  \frac{1}{n}\left\{g_1[1-(1-2x)^2)]+g_2(1-2x)^2\right\} 
\label{EA2}
\end{eqnarray} 
\subsection*{Pressure}
\begin{eqnarray}
P &=& P_k+P_m+P_d = n^2\frac{\partial \mathcal{H}/n}{\partial n}   \label{P1}\\
P_k &=& \frac{2}{3}n\frac{E_k}{A}   \\
P_m &=& \left(\frac{5}{3}-p_4n\right)n\frac{E_m}{A}  \\
P_{dL} &=& n\left\{\frac{E_{dL}}{A}+f_{1L}[1-(1-2x)^2]  \right. \nonumber \\
    && ~~~~\left.+f_{2L}(1-2x)^2\right\}  \\
P_{dH} &=& P_{dL} + (-\delta g_1 + n \delta f_1)[1-(1-2x)^2] \nonumber \\
       && + (-\delta g_2 + n \delta f_2)(1-2x)^2
\label{P2}
\end{eqnarray}
\subsection*{Incompressibility}
\begin{eqnarray}
K &=& K_k + K_m + K_d = 9\frac{\partial P}{\partial n} \\
K_k &=& 10\frac{E_k}{A}  \\
K_m &=& (40-48p_4n+9p_4^2n^2)\frac{E_m}{A}  \\
K_{dL} &=& 18\frac{E_d}{A} + 9\left\{(4f_1+nh_1)[1-(1-2x)^2]\right. \nonumber \\
    && ~~~~~~~~~~~~~~ \left. +(4f_2+nh_2)(1-2x)^2\right\}  \\
K_{dH} &=& K_{dL} + 9n\left(\delta h_1 [1-(1-2x)^2] \right. \nonumber \\
&&   ~~~~~~~~~~~~~\left.+ \delta h_2 (1-2x)^2\right)
\end{eqnarray}
\subsection*{Second derivative of pressure with respect to density}
\begin{eqnarray}
\frac{d^2P}{dn^2} &=& \frac{d^2P_k}{dn^2}+\frac{d^2P_m}{dn^2}+\frac{d^2P_d}{dn^2}  \\
\frac{d^2P_k}{dn^2} &=& \frac{20}{27}\frac{1}{n}\frac{E_k}{A}  \\
\frac{d^2P_m}{dn^2} &=& \left(\frac{200}{27}-\frac{56}{3}p_4n+p_9^2n^2-p_4^3n^3\right)\frac{1}{n}
                             \frac{E_m}{A}  \\
\frac{d^2P_{dL}}{dn^2} &=& \frac{2}{n}\frac{E_{dL}}{A}+\left(\frac{10f_{1L}}{n}+7h_{1L}+nw_{1L}\right)           [1-(1-2x)^2] \nonumber \\
                     && +\left(\frac{10f_{2L}}{n}+7h_{2L}+nw_{2L}\right)(1-2x)^2  \\
\frac{d^2P_{dH}}{dn^2} &=&\frac{d^2P_{dL}}{dn^2} +(\delta h_1 + n\delta w_1)[1-(1-2x)^2]  \nonumber \\
                       && +(\delta h_2 + n\delta w_2)(1-2x)^2
\end{eqnarray}
\subsection*{Symmetry energy}
\begin{eqnarray}
S_2 &=&S_{2k}+S_{2m}+S_{2d}=\frac{1}{8}\left.\frac{\partial^2\mathcal{H}/n}{\partial x^2}\right|_{x=1/2}\\
S_{2k} &=& \frac{10}{9}\frac{1}{2^{5/3}}\frac{(3\pi^2)^{5/3}}{5\pi^2}\frac{\hbar^2}{2m}n^{2/3}  \\
S_{2m} &=& \frac{10}{9}\frac{1}{2^{5/3}}\frac{(3\pi^2)^{5/3}}{5\pi^2}\frac{\hbar^2}{2m}n^{5/3}
               e^{-p_4n}(p_3+2p_5) \nonumber \\ \\
S_{2d} &=& \frac{1}{n}(-g_1+g_2) 
\end{eqnarray}
\subsection*{First derivative of symmetry energy with respect to density}
\begin{eqnarray}
\frac{dS_2}{dn} &=& \frac{dS_{2k}}{dn}+\frac{dS_{2m}}{dn}+\frac{dS_{2d}}{dn}  \\
\frac{dS_{2k}}{dn} &=& \frac{2}{3}\frac{S_{2k}}{n}  \\
\frac{dS_{2m}}{dn} &=& \frac{S_{2m}}{n}\left(\frac{5}{3}-p_4n\right)  \\
\frac{dS_{2dL}}{dn} &=& \frac{S_{2dL}}{n}+\frac{1}{n}(-f_{1L}+f_{2L}) \\
\frac{dS_{2dH}}{dn} &=& \frac{dS_{2dL}}{dn}+\frac{1}{n^2}(\delta g_1 - \delta g_2) \nonumber \\
                    && -\frac{1}{n}(\delta f_1 - \delta f_2) 
\end{eqnarray}
\subsection*{Second derivative of symmetry energy with respect to density}
\begin{eqnarray}
\frac{d^2S_2}{dn^2} &=& \frac{d^2S_{2k}}{dn^2}+\frac{d^2S_{2m}}{dn^2}+\frac{d^2S_{2d}}{dn^2}  \\
\frac{d^2S_{2k}}{dn^2} &=& -\frac{2}{9}\frac{S_{2k}}{n^2}  \\
\frac{d^2S_{2m}}{dn^2} &=& \frac{S_{2m}}{n^2}\left(\frac{10}{9}-\frac{10}{3}p_4n+p_4^2n^2\right) \\
\frac{d^2S_{2dL}}{dn^2} &=& \frac{1}{n^2}(-2f_{1L}+2f_{2L}-nh_{1L}+nh_{2L})  \\
\frac{d^2S_{2dH}}{dn^2} &=& \frac{d^2S_{2dL}}{dn^2} -\frac{2}{n^3}(\delta g_1 - \delta g_2) \nonumber \\
  && +\frac{2}{n^2}(\delta f_1 - \delta f_2) - \frac{1}{n}(\delta h_1 - \delta h_2)
\end{eqnarray}
\subsection*{Chemical potentials}
\begin{eqnarray}
\mu_i &=& \mu_{ik}+\mu_{im}+\mu_{id} = \frac{\partial \mathcal{H}}{\partial n_i} \label{MU1} \\
\mu_{ik} &=& \frac{5}{3}\frac{(3\pi^2)^{5/3}}{5\pi^2}\frac{\hbar^2}{2m}n_i^{2/3}  \\
\mu_{im} &=& \frac{(3\pi^2)^{5/3}}{5\pi^2}e^{-p_4n} \nonumber \\
         &*& \left\{p_5\left[\frac{8}{3}n_i^{5/3}-p_4\left(n_i^{8/3}+n_j^{8/3}\right)\right]  \right. 
\nonumber \\
&&  \hspace{5pt} +p_3\left[\frac{8}{3}n_i^{5/3}+\frac{5}{3}n_i^{2/3}n_j+n_j^{5/3}\right. \nonumber \\
&& \left.\left. \hspace{23pt}-p_4\left(n_i^{8/3}+n_i^{5/3}n_j+n_in_j^{5/3}+n_j^{8/3}\right)\right]\right\}  
\nonumber \\ \\
\mu_{idL} &=& \frac{1}{n^2}\left[4n_jg_{1L}+4n_in_jf_{1L} \right. \nonumber \\
          &&~~~~~\left.  +2(n_i-n_j)g_{2L}+(n_i-n_j)^2f_{2L}\right]   \\
\mu_{idH} &=& \mu_{idL}-\frac{4}{n^3}n_j(n_i-n_j)(\delta g_1 - \delta g_2) \nonumber \\
&& +\frac{1}{n^2}[4n_in_j\delta f_1 + (n_i-n_j)^2 \delta f_2]
\label{MU2}
\end{eqnarray}
\subsection*{Inverse susceptibilities}
\begin{eqnarray}
\chi_{ii} &=& \chi_{iik}+\chi_{iim}+\chi_{iid}=\frac{\partial \mu_i}{\partial n_i}  
\label{CHI1} \\
\chi_{iik} &=& \frac{2}{3}\frac{\mu_{ik}}{n_i}  \\
\chi_{iim} &=& -p_4\mu_{im} + \frac{(3\pi^2)^{5/3}}{5\pi^2}e^{-p_4n} \nonumber \\
           &*& \left\{
           p_5\left[\frac{40}{9}n_i^{2/3}-\frac{8}{3}p_4n_i^{5/3}\right]   \right.\nonumber \\
&&   +p_3\left[\frac{40}{9}n_i^{2/3}+\frac{10}{9}n_i^{-1/3}n_j \right. \nonumber \\
&& \left.\left. \hspace{13pt}-p_4\left(\frac{8}{3}n_i^{5/3}+\frac{5}{3}n_i^{2/3}n_j+n_j^{5/3}\right)\right]\right\}  \\
\chi_{iidL} &=& \frac{1}{n^2}\left[8n_jf_{1L}+4n_in_jh_{1L} \right. \nonumber \\
           && ~~~~~ \left.+4(n_i-n_j)f_{2L}+(n_i-n_j)^2h_{2L}\right]   \\
\chi_{iidH} &=& \chi_{iidL} + \frac{8}{n^4}n_j(n_i-2n_j)(\delta g_1-\delta g_2) \nonumber  \\
&& - \frac{8}{n^3}n_j(n_i-n_j)(\delta f_1 - \delta f_2)   \nonumber  \\
&& +\frac{4n_in_j}{n^2}\delta h_1 + \frac{(n_i-n_j)^2}{n^2}\delta h_2  \\
\chi_{ij} &=& \chi_{ijk}+\chi_{ijm}+\chi_{ijd}=\frac{\partial \mu_i}{\partial n_j}  \\
\chi_{ijk} &=& 0  \\
\chi_{ijm} &=& -p_4\mu_{im} + \frac{(3\pi^2)^{5/3}}{5\pi^2}e^{-p_4n} \nonumber \\
           &*& \left\{
           -\frac{8}{3}p_4p_5n_j^{5/3} \right.   \nonumber  \\
&& +p_3\left[\frac{5}{3}n_i^{2/3}+\frac{5}{3}n_j^{2/3} \right. \nonumber \\
&&\left.\left.\hspace{13pt}-p_4\left(n_i^{5/3}+\frac{5}{3}n_i^{2/3}n_j
                +\frac{8}{3}n_j^{5/3}\right)\right]\right\}  \\
\chi_{ijdL} &=& \frac{1}{n^2}\left[4g_{1L}+4nf_{1L}+4n_in_jh_{1L} \right. \nonumber \\
           && ~~~~~ \left. -2g_{2L}+(n_i-n_j)^2h_{2L}\right]   \\
\chi_{ijdH} &=& \chi_{ijdL}-\frac{4}{n^4}[(n_i-n_j)^2-2n_in_j](\delta g_1 - \delta g_2) \nonumber \\
&& +\frac{4}{n^3}(n_i-n_j)^2(\delta f_1 - \delta f_2)  \nonumber  \\
&& +\frac{4n_in_j}{n^2}\delta h_1 + \frac{(n_i-n_j)^2}{n^2}\delta h_2  
\label{CHI2}
\end{eqnarray}
\subsection*{Speed of Sound}
\begin{eqnarray}
\left(\frac{c_s}{c}\right)^2 &=& \frac{dP}{d\varepsilon}  \\
                             &=& \frac{1}{(1-x)\mu_n+x\mu_p+m}\frac{K}{9}  \\
                             &=& \frac{n}{(1-x)\mu_n+x\mu_p+m}  \\
&*&\left[\chi_{nn}(1-x)^2+x(1-x)(\chi_{np}+\chi_{pn}) + \chi_{pp}x^2\right] \nonumber
\end{eqnarray}
Here, $\varepsilon$ includes the nucleon rest mass.
\subsection*{Landau effective mass}
\begin{equation}
m_i^* = \left[\frac{1}{m}+\frac{2}{\hbar^2}\left(np_3+n_ip_5\right)e^{-p_4n}\right]^{-1}
\end{equation}
\subsection*{Derivatives of $m_i^*$ with respect to $n$, $x$, $n_i$, and $n_j$}
\begin{eqnarray}
\frac{dm_i^*}{dn} &=& -\frac{m_i^*}{n}\left(1-\frac{m_i^*}{m}\right)(1-np_4)  \\
\frac{dm_i^*}{dx} &=& \pm_{(p)}^{(n)}\frac{2}{\hbar^2}p_5m_i^{*2}ne^{-p_4n}  \\
\frac{dm_i^*}{dn_i} &=&-\frac{2}{\hbar^2}m_i^{*2}\left[p_3(1-np_4)+p_5(1-n_ip_4)\right]e^{-p_4n} \nonumber \\ \\
\frac{dm_i^*}{dn_j} &=&-\frac{2}{\hbar^2}m_i^{*2}\left[p_3(1-np_4)-n_ip_4p_5)\right]e^{-p_4n}  \\
\frac{d^2m_i^*}{dn^2} &=&\frac{m_i^*}{n^2}\left(1-\frac{m_i^*}{m}\right)
                         -\frac{1}{n}\frac{dm_i^*}{dn}(1-np_4)   \\
\frac{d^2m_i^*}{dndn_i} &=&\frac{m_i^*}{n^2}\left(1-\frac{m_i^*}{m}\right)
                         -\frac{1}{n}\frac{dm_i^*}{dn_i}(1-np_4)   \\
\frac{d^2m_i^*}{dndn_j} &=&\frac{m_i^*}{n^2}\left(1-\frac{m_i^*}{m}\right)
                         -\frac{1}{n}\frac{dm_i^*}{dn_j}(1-np_4)  
\end{eqnarray}
\subsection*{Single-particle energy spectrum}
\begin{eqnarray}
\epsilon_{ki} &=& k_i^2T_i + V_i  \\
T_i &=& \frac{\partial\mathcal{H}}{\partial\tau_i}=\frac{\hbar^2}{2m_i^*}  \\
V_i &=& \frac{\partial\mathcal{H}}{\partial n_i}
     = \frac{\partial\mathcal{H}_m}{\partial n_i}+\frac{\partial\mathcal{H}_d}{\partial n_i}  \\
\frac{\partial\mathcal{H}_m}{\partial n_i} &=& 
  \left\{\left[p_3+p_5-p_4(np_3+n_ip_5)\right]\tau_i  \right. \nonumber \\
 && \left.+\left[p_3-p_4(np_3+n_jp_5)\right]\tau_j\right\}e^{-p_4n}  \\
\frac{\partial\mathcal{H}_d}{\partial n_i} &=& \mu_{id}  
\end{eqnarray}
\subsection*{Derivatives of $V_i$ with respect to $n_i$ and $n_j$}
(for use in the finite-T susceptibilities)
\begin{eqnarray}
\frac{\partial V_{im}}{\partial n_i} &=&
\left\{\left[p_3+p_5-p_4(np_3+n_ip_5)\right]\left(\frac{\partial\tau_i}{\partial n_i}-p_4\tau_i\right)
                \right.  \nonumber \\
&&  -p_4(p_3+p_5)\tau_i -p_4p_3\tau_j   \nonumber \\
&& + \left. \left[p_3-p_4(np_3+n_jp_5)\right]\left(\frac{\partial\tau_j}{\partial n_i}-p_4\tau_j\right)
\right\}e^{-p_4n} \nonumber \\  \\
\frac{\partial V_{id}}{\partial n_i} &=& \chi_{iid} \\
\frac{\partial V_{im}}{\partial n_j} &=&
\left\{\left[p_3+p_5-p_4(np_3+n_ip_5)\right]\left(\frac{\partial\tau_i}{\partial n_j}-p_4\tau_i\right)
\right. \nonumber \\
&&  -p_4p_3\tau_i -p_4(p_3+p_5)\tau_j  \nonumber \\
&&\left. +\left[p_3-p_4(np_3+n_jp_5)\right]\left(\frac{\partial\tau_j}{\partial n_j}-p_4\tau_j\right)
\right\}e^{-p_4n}  \nonumber \\  \\
\frac{\partial V_{id}}{\partial n_j} &=& \chi_{ijd}
\end{eqnarray}
\subsection*{Derivatives of $Q_i$ with respect to $n$, $n_i$, and $n_j$}
\ba
\frac{dQ_i}{dn} &=&-\frac{3}{2m_i^*}\left[\frac{dm_i^*}{dn} \right.  \nonumber \\
&& ~~~~~~~~~~\left.-\frac{n}{m_i^*}\left(\frac{dm_i^*}{dn}\right)^2+n\frac{d^2m_i^*}{dn^2}\right]         \\
\frac{dQ_i}{dn_i} &=&-\frac{3}{2m_i^*}\left[\frac{dm_i^*}{dn} \right.  \nonumber \\
&& ~~~~~~~~~~\left.-\frac{n}{m_i^*}\frac{dm_i^*}{dn}\frac{dm_i^*}{dn_i}+n\frac{d^2m_i^*}{dndn_i}\right]    \\
\frac{dQ_i}{dn_j} &=&-\frac{3}{2m_i^*}\left[\frac{dm_i^*}{dn} \right.  \nonumber \\
&& ~~~~~~~~~~\left.-\frac{n}{m_i^*}\frac{dm_i^*}{dn}\frac{dm_i^*}{dn_j}+n\frac{d^2m_i^*}{dndn_j}\right]   
\ea

\section{CONTRIBUTIONS FROM LEPTONS AND PHOTONS}

Charge neutrality requires that the total charge of the protons be exactly cancelled by that of the electrons.  
At $T=0$, this can be stated in terms of the number densities as
$n_p = n_{e^-}$, where the electron (with its 2 spin degrees of freedom) number density $n_{e^-}$ is given by
\be
n_{e^-} = 2\int_0^{k_{Fe^-}}\frac{d^3k}{(2\pi)^3} = \frac {k_{Fe^-}^3}{3\pi^2}
\ee
so that the electron Fermi momentum is $k_{Fe^-} = (3\pi^2n_{e^-})^{1/3}$.
The chemical potential of the electrons is equal to their energy on the Fermi surface:
\be
\mu_{e^-} = \epsilon_{Fe^-} = (k_{Fe^-}^2+m_e^2)^{1/2}.
\ee
Because electromagnetic interactions yield negligible corrections ~\cite{kapusta}, electrons can be 
treated as a free Fermi gas and hence their contributions to the energy density and the pressuse of the 
system are
\ba
\varepsilon_{e^-} &=& 2\int_0^{k_{Fe^-}}\frac{d^3k}{(2\pi)^3}(k^2+m_e^2)^{1/2} \nonumber \\
         &=& \frac{1}{8\pi^2}\left[k_{Fe^-}\epsilon_{Fe^-}(2k_{Fe^-}^2+m_e^2)\right.  \nonumber \\ 
         && \left. +m_e^4\ln\left(\frac{m_e}{k_{Fe^-}+\epsilon_{Fe^-}}\right)\right]  \\
p_{e^-} &=& \frac{2}{3}\int_0^{k_{Fe^-}}\frac{d^3k}{(2\pi)^3}\frac{k^2}{(k^2+m_e^2)^{1/2}} \nonumber \\
         &=& \frac{1}{24\pi^2}\left[k_{Fe^-}\epsilon_{Fe^-}(2k_{Fe^-}^2-3m_e^2)\right.  \nonumber \\ 
         && \left. +3m_e^4\ln\left(\frac{k_{Fe^-}+\epsilon_{Fe^-}}{m_e}\right)\right]  
\ea
At finite $T$, one must consider the net electric charge of electrons and positrons because in 
supernovae temperature rises well above the 1 MeV threshold for $e^-e^+$ pair production. 
Accordingly, the charge neutrality condition becomes
$n_p = n_{e^-} - n_{e^+} \equiv n_e$,  where the net lepton density is given by
\be
n_{e} = 2\int_0^{k_{Fe^-}}\frac{d^3k}{(2\pi)^3}\left[\frac{1}{1+e^{\frac{k-\mu_e}{T}}}
                                  -\frac{1}{1+e^{\frac{k+\mu_e}{T}}}\right]
\label{ne}
\ee
with the chemical potentials of electrons and positrons being equal in magnitude, but  opposite in sign.
In the range of densities and temperatures pertaining to supernovae $\mu_e,~T \gg m_e$ and thus 
the relativistic limit applies:
\ba
\epsilon_k &=& (k^2+m_e)^{1/2} \simeq k\left(1+\frac{m_e^2}{2k^2}\right)  \\
\frac{1}{1+e^{\frac{\epsilon_k\pm\mu_e}{T}}} &\simeq& \frac{1}{1+e^{\frac{k\pm\mu_e}{T}}} \nonumber \\
   &&  \pm \frac{\partial}{\partial \mu_e}\left(\frac{m_e^2}{2k}\frac{1}{1+e^{\frac{k\pm\mu_e}{T}}}\right)
\ea
Then, Eq. (\ref{ne}) can be integrated analytically with the result
\be
n_e = \frac{\mu_e^3}{3\pi^2}\left[1+\mu_e^{-2}(\pi^2T^2-\frac{3}{2}m_e^2)\right]
\ee
which can be solved for the chemical potential
\ba
\mu_e &=& \left(\frac{3\pi^2n_e}{2}\right)^{1/3}   \nonumber \\
      &*& \left\{\left(1-\left[1+\left(\frac{\pi^2T^2}{3}-\frac{m_e^2}{2}\right)^3
                  \left(\frac{2}{3\pi^2n_e}\right)^2\right]^{1/2}\right)^{1/3} \right. \nonumber \\
       &+&   \left. \left(1+\left[1+\left(\frac{\pi^2T^2}{3}-\frac{m_e^2}{2}\right)^3
                  \left(\frac{2}{3\pi^2n_e}\right)^2\right]^{1/2}\right)^{1/3} \right\}  \nonumber \\
\label{muerel}
\ea
The total energy density, total pressure, and total entropy density of the leptons in the relativistic 
regime are 
\ba
\varepsilon_e &=& \varepsilon_{e^-} + \varepsilon_{e^+}   \nonumber  \\
             &=& \frac{\mu_e^4}{4\pi^2}\left[1+\mu_e^{-2}(2\pi^2T^2-m_e^2) \right.  \nonumber \\
             &&  \left. +\pi^2T^2\mu_e^{-4}\left(\frac{7\pi^2T^2}{15}-\frac{m_e^2}{3}\right)\right] 
\label{eerel} \\
p_e &=& p_{e^-} + p_{e^+}  \nonumber   \\
    &=& \frac{\mu_e^4}{12\pi^2}\left[1+\mu_e^{-2}(2\pi^2T^2-3m_e^2) \right.\nonumber \\
    &&    \left. +\pi^2T^2\mu_e^{-4}\left(\frac{7\pi^2T^2}{15}-m_e^2\right)\right]  \label{perel}\\
s_e &=& \frac{\varepsilon_e+p_e-\mu_en_e}{T}  \nonumber \\  
    &=& \frac{\mu_e^2T}{3}\left[1+\mu_e^{-2}\left(\frac{7\pi^2T^2}{15}-\frac{m_e^2}{2}\right)\right]  
\label{serel}     
\ea
In the limit $m_e\rightarrow 0$, $p_e = \frac{1}{3} \varepsilon_e$.                     
The specific heats at constant volume and constant pressure can be obtained by
\ba
C_{Ve} &=& \frac{1}{n_e}\left.\frac{\partial \varepsilon_e}{\partial T}\right|_{n_e}  \nonumber \\
       &=& \frac{1}{n_e}\left(\left.\frac{\partial \varepsilon_e}{\partial \mu_e}\right|_{T}
                             \left.\frac{\partial \mu_e}{\partial T}\right|_{n_e}
                            +\left.\frac{\partial \varepsilon_e}{\partial T}\right|_{\mu_e}\right)  \\
C_{Pe} &=& \left.\frac{\partial}{\partial T}\left(\frac{\varepsilon_e+p_e}{n_e}\right)\right|_{p_e}
        \nonumber \\
       &=& \frac{1}{n_e}\left(\left.\frac{\partial \varepsilon_e}{\partial \mu_e}\right|_{T}
                             \left.\frac{\partial \mu_e}{\partial T}\right|_{p_e}
                            +\left.\frac{\partial \varepsilon_e}{\partial T}\right|_{\mu_e}\right) \nonumber \\
       &-& \frac{(\varepsilon_e+p_e)}{n_e^2}
                     \left(\left.\frac{\partial n_e}{\partial \mu_e}\right|_{T}
                             \left.\frac{\partial \mu_e}{\partial T}\right|_{p_e}
                            +\left.\frac{\partial n_e}{\partial T}\right|_{\mu_e}\right)  
\ea
where
\ba
\left.\frac{\partial \varepsilon_e}{\partial \mu_e}\right|_{T} &=&
      \frac{\mu_e^3}{\pi^2}\left[1+\mu_e^{-2}\left(\pi^2T^2-\frac{m_e^2}{2}\right)\right]  \\
\left.\frac{\partial \mu_e}{\partial T}\right|_{n_e} &=&
     -\frac{2\pi^2T}{3\mu_e\left[1+\pi^2\mu_e^{-2}\left(\frac{T^2}{3}-\frac{m_e^2}{2\pi^2}\right)\right]}
       \\
\left.\frac{\partial \varepsilon_e}{\partial T}\right|_{\mu_e} &=&
       T\mu_e^2\left[1+\mu_e^{-2}\left(\frac{7\pi^2T^2}{15}-\frac{m_e^2}{6}\right)\right]  \\
\left.\frac{\partial \mu_e}{\partial T}\right|_{p_e} &=&
   -\frac{\mu_e^2T}{3\pi^2n_e}\left[
               1+\frac{3\pi^2}{\mu_e^2}\left(\frac{7\pi^2T^2}{15}-\frac{m_e^2}{2}\right)\right]  \\
\left.\frac{\partial n_e}{\partial \mu_e}\right|_{T} &=&
    \frac{\mu_e^2}{\pi^2}\left[1+\pi^2\mu_e^{-2}\left(\frac{T^2}{3}-\frac{m_e^2}{2\pi^2}\right)\right] 
           \\
\left.\frac{\partial n_e}{\partial T}\right|_{\mu_e} &=& \frac{2\mu_eT}{3} .
\ea
Finally, we present the derivatives of the electron chemical potential with respect to the proton and 
neutron number densities. These are essential for our subsequent discussion of the low-to-high-density 
phase transition of $\mathcal{H}_{APR}$ and of our treatment of it by means of a Maxwell construction. 
At $T=0$, we have
\ba
\frac{\partial \mu_e}{\partial n_p} = \frac{k_{Fe^-}^2}{3n_{e^-}\mu_e}  \quad {\rm and} \quad
\frac{\partial \mu_e}{\partial n_n} = 0 \,, 
\ea
whereas at finite temperature $(T>1~\mbox{MeV})$
\ba
\frac{\partial \mu_e}{\partial n_p} = \frac{3\pi^2}{\pi^2T^2-\frac{m^2}{2}+3\mu_e^2}  \quad {\rm and} \quad 
\frac{\partial \mu_e}{\partial n_n} = 0 \,.
\ea
When $T<1$ MeV, numerical evaluation of the relevant FD integrals is required. The numerical methods 
adopted in this work are outlined in Appendix D.  

The contributions from photons are adequately given by the standard blackbody relations for the energy density, 
the pressure, and the entropy density:
\be
\varepsilon_{\gamma} = \frac{\pi^2}{15}\frac{T^4}{(\hbar c)^3}\,,  \quad
p_{\gamma} = \frac{\varepsilon_{\gamma}}{3}\,,  \quad {\rm and} \quad 
s_{\gamma} = \frac{4}{3}\frac{\varepsilon_{\gamma}}{T}, 
\ee
respectively. These remain very small compared to the baryonic and leptonic contributions for all 
temperatures relevant to the supernova problem and, for most practical purposes, can be ignored with 
no repercussions.

\section{NUMERICAL NOTES}

The electronic state variables involve relativistic Fermi-Dirac integrals, the general form of which is
\be
F_{\lambda}(\psi,x) = \int_0^{\infty}\frac{\alpha^{\lambda}\left(\frac{\alpha}{x}+1\right)^{1/2}}
                                             {1+e^{\alpha-\psi}} d\alpha
\ee
where
\ba
x &=& \frac{m_e}{T}  \\
\alpha &=& \frac{(k^2+m_e^2)^{1/2}}{T} + x  \\
\psi &=& \frac{\mu_e-m_e}{T} 
\ea
In particular, the number density, the energy density, and the pressure are given by
\ba
n_e &=& \frac{\sqrt{2}}{\pi^2}T^{5/2}m_e^{1/2}(F_{3/2}+xF_{1/2})  \\
\varepsilon_e &=& \frac{\sqrt{2}}{\pi^2}T^{7/2}m_e^{1/2}(F_{5/2}+2xF_{3/2}+x^2F_{1/2})  \\
p_e &=&  \frac{\sqrt{2}}{3\pi^2}T^{7/2}m_e^{1/2}(F_{5/2}+2xF_{3/2})  
\ea
respectively. \\
We evaluate these quantities numerically, using the JEL method ~\cite{jel} whereby they are expressed 
algebraically in terms of the mass, the temperature, and the chemical potential:
\ba
n_e & = &  \frac{m_e^3}{\pi^2}\frac{fg^{3/2}(1+g)^{3/2}}{(1+f)^{M+1/2}(1+g)^N
                                                           (1+f/a)^{1/2}} \nonumber  \\
      &*&   \sum_{m=0}^M\sum_{n=0}^Np_{mn}f^mg^n
             \left[1+m+\left(\frac{1}{4}+\frac{n}{2}-M\right)\frac{f}{1+f}\right. \nonumber \\
&&~~~~~~+\left.\left(\frac{3}{4}-\frac{N}{2}\right)\frac{fg}{(1+f)(1+g)}\right]  \\
U_e &=& \varepsilon_e-m_en_e  \nonumber \\
    &=& \frac{m_e^4}{\pi^2} \frac{fg^{5/2}(1+g)^{3/2}}{(1+f)^{M+1}(1+g)^N}
              \sum_{m=0}^M\sum_{n=0}^Np_{mn}f^mg^n        \nonumber \\
&&~~~~~\times\left[\frac{3}{2}+n+\left(\frac{3}{2}-N\right)\frac{g}{1+g}\right] \label{eejel} \\
p_e & = & \frac{m_e^4}{\pi^2} \frac{fg^{5/2}(1+g)^{3/2}}{(1+f)^{M+1}(1+g)^N}
              \sum_{m=0}^M\sum_{n=0}^Np_{mn}f^mg^n
\label{pejel}
\ea
where
\ba
\psi  &=&  \frac{\mu_e-m_e}{T}=2(1+f/a)^{1/2}
           \ln\left[\frac{(1+f/a)^{1/2}-1}{(1+f/a)^{1/2}+1}\right]  \nonumber \\ \label{muejel} \\
g  &=&  \frac{T}{m_e}(1+f)^{1/2}\equiv t(1+f)^{1/2}.
\ea
The coefficients $p_{mn}$ for $M=N=3$ and $a=0.433$ are displayed in table \ref{jelpmn}. \\
\begin{table}[h]
\begin{center}
\begin{tabular}{|l|l|l|l|l|}
\hline
$p_{mn}$ & $n=0$ & $n=1$ & $n=2$ & $n=3$ \\
\hline
$m=0$ & 5.34689 & 18.0517 & 21.3422 & 8.53240 \\
$m=1$ & 16.8441 & 55.7051 & 63.6901 & 24.6213 \\
$m=2$ & 17.4708 & 56.3902 & 62.1319 & 23.2602 \\
$m=3$ & 6.07364 & 18.9992 & 20.02285 & 7.11153 \\
\hline
\end{tabular}
\caption[JEL Coefficients]{JEL coefficients $p_{mn}$ for $M=N=3$ and $a=0.433$}
\label{jelpmn}
\end{center}
\end{table}
The entropy density and the free energy density follow from standard thermodynamic relations:
\ba
s_e &=& \frac{1}{T}(\varepsilon_e+p_e-\mu_en_e) \label{se}\\
\mathcal{F}_e &=& \varepsilon_e -Ts_e 
\ea
Furthermore, by taking derivatives of $n_e$, $U_e$, and $p_e$ with respect to $\psi$ and $t$ we can 
get the susceptibilities and the specific heats:
\ba
\left.\frac{\partial \mu_e}{\partial n_p}\right|_{n_n} &=& 
      T\left(\left.\frac{\partial n_e}{\partial \psi}\right|_t
              -t^2\left.\frac{\partial n_e}{\partial t}\right|_{\psi}\right)^{-1}  \\
\left.\frac{\partial \mu_e}{\partial n_n}\right|_{n_p} &=& 0     \\
C_{Ve} &=& \frac{1}{n_em_e}\left(\left.\frac{\partial U_e}{\partial t}\right|_{\psi}
            -\left.\frac{\partial U_e}{\partial \psi}\right|_t
               \frac{\left.\frac{\partial n_e}{\partial t}\right|_{\psi}}
                     {\left.\frac{\partial n_e}{\partial \psi}\right|_t}\right) \label{cve} \\
C_{Pe} &=& \frac{1}{n_em_e}\left(\left.\frac{\partial U_e}{\partial t}\right|_{\psi}
            -\left.\frac{\partial U_e}{\partial \psi}\right|_t
               \frac{\left.\frac{\partial p_e}{\partial t}\right|_{\psi}}
                     {\left.\frac{\partial p_e}{\partial \psi}\right|_t}\right)  \nonumber \\
       &-&\frac{U_e+p_e}{n_e^2m_e}
          \left(\left.\frac{\partial n_e}{\partial t}\right|_{\psi}
            -\left.\frac{\partial n_e}{\partial \psi}\right|_t
               \frac{\left.\frac{\partial p_e}{\partial t}\right|_{\psi}}
                     {\left.\frac{\partial p_e}{\partial \psi}\right|_t}\right)  \label{cpe}
\ea
where
\ba
\left.\frac{\partial}{\partial \psi}\right|_t &=& \frac{f}{1+f/a}
   \left(\left.\frac{\partial}{\partial f}\right|_g
   +\frac{t^2}{2g}\left.\frac{\partial}{\partial g}\right|_f\right)   \\
\left.\frac{\partial}{\partial t}\right|_{\psi} &=& 
         \frac{g}{t}\left.\frac{\partial}{\partial g}\right|_f.
\ea
The non-relativistic Fermi-Dirac integrals 
\ba
F_{\lambda}(\psi) &=& \int_0^{\infty}\frac{x^{\lambda}}{1+e^{x-\psi}}dx \\
x &=& \frac{1}{T}\frac{\hbar^2k^2}{2m^*}, ~~~~\psi = \frac{\mu-V(n)}{T} 
\ea
that are relevant to the thermodynamics of the nucleons are treated by the method developed in \cite{cody}. 
There, three different approximations and corresponding intervals are given for each of $F_{3/2}$, 
$F_{1/2}$, and $F_{-1/2}$:
\ba
F_{\lambda}(\psi) &=& e^{\psi}\left[\Gamma(\lambda+1)+e^{\psi}
       \frac{\displaystyle{\sum_{s=0}^np_se^{s\psi}}}{\displaystyle{\sum_{s=0}^nq_se^{s\psi}}}\right],
       ~~~-\infty<\psi\le1  \nonumber \\ \\
F_{\lambda}(\psi) &=& \frac{\displaystyle{\sum_{s=0}^np_s\psi^s}}{\displaystyle{\sum_{s=0}^nq_s\psi^s}},
    ~~~1\le\psi\le4   \\
F_{\lambda}(\psi) &=& \psi^{\lambda+1}\left[\frac{1}{\lambda+1}+\frac{1}{\psi^2}
        \frac{\displaystyle{\sum_{s=0}^np_s\psi^{-s}}}{\displaystyle{\sum_{s=0}^nq_s\psi^{-2s}}}\right],
   ~~~4\le\psi<\infty  \nonumber \\
\ea
In our code, we have used the coefficients of the $n=4$ case as they appear in \cite{cody}.\\
These integrals have also been computed using the non-relativistic version of the JEL approach:
\ba
F_{3/2} & = & \frac{3f(1+f)^{1/4-M}}{2\sqrt{2}}\sum_{m=0}^Mp_mf^m   \\  
F_{1/2} & = & \frac{f(1+f)^{1/4-M}}{\sqrt{2(1+f/a)}}  \nonumber \\
          &*&  \sum_{m=0}^Mp_mf^m
             \left[1+m-\left(M-\frac{1}{4}\right)\frac{f}{1+f}\right]   \\ 
F_{-1/2} & = & -\frac{f}{a(1+f/a)^{3/2}}F_{1/2} \nonumber \\
         &+& \frac{\sqrt{2}f(1+f)^{1/4-M}}{1+f/a}
         \sum_{m=0}^M p_mf^m \left[(1+m)^2 \right. \nonumber \\
         &-& \left.\left(M-\frac{1}{4}\right)\frac{f}{1+f}
             \left(3+2m-\left[M+\frac{3}{4}\right]\frac{f}{1+f}\right)\right] \nonumber \\
\ea
with 
\be
\psi  =  \frac{\mu-V(n)}{T} 
      = 2(1+f/a)^{1/2}+\mbox{ln}\left[\frac{(1+f/a)^{1/2}-1}{(1+f/a)^{1/2}+1}\right]  
\ee
The coefficients $M$, $a$, and $p_m$ in the above equations are contained in Table VI under the n=0 column. 
The agreement between the two methods is excellent.

\section{CAUSAL EQUATIONS OF STATE}

It is not unusual for equations of state from non-relativstic potential models to become acausal at some 
high density.  Causality is preserved as long as the speed of sound $c_s$ is less than or 
equal to the speed of light $c$. 
In this appendix, we present a thermodynamically consistent method by which 
an EOS based on a non-relativistic potential model can be modified so that 
it remains causal at arbitrary high densities, both at zero temperature and at finite temperature.

\section*{Zero temperature case}

In terms of the pressure $P$ and energy density $\epsilon$ , the condition for an EOS to remain casual is 
\be
\left(\frac{c_s}{c}\right)^2 \equiv \beta = \frac{dP}{d\epsilon}
      = \frac{dP}{dn}\left(\frac{d\epsilon}{dn}\right)^{-1}\le 1 \,.
\label{beta}
\ee
Including the rest-mass energy density $mn$, the total energy density is
\be
\epsilon = \varepsilon + mn\,,
\ee
where $\varepsilon$ is the internal (or specific) energy density of matter.
The pressure and its density derivative are then
\be
P = n\frac{d\varepsilon}{dn}-\varepsilon = n\mu - \varepsilon \qquad 
{\rm and} \qquad
\frac{dP}{dn} = n\frac{d\mu}{dn}\,.
\label{ident}\\
\ee
\noindent We can thus write (\ref{beta}) as a first order differential equation (DE):
\be
\frac{d\mu}{dn} - \frac{\beta}{n}\mu = \frac{\beta m}{n}\,. \label{1DE}
\ee
The integrating factor of Eq. (\ref{1DE}) is given by
\be
f(n) = \exp\left\{-\beta \int\frac{dn}{n}\right\} = n^{-\beta}\,,
\ee
and has the property 
\be
\frac{d}{dn}[n^{-\beta}\mu] = n^{-\beta}~ \frac{\beta m}{n}. \label{ifac}
\ee
Integration of Eq. (\ref{ifac}) leads to 
\be
\mu = \frac{d\varepsilon}{dn} = -m+c_1n^{\beta}\,,
\ee
where $c_1$ is a constant of integration. 
A second integration results in
\be
\varepsilon = -mn + \frac{c_1n^{\beta+1}}{\beta+1} + c_2
\ee
with another constant of integration $c_2$, 
and therefore
\be
P = c_1\frac{\beta}{\beta+1}n^{\beta+1} - c_2\,.
\ee
The integration constants $c_1$ and $c_2$ are determined by the boundary 
conditions
\be
\varepsilon(n_f) = \varepsilon_f 
\qquad {\rm and} \qquad 
P(n_f) = P_f \,,
\label{pf}
\ee
where $n_f$ is the causality 
fixing density, about 0.9-0.95 $n_a$ (at which the EOS becomes acausal), which is chosen such that 
\be
\left.\frac{dP}{d\epsilon}\right|_{n_a} = 1\,,
\ee
and the functional forms of $\varepsilon(n)$ and $P(n)$ are those obtained 
from the original Hamiltonian density.

From Eqs. (\ref{pf}),  we get 
\be
c_1 = \frac{\epsilon_f+P_f}{n_a^{\beta+1}}   \quad {\rm and} \quad
c_2 = \frac{1}{\beta+1}(\beta\epsilon_f-P_f). \label{c12}
\ee 
Thus the energy density and the pressure are given by
\ba
\varepsilon &=& -mn + \frac{(\epsilon_f+P_f)}{\beta+1}
    \left(\frac{n}{n_f}\right)^{\beta+1}
    + \frac{\beta\epsilon_f-P_f}{\beta+1} \label{veps} \\
P &=& \frac{\beta}{\beta+1}(\epsilon_f+P_f)
       \left(\frac{n}{n_f}\right)^{\beta+1}
      -\frac{\beta\epsilon_f-P_f}{\beta+1}. 
      \label{pp}
\ea
Equations (\ref{veps})-(\ref{pp}) can be used for $n \ge n_a$ with $\beta \le 1$ 
so that causality is never violated. Thermodynamic consistency is built-in, 
because Eqs. (\ref{veps})-(\ref{pp}) obey the general identity (\ref{ident}).

\section*{Finite temperature case}

At finite temperature, the causality condition becomes 
\be
\beta = \left.\frac{dP}{d\epsilon}\right|_s
      = \left.\frac{dP}{dn}\right|_s\left(\left.\frac{d\epsilon}{dn}\right|_s\right)^{-1}\le 1 \,.
\label{betat}
\ee
We transform the first term to the variables $n$ and $T$ by the use of Jacobians to get
\be
\left.\frac{dP}{dn}\right|_s = \gamma \left.\frac{dP}{dn}\right|_T   \qquad {\rm with} \qquad
\gamma = \frac{C_P}{C_V}.
\ee  
The second term of (\ref{betat}) can be written as 
\be
\left.\frac{d\epsilon}{dn}\right|_s = \left.\frac{d(\varepsilon+mn)}{dn}\right|_s = \mu + m        
\ee
by employing the identity
\be
\mu = \left.\frac{d\varepsilon}{dn}\right|_s = \left.\frac{d\mathcal{F}}{dn}\right|_T
\ee
where $\mathcal{F}$ is the free energy density.

The pressure and its density derivative at finite temperature change to 
\be
P = n\left.\frac{d\mathcal{F}}{dn}\right|_T-\mathcal{F} = n\mu - \mathcal{F} \qquad 
{\rm and} \qquad
\left.\frac{dP}{dn}\right|_T = n\left.\frac{d\mu}{dn}\right|_T \,.
\ee
Thus the finite-T equivalent of (\ref{1DE}) is:
\be
\left.\frac{d\mu}{dn}\right|_T - \frac{\beta/\gamma}{n}\mu = \frac{\beta/\gamma m}{n}\,
\ee
which leads to (by full analogy with the zero-T case)
\ba
c_1 &=& \frac{\mathcal{F}_f+mn_f+P_f}{n_a^{\beta/\gamma+1}}   \\
c_2 &=& \frac{1}{\beta/\gamma+1}\left[\frac{\beta}{\gamma}(\mathcal{F}_f+mn_f)-P_f\right] \\
\mathcal{F} &=& -mn + \frac{(\mathcal{F}_f+mn_f+P_f)}{\beta/\gamma+1}
    \left(\frac{n}{n_f}\right)^{\beta/\gamma+1} \nonumber \\
    &+& \frac{\beta/\gamma(\mathcal{F}_f+mn_f)-P_f}{\beta/\gamma+1}  \\
P &=& \frac{\beta/\gamma}{\beta/\gamma+1}(\mathcal{F}_f+mn_f+P_f)
       \left(\frac{n}{n_f}\right)^{\beta/\gamma+1}  \nonumber  \\
      &-&\frac{\beta/\gamma(\mathcal{F}_f+mn_f)-P_f}{\beta/\gamma+1}. 
\ea
Note that $\beta$ and $\gamma$ should be evaluated at $n_f$.

\bibliographystyle{h-physrev3}
\bibliography{references}

\end{document}